\documentclass[prc,twocolumn,nofootinbib,superscriptaddress,showpacs]{revtex4}
\usepackage{graphicx,amsmath,amssymb,bm,multirow}
\usepackage{amscd}

\newcommand{\be}[1]{\begin{equation}\label{#1}}
\newcommand{\ee}{\end{equation}}

\newcommand{\fmi}{\, \text{fm}^{-1}}
\newcommand{\fmicube}{\, \text{fm}^{-3}}
\newcommand{\mev}{\, \text{MeV}}

\newcommand{\partialwave}[3]{\ensuremath{{}^{#1}}\textrm{#2}_{#3}}

\begin{document}

\title{Pairing in exotic neutron rich nuclei around the drip line
and in the crust of neutron stars}
\author{A. Pastore}
\email{pastore@ipno.in2p3.fr}
\affiliation{Universit\'e de Lyon, Institut de Physique Nucl\'eaire de Lyon, IN2P3-CNRS, F-69622 Villeurbanne, France}
\affiliation{Institut de Physique Nucl\'eaire, Universit\'e Paris-Sud, IN2P3-CNRS, F-91406 Orsay, France}
\author{J. Margueron}
\affiliation{Institut de Physique Nucl\'eaire, Universit\'e Paris-Sud, IN2P3-CNRS, F-91406 Orsay, France}
\affiliation{Universit\'e de Lyon, Institut de Physique Nucl\'eaire de Lyon, IN2P3-CNRS, F-69622 Villeurbanne, France}
\author{P. Schuck}
\affiliation{Institut de Physique Nucl\'eaire, Universit\'e Paris-Sud, IN2P3-CNRS, F-91406 Orsay, France}
\author{X. Vi\~nas}
\affiliation{Departament de structura i Constituents de la Mat\`eria and Institut de Ci\`encies del Cosmos,
Facultat de Fisica, Universitat de Barcelona, Diagonal 647, E-08028 Barcelona, Spain}

\date{\today}

\begin{abstract}
Exotic and drip-line nuclei as well as nuclei immersed in a low density gas of neutrons in the outer crust 
of neutron stars are systematically investigated with respect to their neutron pairing properties. 
This is done using Skyrme density-functional and different pairing forces such as a density-dependent
contact interaction and a separable form of a finite-range Gogny interaction. 
Hartree-Fock-Bogoliubov and BCS theories are compared.
It is found that neutron pairing is reduced towards the drip line while overcast by strong shell effects. 
Furthermore resonances in the continuum can have an important effect counterbalancing 
the tendency of reduction and leading to a persistence of pairing at the drip line.
It is also shown that in these systems the difference between HFB and BCS 
approaches can be qualitatively large.
\end{abstract}

\pacs{XXXXX}% PACS, the Physics and Astronomy Classification Scheme.

\maketitle

%%%%%%%%%%%%%%%%%%%%%%%%%%%%%%%%%%%%%%%%%%%%%%%%%
%%%%%%%%%%%%%%%%%%%%%%%%%%%%%%%%%%%%%%%%%%%%%%%%%

\section{\label{sec:intro}Introduction}

%%%%%%%%%%%%%%%%%%%%%%%%%%%%%%%%%%%%%%%%%%%%%%%%%
%%%%%%%%%%%%%%%%%%%%%%%%%%%%%%%%%%%%%%%%%%%%%%%%%

Superfluid Fermionic systems, in most of their realizations in physics, are 
either infinite-size and uniform, 
like, e.g., superfluid $^{3,}$He or neutron matter in stars, or confined in a 
finite volume like, e.g., nuclei, cold atoms in traps, or metallic grains.
In the former case, the pairing gap at the Fermi energy is a function of the 
density of matter while
in the latter case, confinement induces a variation of the density on a scale 
which may be smaller or of the order of the
coherence length of the Cooper pairs. 
For example the size of the Cooper pair in nuclei may vary locally by a big factor going from 
the size of the order of the nucleus in the interior to something like 2~fm in the surface 
region~\cite{Pillet2010,Vinas2010,Pastore2011,Pastore2008}. 
In cold atoms, the size of Cooper pairs can be varied with the help of Feshbach 
resonances and one can cover the whole range from the BCS to Bose-Einstein condensation (BEC) regimes.
In nuclear physics, the local density approximation (LDA) is at its very limit 
of applicability because the size of the Cooper pairs is at the best, locally, of the size 
of the surface thickness~\cite{Vinas2010}. Nevertheless  pairing correlations at the surface of 
finite nuclei  may be expected to show some remembrance of what happens in 
infinite nuclear matter~\cite{Margueron2008b,Khan2009,Pastore2011} where the $^1$S$_0$ 
pairing gap is strongly peaked at a density close to $\rho_0/5$, where $\rho_0$ is the 
saturation density of nuclear matter.
Pairing correlations are, therefore, expected to be slightly enhanced at the surface of nuclei. 
In particular, one may naively think that in exotic or
drip line nuclei neutron pairing is enhanced in such situations.
This would be due to the fact that the neutron density extends more smoothly 
out to low density and forms a more or less thick neutron skin which 
may resemble a piece of low density neutron matter. 
This, however, is not the case as we will see in this work. 
The reason lies in the fact that the relation between infinite matter and surface of nuclei is, 
as just mentioned, not one-to-one, because of the long coherence length of the Cooper 
pairs~\cite{Pillet2010}.
As a consequence, finite-nuclei mostly reflect the pairing properties at a density that 
is the average density of the systems~\cite{Khan2010}.
The effect of the change in density on  the pairing gap at the surface of nuclei is still 
difficult to analyze and to pin down from experiment~\cite{Pllumbi2011,Potel2011}. 
This, for instance, concerns the question whether and to what extent the pairing force 
may show a surface peaking~\cite{Sandulescu2005}.

The main objective of the present work is, thus, the study of what happens to neutron 
pairing of nuclei around the neutron drip. 
This also concerns neutron rich nuclei at the frontier of stability, with a large neutron skin,
as well as neutron rich nuclei embedded in a low density neutron gas as it can be in the 
inner crust of neutron stars. 
In this paper we are interested in what happens just at the overflow point where the 
neutron gas is about to appear, or has appeared at a very low density.
Pairing correlations are built there in two rather different systems, the nuclear cluster 
and the shallow superfluid gas. 
Considered alone, these systems are paired, and put together, a mutual effect of pairing 
between the two components of the system might eventually change the pairing properties 
of the whole system.
In cold atomic gases, it is possible to fabricate a trapping potential which goes over from 
a narrow container to a much wider one at a certain energy~\cite{Schuck2011,Viverit2001}.
Filling up this potential with atoms one may also reach drip and overflow situations. 
Our study may, therefore, be of interest for other fields of physics as well.

Since the limit of a nucleus embedded in a vanishing dilute gas is a drip-line 
nucleus, some properties of the latter type of nuclei, such as the existence 
of resonant states in the continuum, can have an important impact on nuclei 
immersed in a low density gas~\cite{Margueron2011}.
On the experimental side, nuclei at the border of stability cannot be created at the
present time, but systematics towards neutron rich systems can be extracted from known 
nuclei masses.
These systematics will therefore also be analyzed in the first part of the work.

In this work an important issue is related to the extrapolation of the pairing properties of
nuclear systems towards the limits of a very dilute external gas of neutrons.
This limit can be obtained either by increasing the size of the box at a fixed number of 
neutrons, or by varying the number of neutrons at a fixed size of the box. 
However, at the end, we will also study realistic Wigner Seitz cell scenarios
in the context of the inner crust of neutron stars, where the neutron gas reaches
non negligible densities.

In addition, we also systematically investigate the mentioned pairing properties 
using two different forces: a density-dependent contact interaction (DDCI) and a 
separable finite-range interaction (SFRI) within the Hartree-Fock-Bogoliubov 
framework which is appropriate for inhomogeneous systems.
The appropriateness of these two pairing interactions will be assessed.
In inhomogeneous systems where the change in density is of the order of the 
coherence length or smaller, DDCI might be at its limit, since the interaction 
depends on the density via a local density approximation.
It has been shown that, given a DDCI that reproduces the gaps 
of SFRI in both symmetric nuclear matter (SNM) and pure neutron matter (PNM),   
the two interactions behave in a similar way also in inhomogeneous systems as  
nuclear clusters in the inner crust of neutron stars~\cite{Pastore2012}.

In the present article, we discuss the signature of the superfluid state comparing two 
different theoretical prescriptions: we take  the pairing gap, in canonical basis, closer to 
the Fermi energy, also called pairing gap of the Lowest Canonical State (LCS) 
$\Delta_{LCS}$~\cite{Lesinski2009}, and the pairing gap $\Delta_{UV}$~\cite{Bender2000}, 
averaged over all the states with the pairing tensor.
We show that in finite systems like nuclei or Wigner-Seitz cells, the pairing gap 
$\Delta_{LCS}$ can be suppressed at overflow while the pairing energy and 
the pairing gap $\Delta_{UV}$ may persists at overflow.
The average pairing gap $\Delta_{UV}$ and the pairing energy can, for certain superfluid 
features, be more appropriate quantities concerning the properties of inhomogeneous systems 
than the pairing gap  $\Delta_{LCS}$.

The paper is organized as follows: in Sec.~\ref{Method} we present the equations we use to 
do our calculations and the methods to solve them; the results concerning nuclei around the 
drip-line are give in Sec.~\ref{beyonddrip}, while in Sec.~\ref{WSsystem} we discuss the 
phenomenon of overflow in the passage outer/inner crust of a neutron star. 
Finally we give our conclusions in Sec.~\ref{conclusion}.

%%%%%%%%%%%%%%%%%%%%%%%%%%%%%%%%%%%%%%%%%%%%%%%%%
%%%%%%%%%%%%%%%%%%%%%%%%%%%%%%%%%%%%%%%%%%%%%%%%%

\section{The Hartree-Fock Bogoliubov (HFB) theory}\label{Method}

%%%%%%%%%%%%%%%%%%%%%%%%%%%%%%%%%%%%%%%%%%%%%%%%%
%%%%%%%%%%%%%%%%%%%%%%%%%%%%%%%%%%%%%%%%%%%%%%%%%

The self-consistent HFB equations, see Eqs.~(\ref{paper:eq:HFBeq}), are solved 
in a box on a spherical mesh with radius $R_{box}$~\cite{Book:Ring1980}.
Using the standard notation $(nlj,q)$ for the spherical single-particle
states with radial quantum number $n$, orbital angular momentum $l$,
total angular momentum $j$, and isospin $q$=n, p, the single-particle wave functions 
$(U,V)^{nlj,q}(r)$, are expanded on a basis of spherical Bessel functions,
\begin{equation}
(U,V)^{nlj,q}(r) = \sum_\alpha (U,V)^{nlj,q}_{\alpha} u_{\alpha,l}(r) ,
\label{eq:uvbasis}
\end{equation}
where $u_{\alpha,l}(r)=C_{\alpha,l}  r j_l(k_{\alpha,l} r)$, $C_{\alpha,l} $ is the normalization 
factor in the box, and $j_l$ are Bessel functions of the first kind with integer index $l$.
The index $\alpha$ runs over a set of zeros of the Bessel function $j_l(k_{\alpha,l} R_{box})$, going
from the lower value $k_{\alpha\equiv1,l}$ up to the momentum cutoff $k_{max}=4\fmi$. 
This corresponds to an HFB model-space energy cutoff of about 
$\hbar^2k_{max}^2/2m \approx 320\mev$ (see Ref.~\cite{Thesis:Lesinski2008} and references 
therein for more details).

We use a Skyrme functional to build the single-particle Hamiltonian $h$
and then we let the particles interact pairwise in the pairing channel.
The two-body matrix elements of the pairing interaction in the $J=0$, $T=1$ channel enter the 
neutron-neutron and proton-proton gap equations, whose solutions provide the matrix elements
of the state-dependent gap matrix $\Delta$.
The latter, in turn, enters the HFB equations,
\begin{eqnarray}
  \sum_{\alpha'}(h_{\alpha'\alpha}^{lj,q}- \mu_{F}^{q})U^{nlj,q}_{\alpha'}+\sum_{\alpha'}\Delta_{\alpha \alpha '}^{lj,q}V^{nlj,q}_{\alpha'}&=&E^{nlj,q}U^{nlj,q}_{\alpha} ,
     \nonumber \\
  \sum_{\alpha'}\Delta^{lj,q}_{\alpha \alpha'}U^{nlj,q}_{\alpha'} -\sum_{\alpha'}(h^{lj,q}_{\alpha'\alpha}- \mu_{F}^{q})V^{nlj,q}_{\alpha'} &=&E^{nlj,q}V^{nlj,q}_{\alpha} ,
      \nonumber \\
\label{paper:eq:HFBeq}
\end{eqnarray}
where $\mu_{F}^{q}$ is the chemical potential and $U^{nlj,q}_{\alpha}$ and $V^{nlj,q}_{\alpha}$ are 
the Bogoliubov amplitudes for the quasiparticle of energy $E^{nlj,q}$.
For the pairing channel we used two pairing interactions: a density dependent contact force
and a finite range  interaction in its separable approximation:
\begin{itemize}
\item[(i)]
The two-body Density-Dependent Contact Interaction (DDCI) 
between particles at positions $\mathbf{r_1}$ and $\mathbf{r_2}$ 
reads~\cite{Bertsch1991,Garrido1999}
\begin{eqnarray}
\label{pairing_int_contact}
\qquad \quad v(\mathbf{r}_{1},\mathbf{r}_{2})=V_{0}\left[ 1- \eta \left( \frac{\rho_b\left(
\mathbf{R}\right)}{\rho_0}\right)^{\alpha}\right]
\delta(\mathbf{r}), & & \nonumber \\
& &
\end{eqnarray} 
$\mathbf{R}=(\mathbf{r_1}+\mathbf{r_2})/2$ is the center of mass of the two
interacting particles and $\mathbf{r}=\mathbf{r_1}-\mathbf{r_2}$ is their mutual distance.
We choose $V_{0}=-530.0$ MeV fm$^3$, $\eta=0.7$, $\alpha=0.45$, $\rho_0=0.16 \fmicube$~\cite{Grill2011}.
We use a smooth cut-off acting in quasiparticle space at  $E^{a}\ge20$ MeV, that is  defined by an 
gaussian factor $\exp(-(E^{a}-20)^{2}/100)$, where $a$ is a shorthand notation for $a=(nlj,q)$.
The parameters of this interaction are adjusted such that it mimics the gaps obtained using  a Gogny force in SNM.
\item[(ii)]
A Separable Finite-Range pairing Interaction (SFRI)~\cite{Duguet2004}
that reproduces the $\partialwave{1}{S}{0}$ Gogny D1S pairing gap at the Fermi surface in infinite 
nuclear matter (INM)~\cite{Tian2009},
\begin{eqnarray}
\label{pairing_int_gogny}
v(\mathbf{r}_1 ,\mathbf{r}_2 ,\mathbf{r}_1 ',\mathbf{r}_2 ') = 
 \gamma P(r)P(r')\delta (\mathbf{R} - \mathbf{R}')
       \frac{1}{2}(1-P^\sigma). \nonumber \\
\end{eqnarray}
The operator $\frac{1}{2}(1-P^\sigma)$ restricts the interaction to total spin $S=0$.
Strength and form factor are $\gamma=-728\mev\fmicube$ and 
$P(r)= 1/(4\pi b^2 )^{3/2} \exp ( - r^2 /(4b^2 ))$, where $b=0.644$.
The finite-range interaction in the pairing channel is added on top of a single-particle spectrum obtained
with Skyrme interaction (see for instance Ref.~\cite{Chabanat1998a}).
A slight correction of the strength of the SFRI is therefore necessary and we fix 
$\gamma\rightarrow 0.9\gamma$~\cite{Tian2009}.
\end{itemize}

Since we are interested in  systems at or beyond the neutron drip ($i.e.$ a nucleus surrounded by a gas), 
we consider the same boundary conditions used in the calculation of Wigner Size cells in the 
inner crust of neutron stars.
We thus  impose the following Dirichlet-Neumann mixed boundary conditions
\cite{Pastore2011}: (i) even-parity wave functions vanish at $R=R_{box}$; (ii) the first derivative
of odd-parity wave functions vanishes at $R=R_{box}$.
When presenting the results for the HFB neutron pairing gap, we
use two different definitions for the pairing gap.
The first one, $\Delta_{LCS}$, is defined as the diagonal pairing matrix element 
corresponding to the canonical single-particle state~\cite{Lesinski2009}, whose quasi-particle energy,
\begin{equation}
\label{gapLCS}
E^{a}=\sqrt{\left( \varepsilon^{a}-\mu^{q}_{F} \right)^{2}+\left(\Delta^{a}\right)^{2}},
\end{equation}
is the lowest. Here $\varepsilon^{a}$ stands for the diagonal matrix element of the 
single-particle field $h^{lj,q}$ in canonical basis and $\Delta^{a}$ the corresponding diagonal 
pairing-field matrix element.
 
The second definition of the pairing gap, $\Delta_{UV}^q$, is related to the average 
of the state dependent gaps over the pairing tensor, i.e.
\begin{equation}
\Delta_{UV}^q=\frac{\sum_{nlj} (2j+1) \sum_{\alpha} U^{nlj,q}_{\alpha}  \Delta^{lj,q}_{\alpha} V^{nlj,q}_{\alpha}}{\sum_{nlj}(2j+1) \sum_{\alpha}U^{nlj,q}_{\alpha}V^{nlj,q}_{\alpha}}.
\label{paper:eq:uvdelta}
\end{equation}
%To simplify the notations and since we will present results only for neutrons, we will omit the exponent $q=n$
%in the following. 

The pairing energy is defined in terms of the pairing tensor as,
\begin{equation}
E^{q}_{pair}=\frac{1}{2}\sum_{nlj} (2j+1)\sum_{\alpha\alpha'}U^{nlj,q}_{\alpha} \Delta^{lj,q}_{\alpha \alpha'} V^{nlj,q}_{\alpha'}.
\label{paper:eq:pe}
\end{equation}

\begin{figure*}
\begin{center}
\includegraphics[clip=,width=0.38\textwidth,angle=-90]{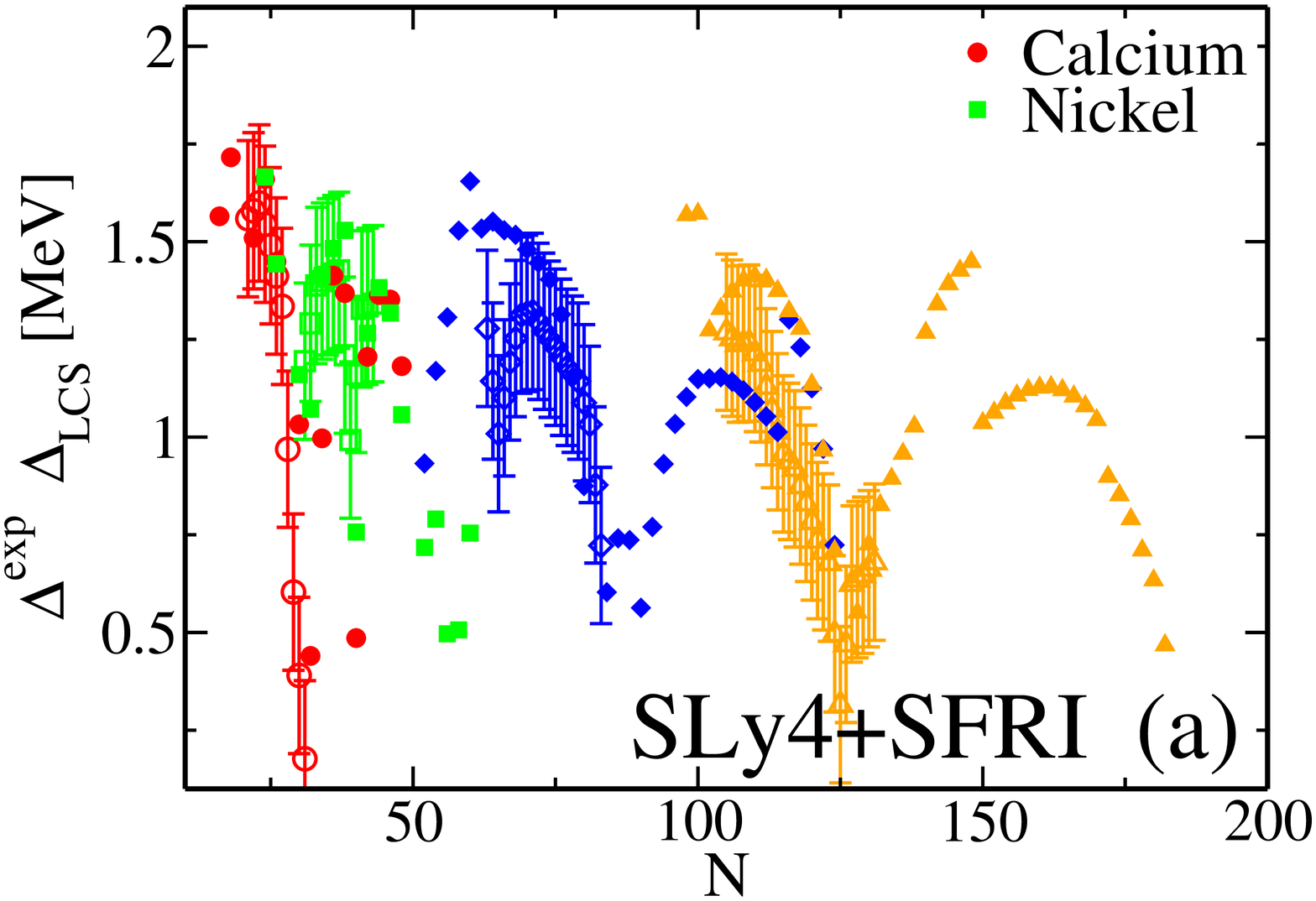}
\includegraphics[clip=,width=0.38\textwidth,angle=-90]{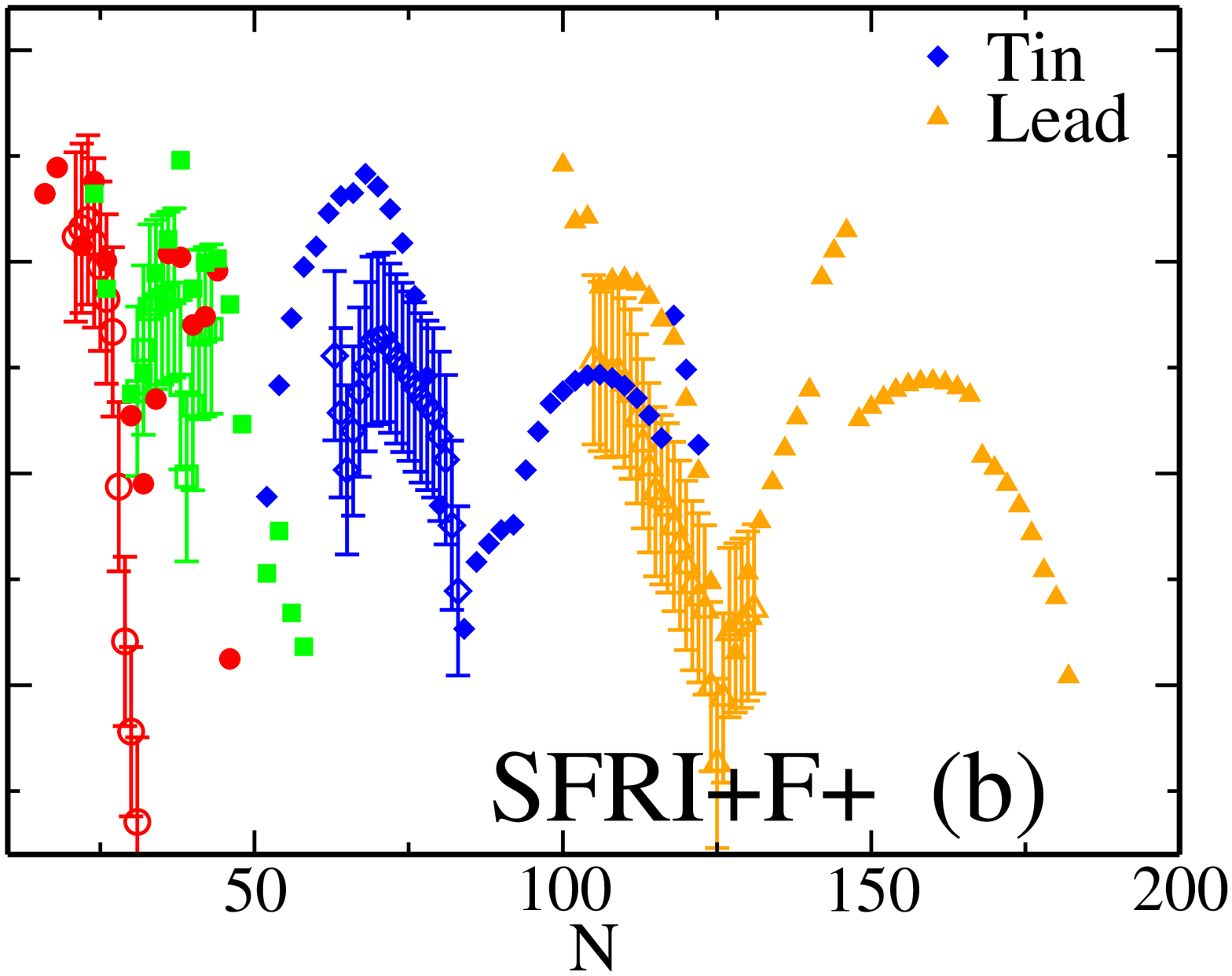}\\
\includegraphics[clip=,width=0.38\textwidth,angle=-90]{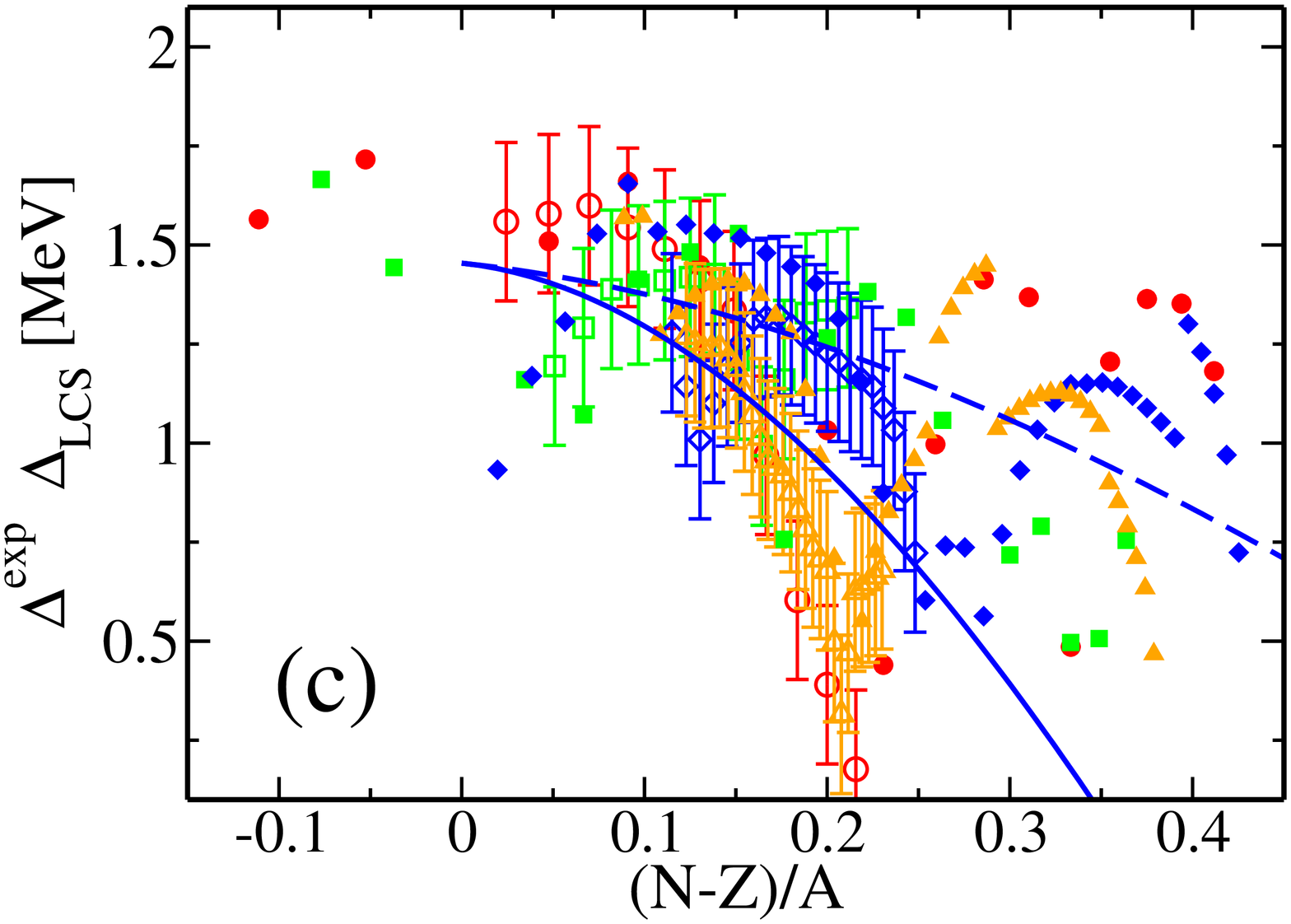}
\includegraphics[clip=,width=0.38\textwidth,angle=-90]{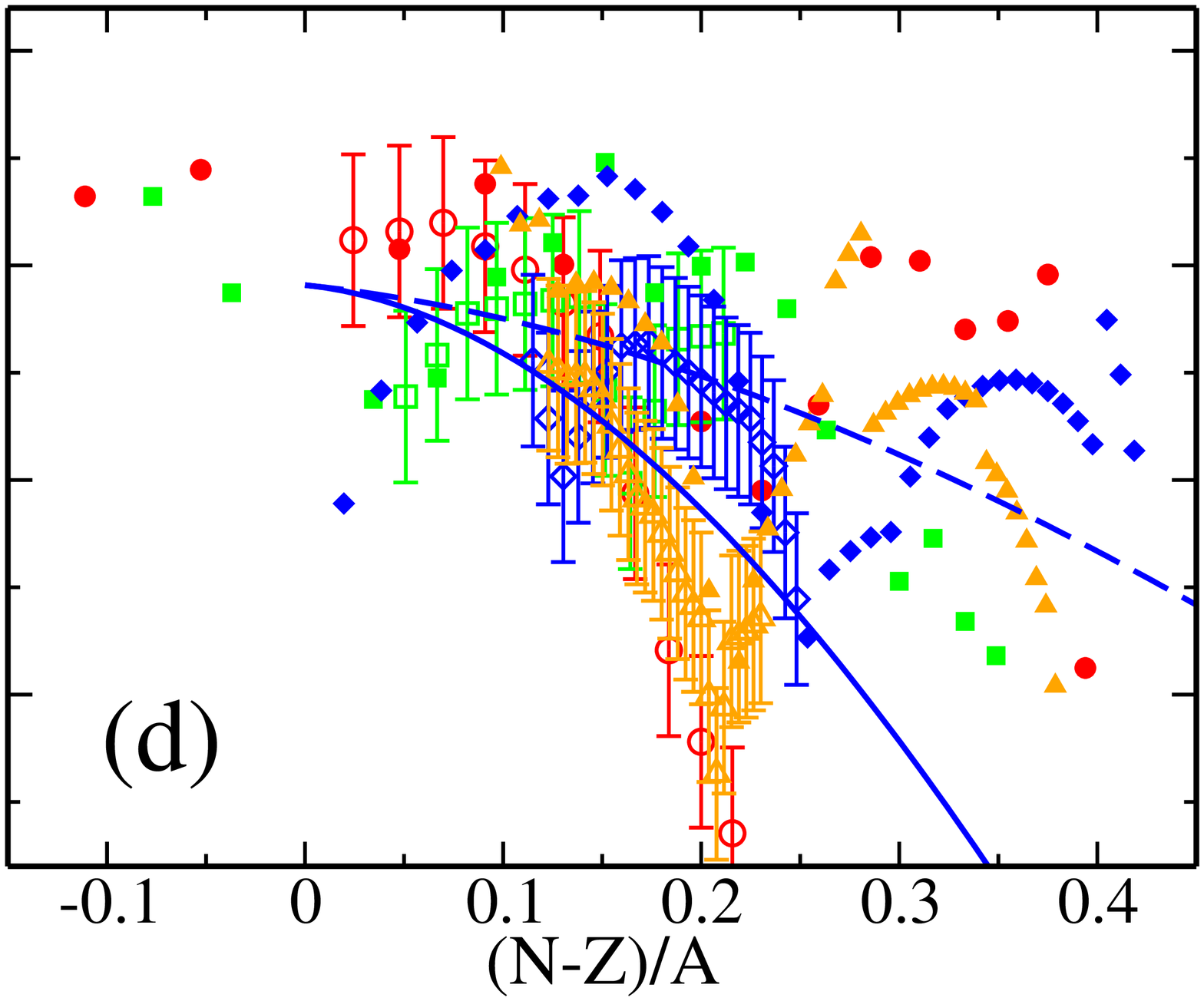}
\end{center}
\caption{ (Colors online) Graphical representation of experimental gaps for isotopes  obtained 
with a 3-point filter and the theoretical gaps $\Delta_{\text{LCS}}$ using the SLy4 (left panel) or 
F+ (right panel) Skyrme force and same separable pairing interaction. 
It should be underlined that the experimental error is obtained simply by the uncertainty of 
masses plus a constant value of 200 KeV that is due to the limits of the theoretical model
(see text for more details). 
In panels (c) and (d) the solid line corresponds to Eq.~(\ref{paper:eq:yamagami}), while the dashed line 
represents Eq.~(\ref{paper:eq:newfit}), both lines been drawn for Sn isotopes.}
\label{paper:fig:ExpDelta}
\end{figure*}

A comparison of the pairing gaps $\Delta_{LCS}$ and $\Delta_{UV}$ with experimental value
can be made considering some limitations due to other effects, such as the one induced
by the time-odd term of the Hamiltonian.
The experimental gap for odd nuclei can, however, be deduced from the binding energies using a three-point 
formula centered on the odd nucleus, 
see the discussion in Ref.~\cite{Duguet2001b} and references therein,
as
\begin{equation}
\Delta^{exp}_{odd}(N)=\frac{1}{2}\left[ E_{b}(N+1)-2E_{b}(N)+E_{b}(N-1)\right] ,
\label{paper:eq:deltaodd}
\end{equation}
where $E_{b}$ is the binding energy taken from Audi's database~\cite{Audi2003a}.
The experimental pairing gap for even nuclei are deduced from the average of the three-point 
formula~(\ref{paper:eq:deltaodd}) applied to the two closest even nuclei as~\cite{Dobaczewski2002b},
\begin{equation}
\Delta^{exp}_{even}(N)=\frac{1}{2}\left[ \Delta^{exp}_{odd}(N-1) + \Delta^{exp}_{odd}(N+1) \right].
\label{paper:eq:delta even}
\end{equation}

Definitions (\ref{paper:eq:deltaodd}) and (\ref{paper:eq:delta even}) are closer to 
$\Delta_{LCS}$~\cite{Lesinski2012} than to $\Delta_{UV}$ and, therefore,  
a comparison of the pairing gap $\Delta_{LCS}$ and the experimental gap~(\ref{paper:eq:delta even}) 
is shown in Fig.~\ref{paper:fig:ExpDelta} for calcium, nickel, tin and lead isotopes.
The error bars on the experimental gaps are estimated to be $\pm$200~keV. 
This takes into account a small contribution coming from the experimental error bars on the masses, and 
a large contribution due to other contributions than the pairing gap on the 
experimental
gap, such as for instance, the time odd terms in the mean 
field~\cite{Bertsch2009,Schunck2010,Margueron2009a}, the use
of the 3-point formula~\cite{Bender2000,Dobaczewski2001b}, 3-body terms in the pairing 
channel~\cite{Hebeler2009,Lesinski2012}, and the particle-vibration 
coupling~\cite{Barranco1999,Pastore2008,Pastore2009}.

On the left side of Fig.~\ref{paper:fig:ExpDelta} are shown the theoretical predictions based on SLy4~\cite{Chabanat1997,Chabanat1998a,Chabanat1998b} Skyrme interaction and SFRI in
the pairing channel while on the right, the results are obtained with the F+ interaction ~\cite{Lesinski2006} and SFRI.
The evolution with the neutron number $N$ of the experimental gap~(\ref{paper:eq:delta even}) 
shown in Fig.~\ref{paper:fig:ExpDelta}  (top panels) is well reproduced by the HFB model with both 
the SLy4 or F+ interaction in the mean-field and same interaction SFRI in the pairing channel.
The theoretical calculations have been performed up to the drip line, beyond the domain where
experimental information are known. 
The limitations of the experimental data is quite visible in Fig.~\ref{paper:fig:ExpDelta}  (bottom panels)
where they are represented versus the asymmetry parameter $(N-Z)/A$.
The experimental data hardly reach $(N-Z)/A\approx 0.25$.
It is interesting to notice that at the edge of the experimental data the gaps (experimental and theoretical) 
tend to decrease with an important slope in the asymmetry direction $(N-Z)/A$, while beyond, the theoretical
gaps go up again for larger $(N-Z)/A \ge 0.3$.
The decrease of the experimental gaps~(\ref{paper:eq:delta even}) is mostly due to the shell closure
at the boundary of experimental measurements.
As a consequence, the asymmetry dependence of the experimental pairing gap, which has been fitted as
\begin{equation}
\Delta=\left[ 1-7.74\left( \frac{N-Z}{A}\right)^{2}\right]\frac{6.75}{A^{1/3}}
\label{paper:eq:yamagami}
\end{equation}
in the work of  Ref.~\cite{yamagami2008,Yamagami2012}, as well as in the former work of Ref.~\cite{Vogel1984}, tends to 
overestimate the asymmetry dependence of the pairing gap.
This is clearly illustrated in Figs.~\ref{paper:fig:ExpDelta}(c) and \ref{paper:fig:ExpDelta}(d) where the 
fit~(\ref{paper:eq:yamagami}) is shown for Sn isotopes, $i.e.$ Z=50 (solid line).
Completing the unknown experimental data for large $(N-Z)/A$ by the theoretical calculations shown
in the upper panels of Fig.~\ref{paper:fig:ExpDelta}, we can obtain a new set of parameters
\begin{equation}
\Delta=\left[ 1-2 \left( \frac{N-Z}{A}\right)^{2}\right]\frac{6.75}{A^{1/3}} ,
\label{paper:eq:newfit}
\end{equation}
where the coefficient in front of the asymmetry is lower than in Eq.~(\ref{paper:eq:yamagami}).
The fit~(\ref{paper:eq:newfit}) is also shown for Sn isotopes in 
Figs~\ref{paper:fig:ExpDelta}(c) and \ref{paper:fig:ExpDelta}(d) 
(dashed lines).

%and calculated for Sn isotopes, are shown in Fig.~\ref{paper:fig:ExpDelta}(c). 
%The average pairing calculated from Thomas-Fermi BCS approximation is systematically smaller, by
%about 40\% to 60\% compared to the gaps obtained from quantal HFB.
%This feature is mostly due to the BCS approximation which neglects non-diagonal pairing couplings.
%It is a known feature in nuclei~\cite{Dobaczewski1996} and it will be discussed for dilute clusters in 
%the following.
%{\bf May be is better The average pairing gaps obtained from a Thomas-Fermi BCS approximation, see Appendix \ref{app:TF},
%and calculated for Sn isotopes, are shown in Fig.~\ref{paper:fig:ExpDelta}(c).
%The average pairing calculated from Thomas-Fermi BCS approximation shows as a function of the neutron-proton
%asymmetry a similar downards trend as Eq.(\ref{paper:eq:yamagami} as it can be seen in Fig.2 of Ref.
%[X.Vi\~nas, P.Schuck and M.Farine, Int.J.Mod.Phys.E20,399(2011).]}
\begin{figure*}
\begin{center}
\includegraphics[clip=,width=0.3\textwidth,angle=-90]{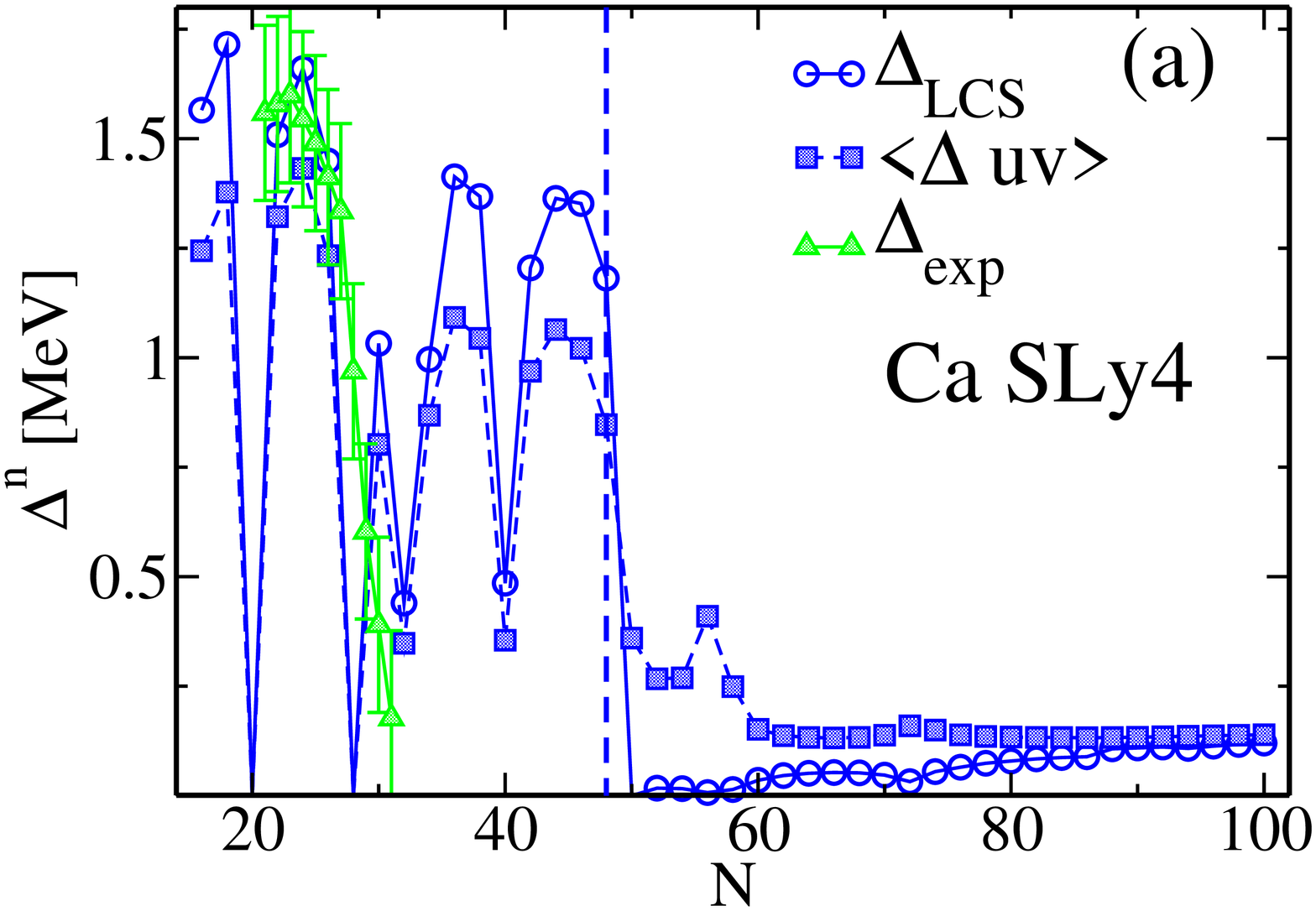}
\includegraphics[clip=,width=0.3\textwidth,angle=-90]{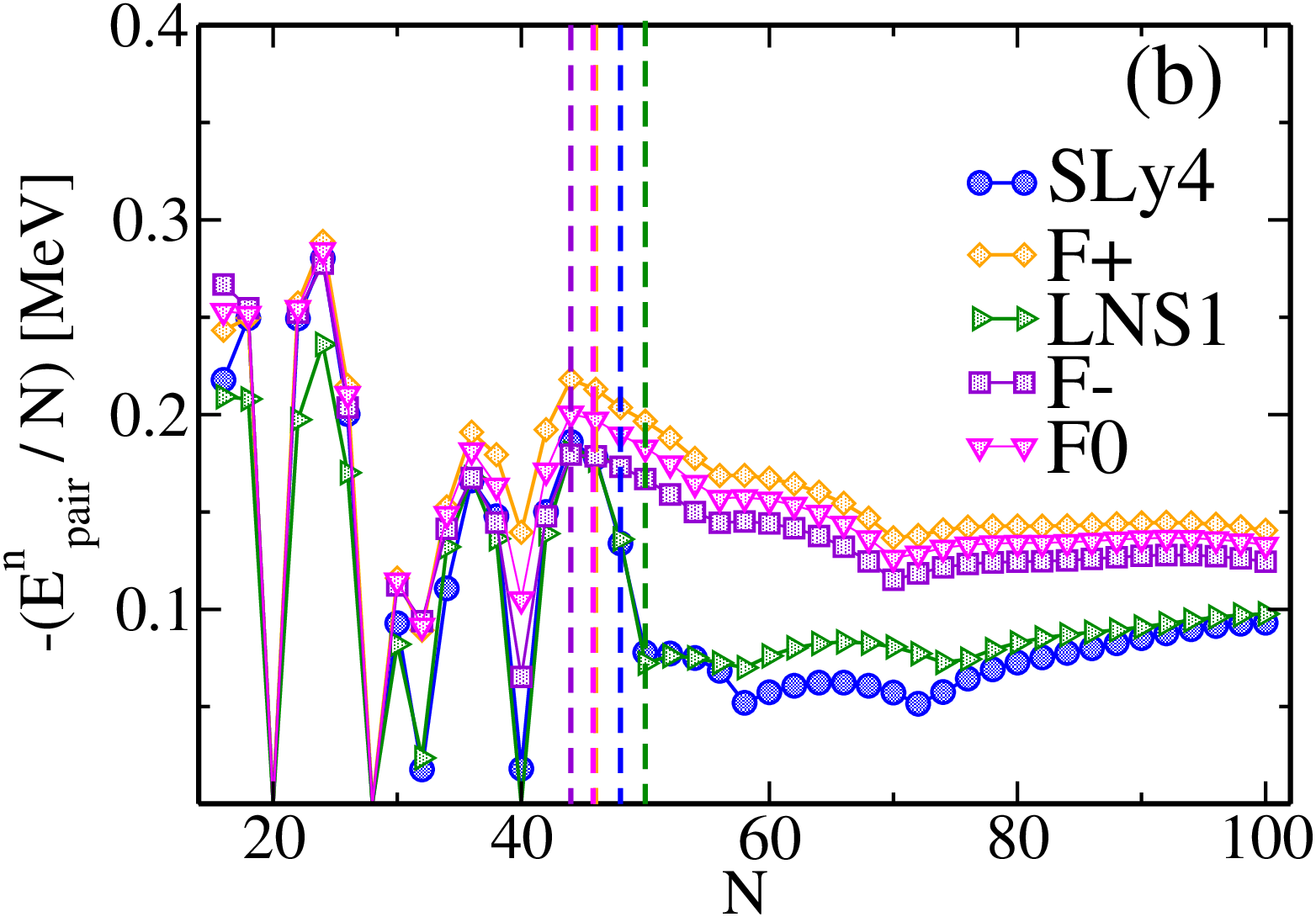}\\
\includegraphics[clip=,width=0.3\textwidth,angle=-90]{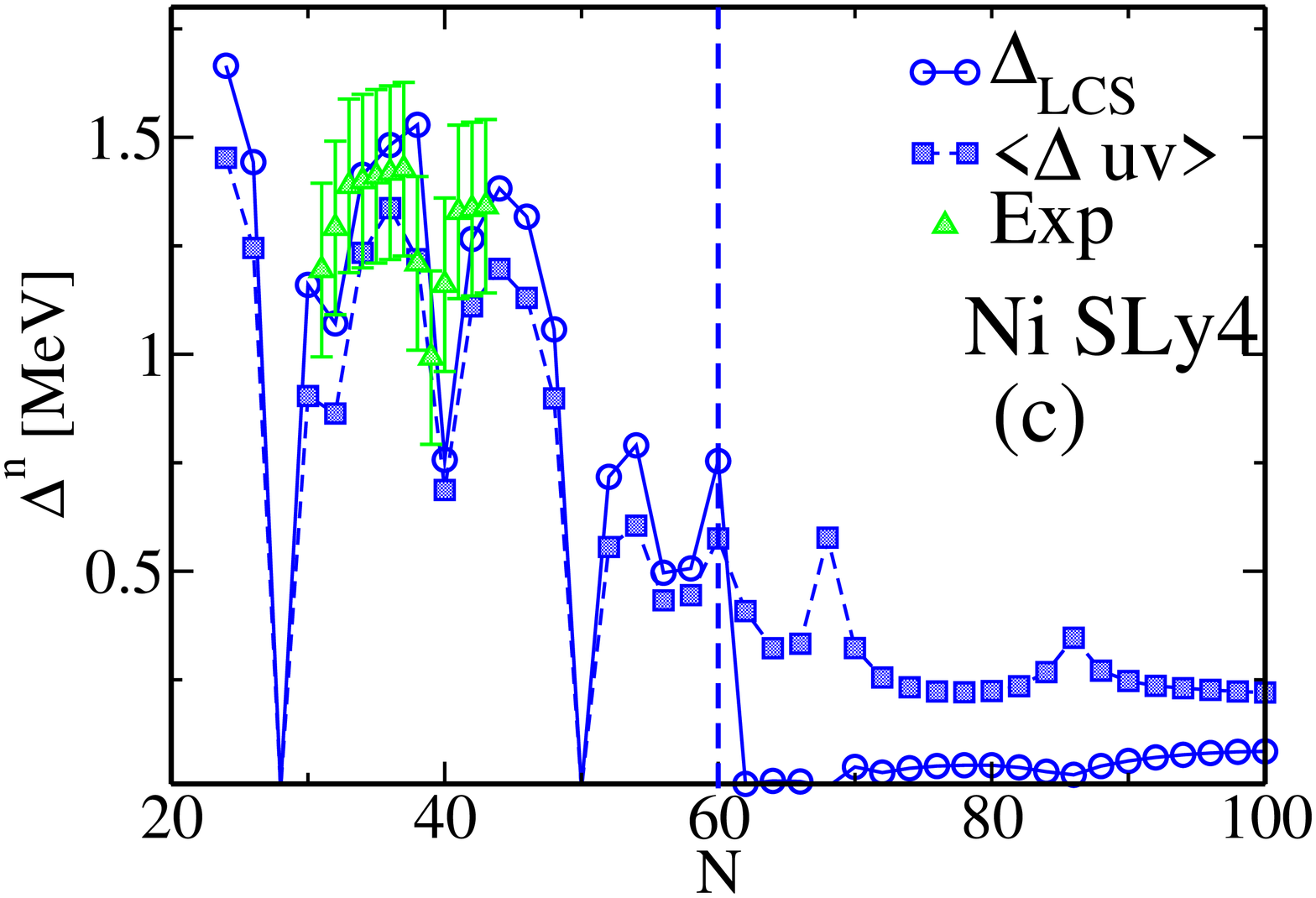}
\includegraphics[clip=,width=0.3\textwidth,angle=-90]{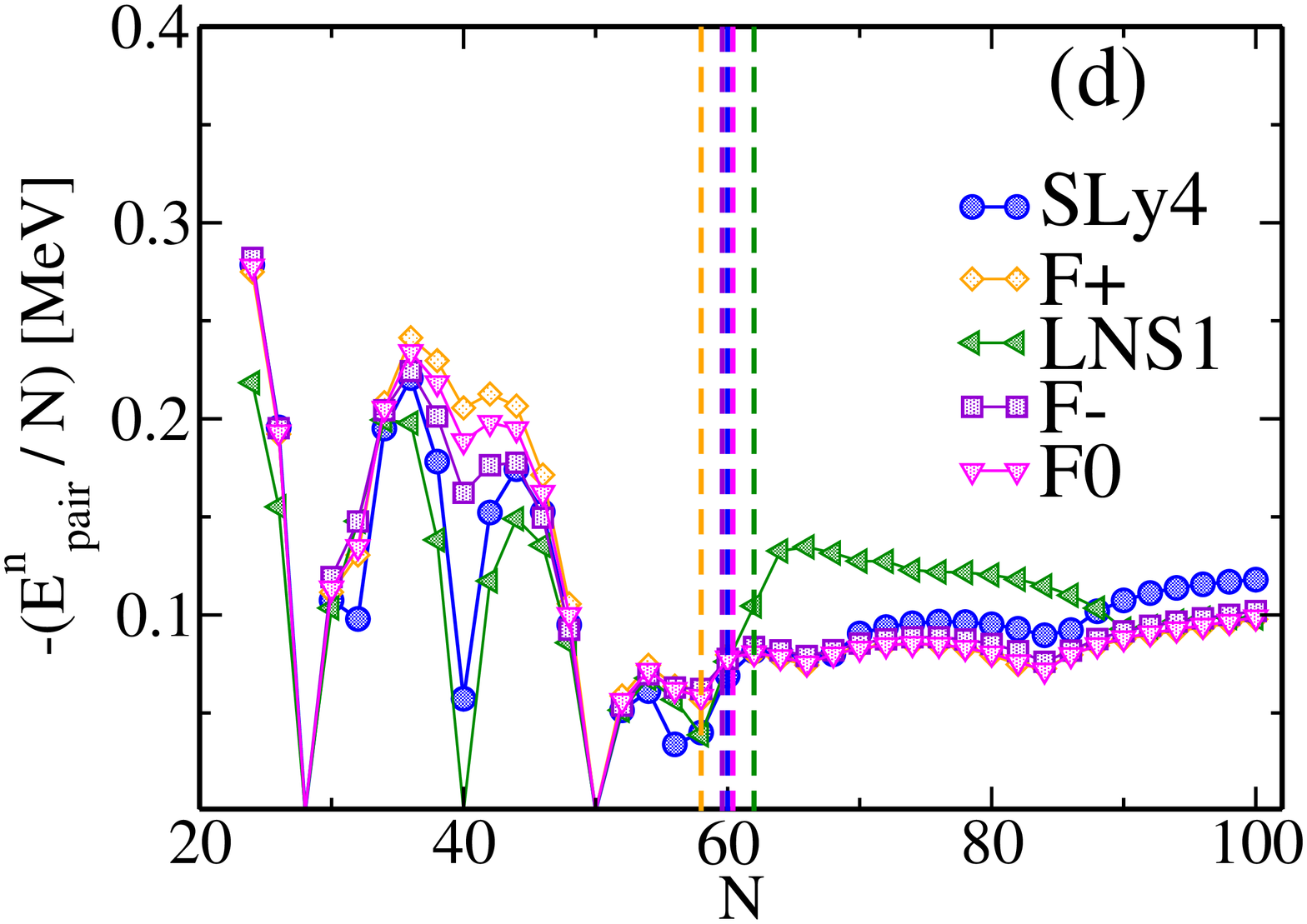}\\
\includegraphics[clip=,width=0.3\textwidth,angle=-90]{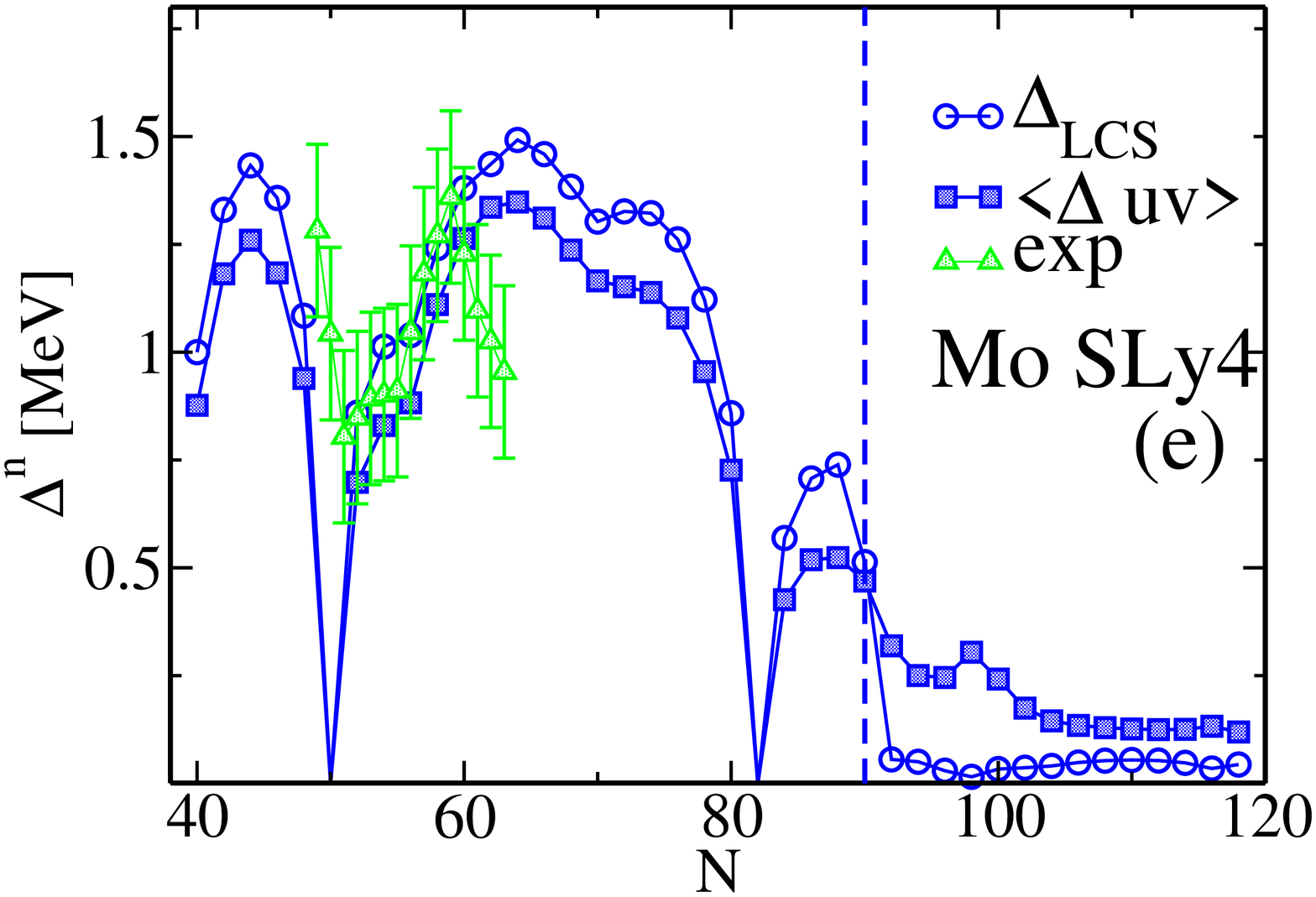}
\includegraphics[clip=,width=0.3\textwidth,angle=-90]{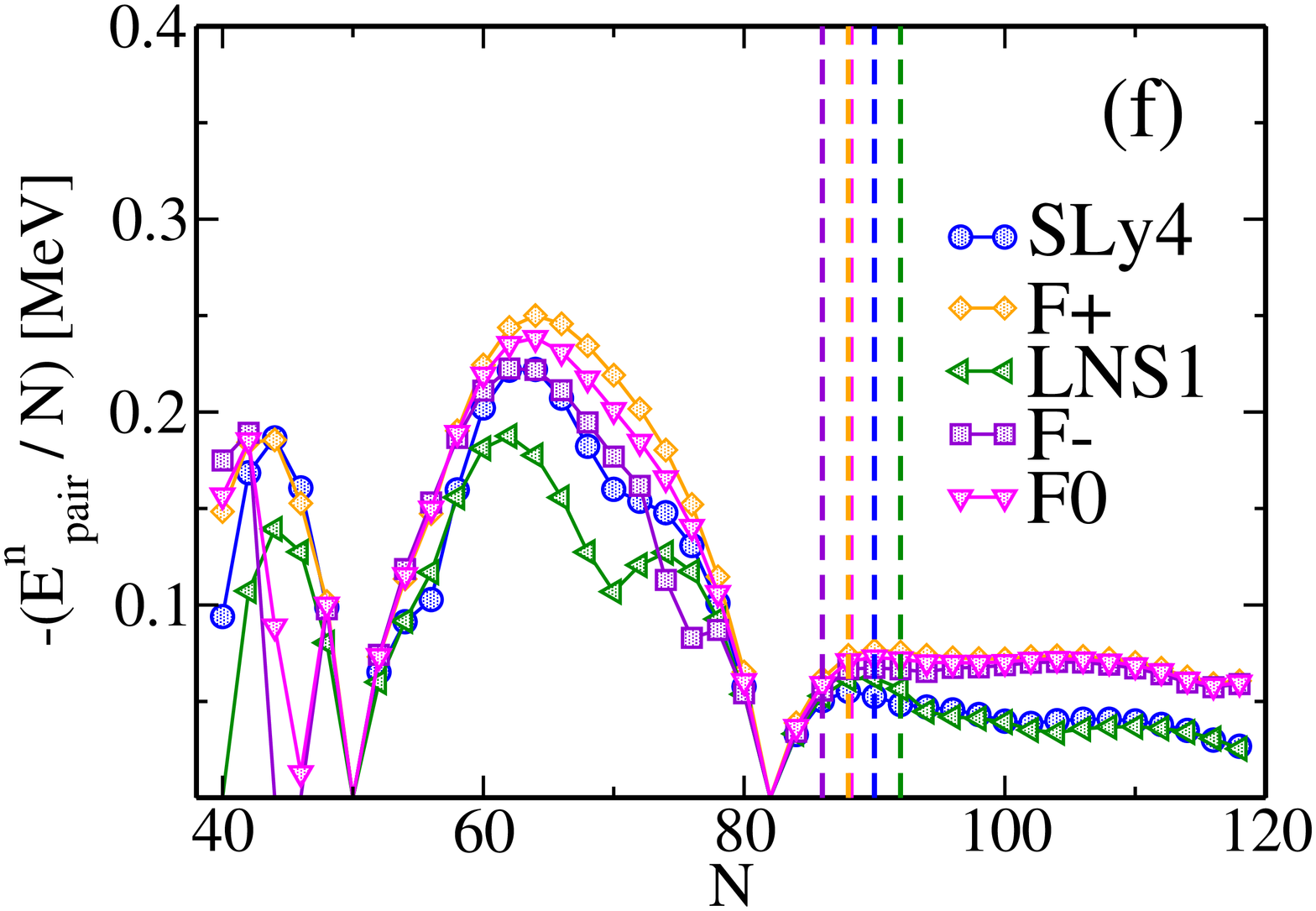}\\
\end{center}
\caption{(Colors online) The evolution of pairing properties is shown along different isotopic chains: 
from top to bottom Calcium, Nickel and Molybdenum isotopes are represented.
On the left are compared the two definitions of pairing gaps for $\Delta_{LCS}$ and $\Delta_{UV}$ 
using Skyrme SLy4 interaction for the $ph$ channel and the SFRI~(\ref{pairing_int_gogny}) for the pairing channel. 
On the right, the pairing energy per neutron~(\ref{paper:eq:pe}) is compared using different interactions 
in the $ph$ channel such as F+, F0, F-, LNS1 and SLy4.
The vertical dashed line stands for the neutron drip-line nucleus using the same color as the associated
interaction.}
\label{paper:fig:IsotopesGognyRed}
\end{figure*}

%%%%%%%%%%%%%%%%%%%%%%%%%%%%%%%%%%%%%%%%%%%%%%%%%
%%%%%%%%%%%%%%%%%%%%%%%%%%%%%%%%%%%%%%%%%%%%%%%%%

\section{Nuclear systems beyond the drip-line}\label{beyonddrip}

%%%%%%%%%%%%%%%%%%%%%%%%%%%%%%%%%%%%%%%%%%%%%%%%%
%%%%%%%%%%%%%%%%%%%%%%%%%%%%%%%%%%%%%%%%%%%%%%%%%

In the following, the predictions for overflowing nuclear systems based on different pairing forces are analyzed.

\begin{figure*}
\begin{center}
\includegraphics[clip=,width=0.3\textwidth,angle=-90]{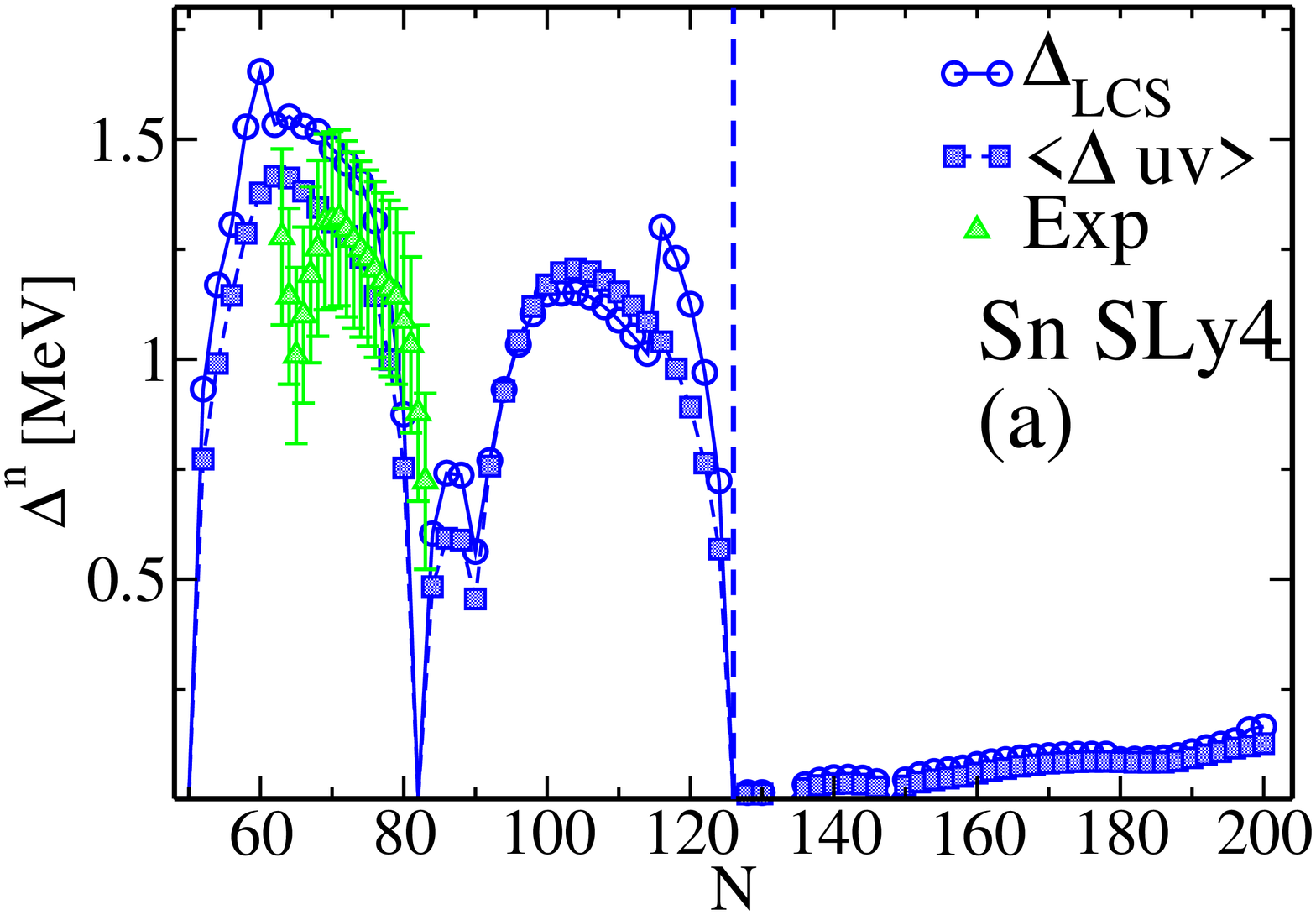}
\includegraphics[clip=,width=0.3\textwidth,angle=-90]{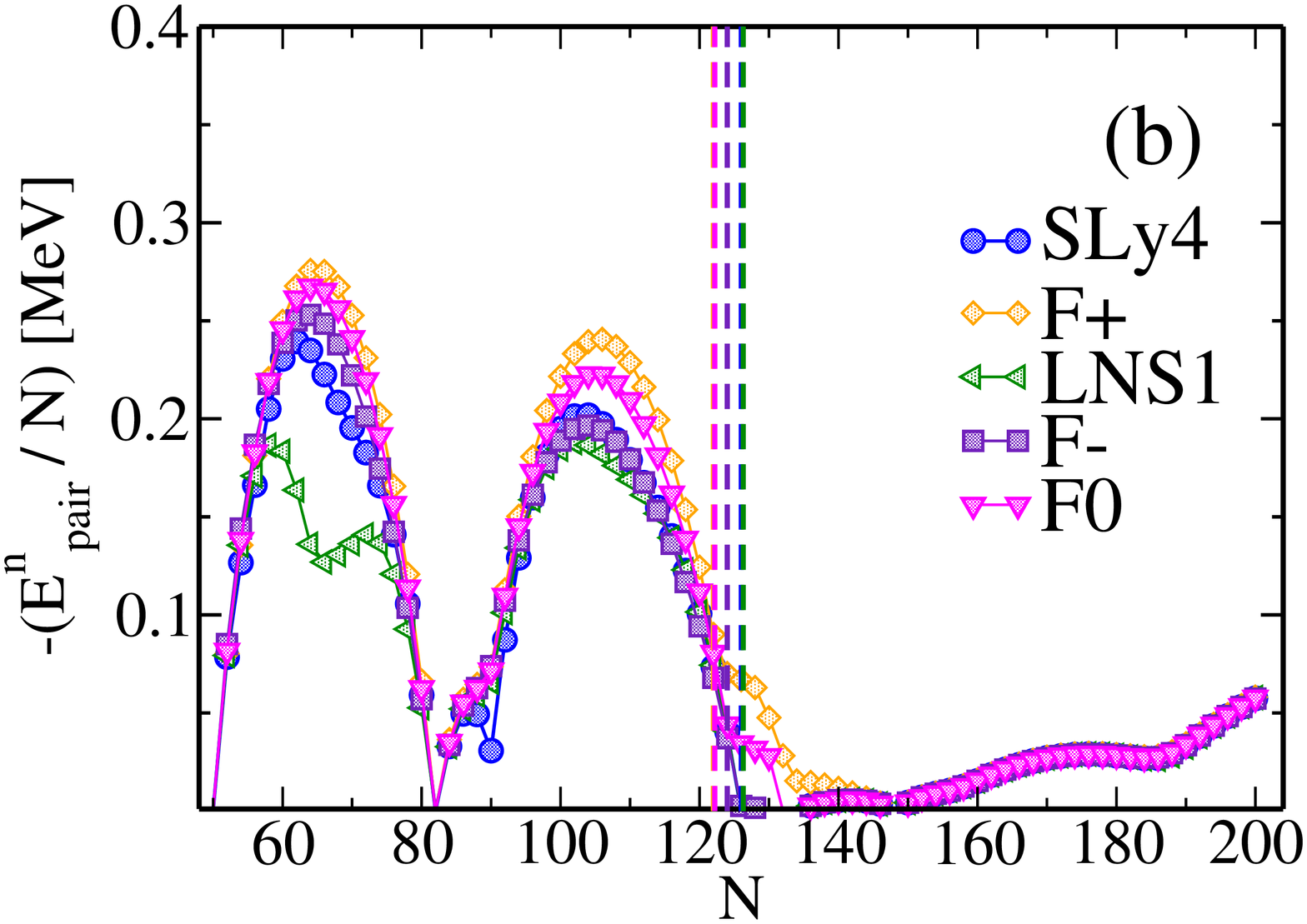}\\
\includegraphics[clip=,width=0.3\textwidth,angle=-90]{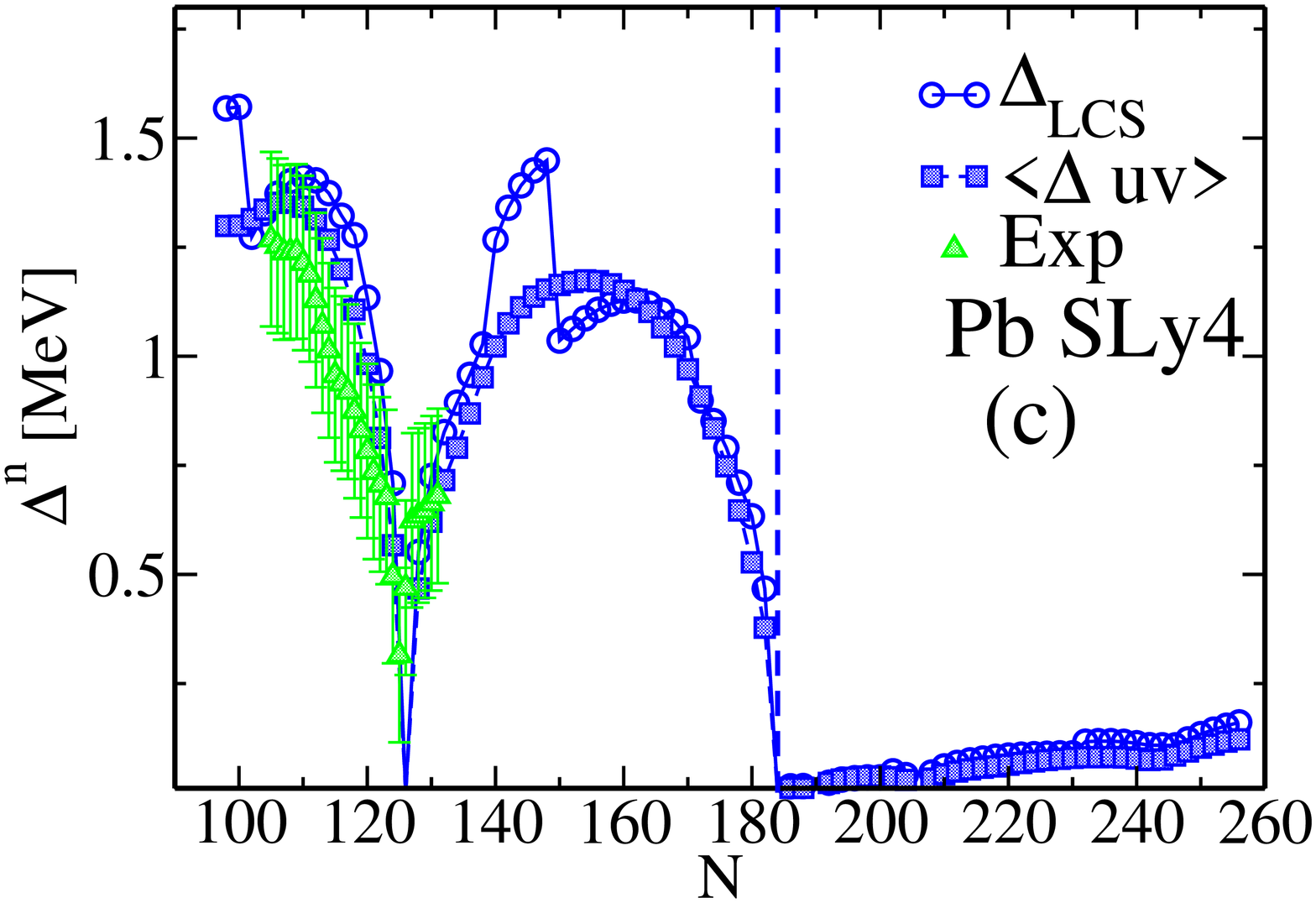}
\includegraphics[clip=,width=0.3\textwidth,angle=-90]{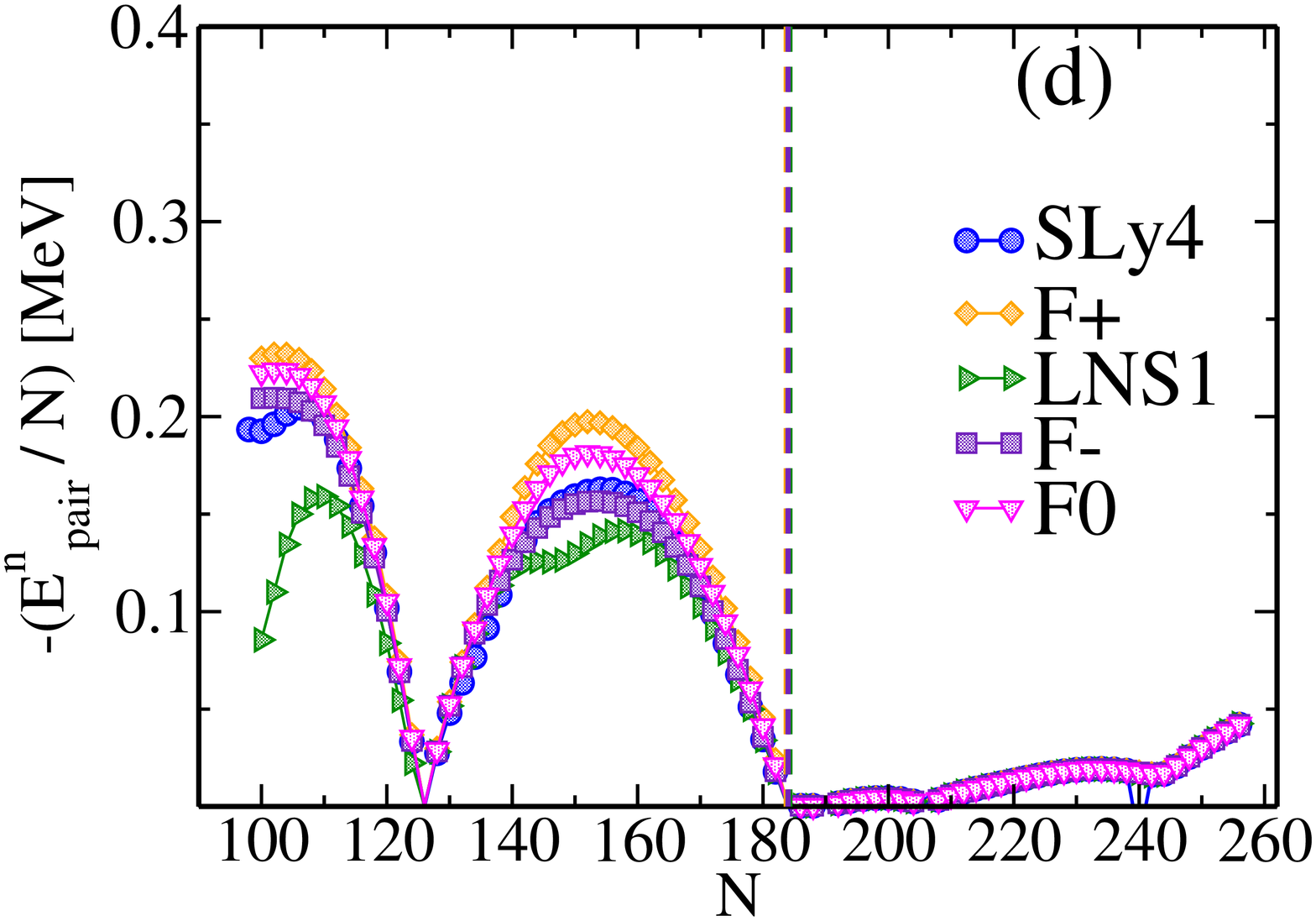}\\
\end{center}
\caption{(Colors online) As Fig.\ref{paper:fig:IsotopesGognyRed}, but for Tin and Lead isotopes. }
\label{paper:fig:IsotopesGognyRedB}
\end{figure*}

\subsection{Global properties around the drip-line}

To investigate the behavior of different systems when crossing the drip line, 
we perform a systematic study of different isotopic chains, namely the 
Calcium, Nickel, Molybdenum,Tin, and Lead isotopes.
The calculations have been done in a box of 40~fm radius and with the 
Dirichlet-Neumann mixed boundary conditions.

In Fig.~\ref{paper:fig:IsotopesGognyRed} we display the representative neutron gaps
$\Delta_{LCS}$ and  $\Delta_{UV}$ (left panel) as well as the pairing energy per neutron 
(right panel) corresponding to Calcium, Nickel, and Molybdenum isotopic chains computed
using the SFRI interaction for the pairing channel  and the mean-field provided by
the SLy4, LNS1~\cite{Gambacurta2011}, F$+$, 
F$-$, and F$0$~\cite{Lesinski2006} Skyrme forces. In the left panel, the experimental 
neutron gaps are also displayed together with their associated error bars.
The same quantities are shown in the two panels of
Fig.~\ref{paper:fig:IsotopesGognyRedB} for Tin and Lead isotopes.
The isotopes represented in Figs.~\ref{paper:fig:IsotopesGognyRed} and \ref{paper:fig:IsotopesGognyRedB}
have been selected from their behavior at the drip line~\cite{Margueron2012}:
in Fig.~\ref{paper:fig:IsotopesGognyRed} the isotopes are not magic at the drip line and some pairing
correlations persist, while in Fig.~\ref{paper:fig:IsotopesGognyRedB}, the isotopes are magic at the
drip line and pairing correlations are strongly reduced at and beyond the drip.

From Fig.~\ref{paper:fig:IsotopesGognyRed} it can be observed that the two definitions of the pairing gap: 
$\Delta_{\text{LCS}}$~ \cite{Lesinski2009} and $\Delta_{UV}$ of  Eq.~(\ref{paper:eq:uvdelta}) give 
quite similar predictions for bound nuclei within the experimental error-bars, while they show a 
noticeable difference at the drip-line and beyond.
In particular it is found 
that the gap $\Delta_{LCS}$ is almost zero at the drip line 
for all nuclei
analyzed in Figs.~\ref{paper:fig:IsotopesGognyRed} and \ref{paper:fig:IsotopesGognyRedB}, while 
the gap $\Delta_{UV}$ can persist with non zero values in some cases, namely for the Ca, Ni,
and Mo isotopic chains shown in Fig.~\ref{paper:fig:IsotopesGognyRed}, 
while it is zero for other cases such as the Sn and Pb chains represented  in 
Fig.~\ref{paper:fig:IsotopesGognyRedB}.
At the drip line, since the LCS-gap is the gap at the Fermi energy, the LCS-gap changes from 
its value in a bound state to the value corresponding to a delocalized unbound
state at low density. 
It is known that the $^1$S$_0$ pairing gap in neutron matter is essentially 
zero at vanishing  density \cite{Book:Lombardo2001}.
Therefore, the LCS-gap is, indeed, expected to be quite 
suppressed at the drip-line.

The pairing energies go to zero at the drip-line for Sn and Pb isotopes, see 
Fig.~\ref{paper:fig:IsotopesGognyRedB}, while they persist 
in the case of Ca, Ni, and Mo as shown
in Fig.~\ref{paper:fig:IsotopesGognyRed}.
In Ref.~\cite{Margueron2011}, it has been argued that these differences are
due to the presence of resonance states lying near the Fermi level of drip-line 
nuclei with 
non-negligible occupancy. In this case pairing correlations can persist and the
pairing energy remains non-zero, as it happens in the Ca, Ni, and Mo isotopic 
chains.
However, if there is a large energy gap between the last fully occupied bound 
state at the 
drip line and the first unoccupied resonant unbound state, as it is the case in the Sn 
and Pb isotopic 
chains, pairing correlations are strongly reduced at the drip-line.

\begin{figure*}
\begin{center}
\includegraphics[width=0.32\textwidth,angle=-90]{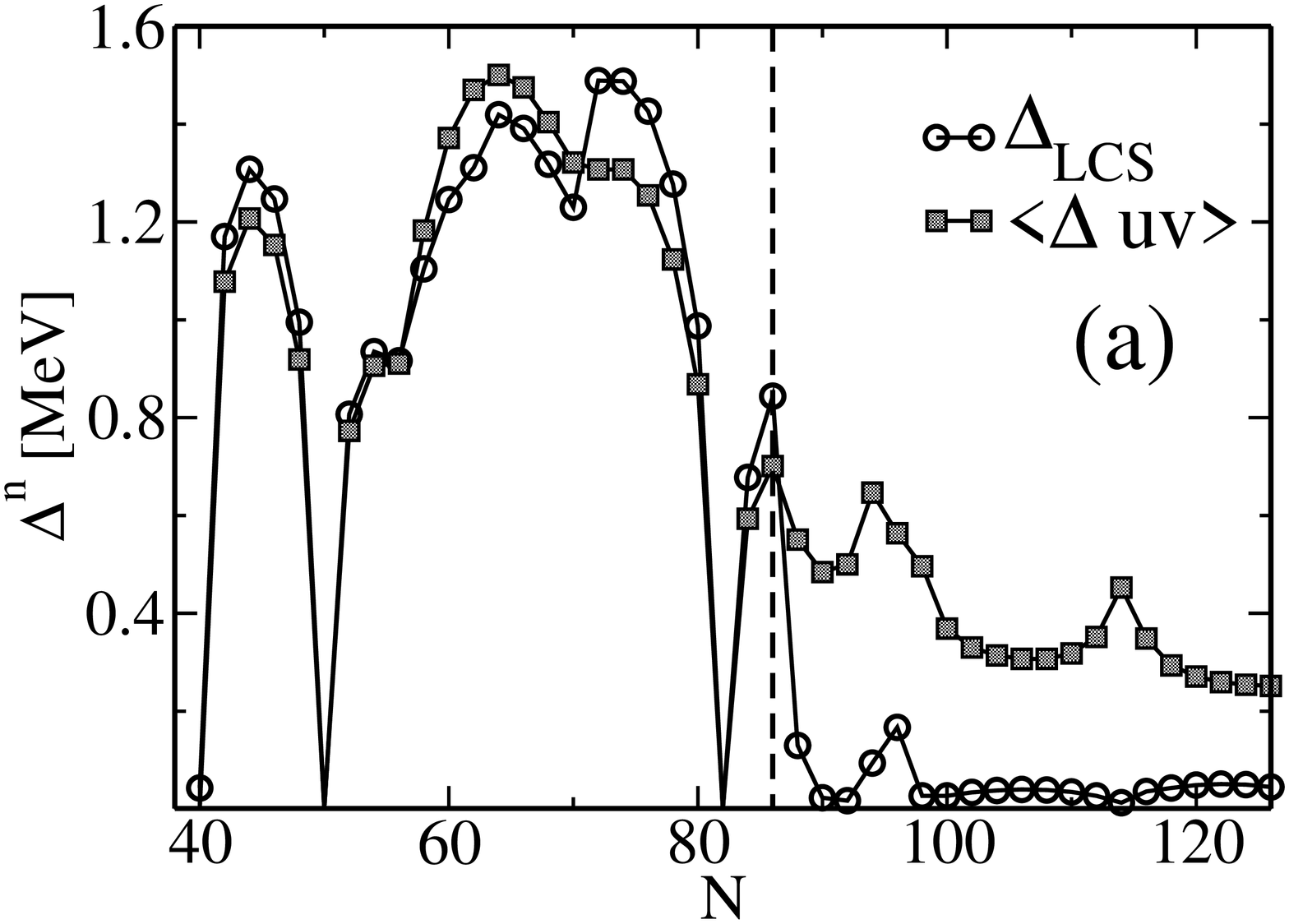}
\includegraphics[width=0.32\textwidth,angle=-90]{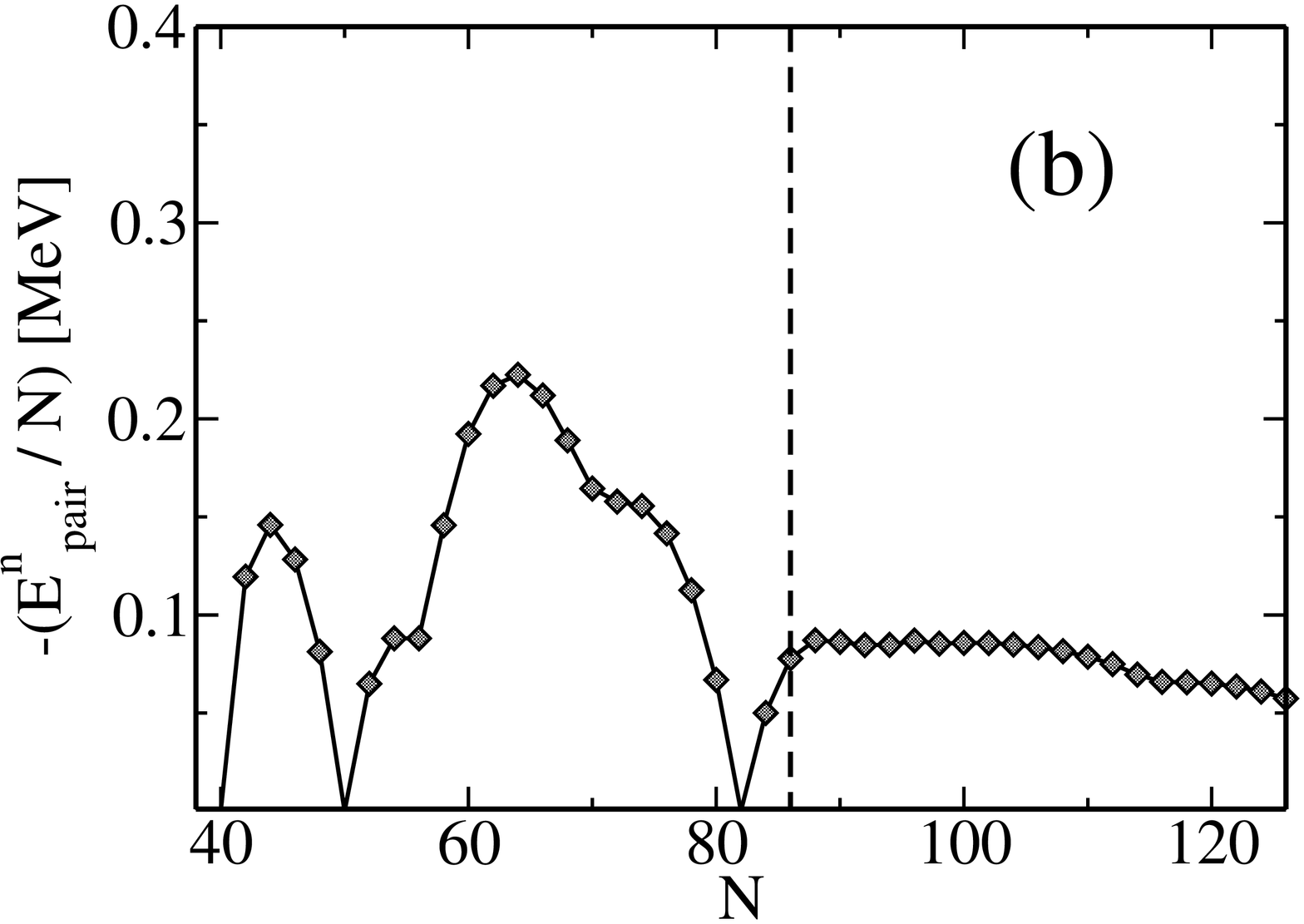}
\includegraphics[width=0.32\textwidth,angle=-90]{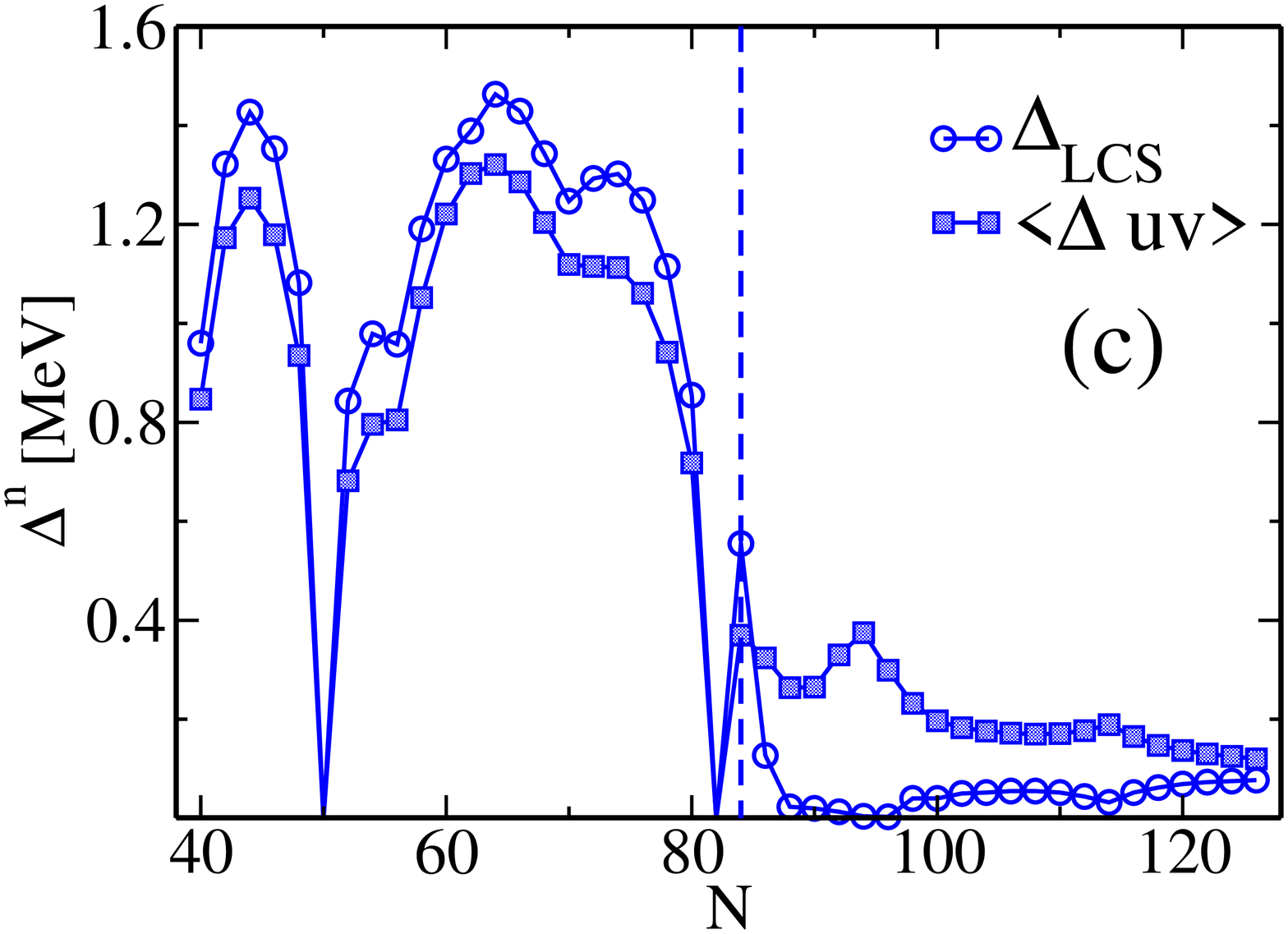}
\includegraphics[width=0.32\textwidth,angle=-90]{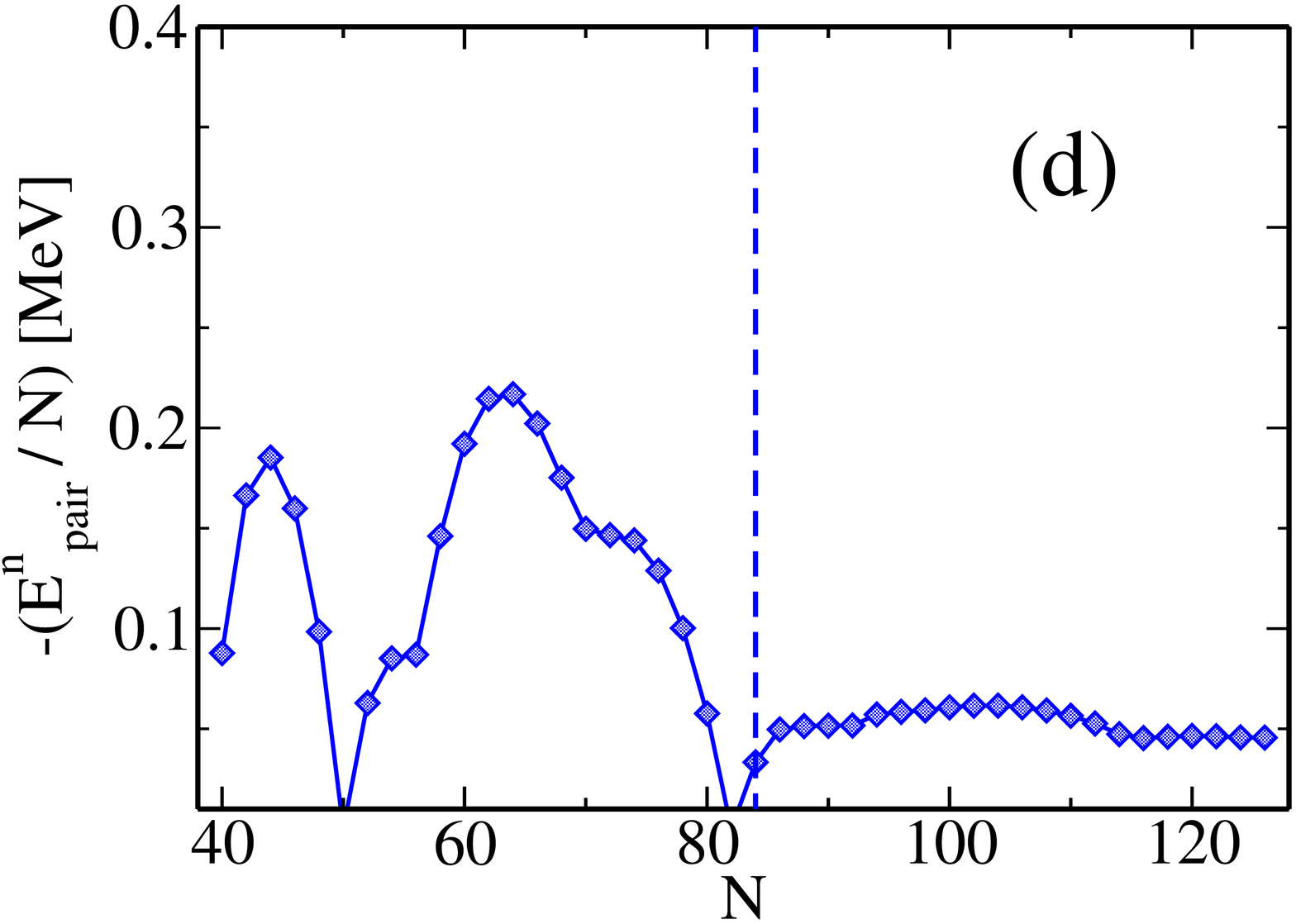}
\end{center}
\caption{ (Colors online) Similar to Fig.\ref{paper:fig:IsotopesGognyRed} but for Zr isotopes. 
On the first four panels the behavior using SLy4 Skyrme interaction and different pairing forces, 
DDCI (panel (a)-(b)) and SFRI  (panel (c)-(d)), are compared.}
\label{paper:fig:ZrGognyRed}
\end{figure*}

Since the pairing gaps $\Delta_{UV}$ displayed on the left of Figs.~\ref{paper:fig:IsotopesGognyRed} 
and \ref{paper:fig:IsotopesGognyRedB} (see also Fig.~\ref{paper:fig:ZrGognyRed}) 
represent an average of the pairing correlations over
all the states, it behaves similarly to the pairing energy see Eqs.~(\ref{paper:eq:uvdelta}) and 
(\ref{paper:eq:pe}).
The qualitative difference between the pairing gaps $\Delta_{LCS}$ and $\Delta_{UV}$ shown in
Fig.~\ref{paper:fig:IsotopesGognyRed} is, therefore, simply related to the 
fact that in 
overflowing systems, there could be a significant difference between the 
average pairing properties and the
pairing gap corresponding to the last occupied state.
The pairing gap which enters the ground state energy is the pairing gap 
$\Delta_{UV}$, since it behaves like
the pairing energy, while the pairing gap $\Delta_{LCS}$ provides 
information on the last occupied state which, e.g. influences the gap in 
the level density and quantities which depend on it.
The strong reduction of the pairing gap $\Delta_{LCS}$, being related to pairing property
of a single state, does not necessary induce the suppression of the pairing energy,
as shown in Fig.~\ref{paper:fig:IsotopesGognyRed}.
Let us notice again that in stable nuclei,
the pairing gaps $\Delta_{LCS}$ 
and $\Delta_{UV}$ are very similar (like in the nuclei represented on the left 
panels of
Figs.~\ref{paper:fig:IsotopesGognyRed} and \ref{paper:fig:IsotopesGognyRedB}).
It was, therefore, at first a surprise to observe a qualitative difference 
between the pairing gaps $\Delta_{LCS}$ and $\Delta_{UV}$ in overflowing 
systems. However, since $\Delta_{UV}$ is an average of the gaps over the 
pairing tensor, 
it is clear that in regions where the individual $\Delta_i$'s vary rapidly, 
as it happens around the drip, an average will be different from the gap 
$\Delta_{LCS}$ of the last occupied level. On the contrary, for stable nuclei, 
the  individual gaps are smoothly varying and then $\Delta_{LCS} \sim 
\Delta_{UV}$.

\begin{table}%\footnotesize
\setlength{\tabcolsep}{.15in}
\renewcommand{\arraystretch}{1.6}
\begin{center}
\begin{tabular}{ccccc}
\hline
\hline
Force &$m^{*}_{s}/m$& $m^{*}_{v}/m$ & $\Delta m^{*}$ & $m^*_n/m$ \\
\hline
F+&0.700 &0.625 &0.170 & 0.795\\
F0&0.700 &0.700 &0.001 & 0.700\\
F-&0.700 &0.870 &-0.284 & 0.586\\
SLy4 & 0.695 & 0.800&  -0.186 & 0.614\\
LNS1 & 0.604 & 0.478  &0.342  & 0.820\\
\hline
\hline
\end{tabular}
\caption{In this table we show the isoscalar and isovector masses $m^{*}_{s}/m$ and $m^{*}_{v}/m$, 
as well as the difference in neutron matter $\Delta m^{*}=m^{*}_{n}/m-m^{*}_{p}/m$ and the
neutron effective mass $m^*_n/m$ for SLy4~\cite{Chabanat1998a,Chabanat1998b}, %SIII~\cite{Beiner75}, 
F+, F-, F0~\cite{Lesinski2006} and LNS1~\cite{Gambacurta2011}. }
\label{paper:tab:forces}
\end{center}
\end{table}

In Figs.~\ref{paper:fig:IsotopesGognyRed} and \ref{paper:fig:IsotopesGognyRedB} are also shown
the pairing energies for  several mean-field models, 
namely SLy4~\cite{Chabanat1998a,Chabanat1998b}, 
F+, F-, F0~\cite{Lesinski2006} and LNS1~\cite{Gambacurta2011}.
It is observed that the reduction of the gaps at the drip-line and beyond is a property which is independent
of the considered models, while the absolute value of the pairing energy can vary from
one model to another.
The main difference among these models is related to the effective mass in symmetric and in
neutron matter. In this case the effective mass is defined as~\cite{Lesinski2006},
\begin{equation}
\frac{m}{m^*_q} = \frac{m}{m^*_s} + q I \left(  \frac{m}{m^*_s}- \frac{m}{m^*_v}\right),
\label{paper:eq:effmass}
\end{equation}
where $I=(\rho_n-\rho_p)/(\rho_n+\rho_p)$ and $q$ is the isospin charge ($q=+1$, $-1$ respectively for neutrons and protons). 
In Eq.~(\ref{paper:eq:effmass}) the effective mass in symmetric matter is given by the
isoscalar effective mass $m^{*}_{s}/m$ and the isovector effective mass $m^{*}_{v}/m$ is related
to the isospin splitting in asymmetric matter.
These quantities, as well as the effective mass splitting $\Delta m^{*}=m^{*}_{n}/m-m^{*}_{p}/m$ and the neutron effective 
mass $m^*_n/m$, both computed in neutron matter, are given in Tab.~\ref{paper:tab:forces} 
for the Skyrme interactions represented in Figs.~\ref{paper:fig:IsotopesGognyRed} and 
\ref{paper:fig:IsotopesGognyRedB}.

\begin{table*}%\footnotesize
\setlength{\tabcolsep}{0.05in}
\renewcommand{\arraystretch}{1.6}
\begin{center}
\begin{tabular}{cccc|cccc|cccc}
\hline
\hline
\multicolumn{12}{c}{F0}\\
\multicolumn{2}{c}{$^{66}\text{Ca}$} & \multicolumn{2}{c}{$e_{F}$ =-0.02 MeV} &\multicolumn{2}{c}{$^{68}\text{Ca}$} & \multicolumn{2}{c}{$e_{F}$ =0.04 MeV}  &\multicolumn{2}{c}{$^{70}\text{Ca}$} & \multicolumn{2}{c}{$e_{F}$ =0.06 MeV} \\
\hline
 $\varepsilon_{nlj }$  [MeV] & $v^{2}_{nlj}$& $\ell$ & 2$\jmath$ &$\varepsilon_{nlj }$  [MeV] & $v^{2}_{nlj}$& $\ell$ & 2$\jmath$  &$\varepsilon_{nlj }$  [MeV] & $v^{2}_{nlj}$& $\ell$ & 2$\jmath$  \\           
                \hline
      1.892    &      0.051   & 2   & 5 & 1.813     &      0.057  &  2   & 5   &1.811 &       0.058  & 2 & 5 \\
      0.994    &      0.054  &  0  &  1 & 0.554     &      0.091   & 0 &   1   &0.466 &      0.105&        0 & 1 \\
     -0.206   &      0.567   & 4  &  9  &      -0.240      &      0.600  &  4 &   9 &-0.231 &       0.600 & 4 & 9\\
     -4.639  &    0.971    &3   & 5  & -4.740      &     0.974    &3    &5      &-4.735 &       0.974 &      3 & 5 \\
\hline
\hline
\multicolumn{12}{c}{SLy4}\\
\multicolumn{2}{c}{$^{66}\text{Ca}$} & \multicolumn{2}{c}{$e_{F}$ =-0.34 MeV} &\multicolumn{2}{c}{$^{68}\text{Ca}$} & \multicolumn{2}{c}{$e_{F}$ =-0.17 MeV}  &\multicolumn{2}{c}{$^{70}\text{Ca}$} & \multicolumn{2}{c}{$e_{F}$ =0.04 MeV}\\
\hline
 $\varepsilon_{nlj }$  [MeV] & $v^{2}_{nlj}$& $\ell$ & 2$\jmath$ &$\varepsilon_{nlj }$  [MeV] & $v^{2}_{nlj}$& $\ell$ & 2$\jmath$  &$\varepsilon_{nlj }$  [MeV] & $v^{2}_{nlj}$& $\ell$ & 2$\jmath$   \\           
                \hline
      1.835    &      0.037  &  2    &5&1.529     &     0.047  &  2  &  5 &1.250      &       0.056 & 2 & 5\\
    1.614      &   0.023   & 0    &1 & 1.127      &   0.035  &  0    &1 &0.360      &      0.101&        0 & 1\\
     -0.573     &      0.583    &4&    9&-0.910   &      0.765 &   4   & 9 &-1.167     &      0.903  & 4 & 9\\
     -4.948    &      0.974    &3   & 5&-5.430   &     0.984    &3   & 5 &-5.799     &       0.993&      3 & 5\\
\hline
\hline
\end{tabular}
\caption{Canonical single particle energies and canonical occupation probabilities for some calcium isotopes }
\label{paper:tab:Levels}
\end{center}
\end{table*}

For calcium isotopes, the pairing energy at the drip-line and beyond shown in Fig.~\ref{paper:fig:IsotopesGognyRed} 
depends on the Skyrme interaction. 
There is indeed a group of Skyrme interactions for which the pairing gap  
is large at the drip line (F+, F0, F-),
and another group for which it is reduced approximatively by a factor 2 at the drip line (SLy4, LNS1).
This qualitative difference is due to the structure of the drip line nuclei around the 
Fermi energy as it is shown
in Tab.~\ref{paper:tab:Levels} for $^{66,68,70}$Ca and for F0 (top) and SLy4 (bottom).
The Skyrme interaction F0 predicts for $^{66}$Ca, the nucleus at the drip-line, a bound state $g_{9/2}$, 
in the canonical basis, close to the continuum with an energy 
$\approx-0.2$~MeV, while
SLy4 predict for $^{68}$Ca, the nucleus at the drip line, a   $g_{9/2}$ state with lower energy $\approx-0.9$~MeV.
Since the $g_{9/2}$ is lower in energy with SLy4 compared to F0, its occupation number is closer to 1 at the
drip point and this state participates to a lower extent to the pairing correlations.
For both interaction, the system gains energy if instead of realizing the shell closure at N=50 and 
becoming non-superfluid, it leaves the $g_{9/2}$ state partially unfilled so to gain extra pairing energy.
This is what is observed in Fig.~\ref{paper:fig:IsotopesGognyRed} not only at the drip line, but also beyond.
The structure of the drip line nuclei is mostly conserved even in the presence of a gas.
Despite these differences, it is interesting to remark that the global level occupation picture around the
drip line is similar for the Skyrme interactions F0 and SLy4: the unbound states are occupied before the 
drip line is reached at variance with the usual claim that the drip occurs when the first unbound 
particle is produced~\cite{Book:Bohr1998}.
In fact, among unbound states there are resonant states which play an important role around the drip line.
They have large spatial overlaps with the single-particle 
levels of the nucleus and give a significant contribution to the pairing correlations.
In most of the cases, resonant states are populated before unbound scattering states.
They are very important to understand the transition from isolated nuclei 
to overflowing systems.

We also analyzed the difference between the two kinds of pairing interactions which we used in this work:
SFRI and DDCI.
It is important to analyze the influence of the density dependence of the DDCI in inhomogeneous
systems such as Wigner-Seitz cells at neutron overflow.
A comparison of the pairing properties of isolated zirconium isotopes and in 
Wigner-Seitz cells is shown 
in Fig.~\ref{paper:fig:ZrGognyRed} for SFRI (left) and DDCI (right).
The drip-line nucleus predicted by SLy4+SFRI is $N=84$ while it is $N=88$ for SLy4+DDCI.
Despite this small difference, the behavior of the pairing gaps $\Delta_{LCS}$ and $\Delta_{UV}$, as
well as the pairing energy is very similar at the drip line and beyond.
We conclude that provided that the DDCI reproduce the same pairing gaps in symmetric and neutron
matter, DDCI give similar results compared to SFRI in inhomogeneous systems such as
the Wigner-Seitz cells~\cite{Pastore2012}. Otherwise the same scenario as in Figs. 2 and 3 is 
recovered also for the Zirconium isotopes.

\subsection{The limit of nuclei immersed in a vanishing dilute gas}

Below, we will explore what happens in the outer crust of neutron stars around the point where the neutrons start to drip 
into the free space between the lattice sites built by the nuclei. The question we want to answer is whether a very low density gas of superfluid neutrons in a large container can have any major influence on the superfluidity of the nuclei at the lattice sites.
In this section, we, thus, explore, in a schematic study, the limit of nuclei immersed in a  dilute gas 
with the density of the gas going to zero.

\begin{figure}
\begin{center}
\includegraphics[clip=,width=0.38\textwidth,angle=-90]{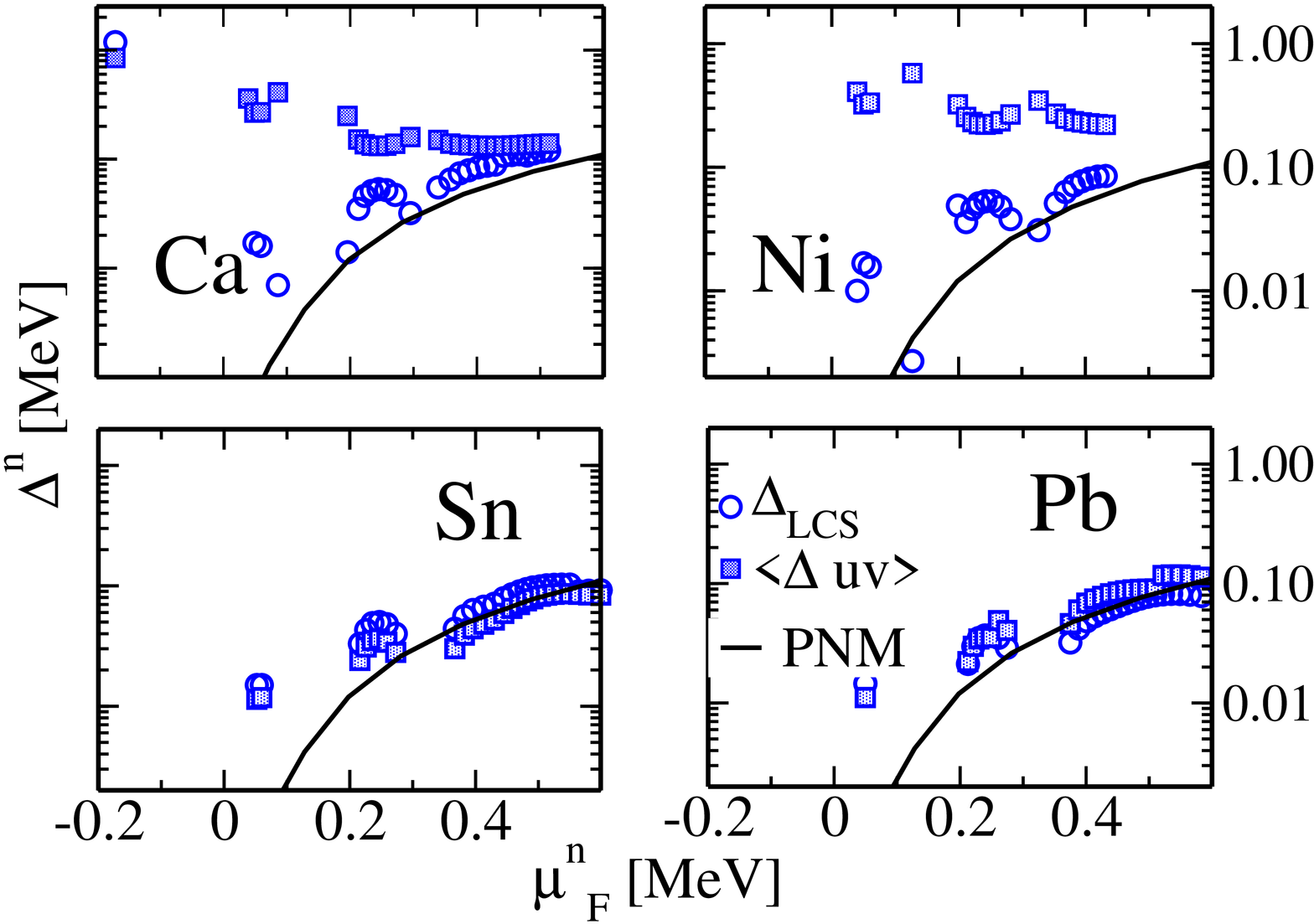}
\end{center}
\caption{(Colors online) The gaps $\Delta_{LCS}$ and $\Delta_{UV}$  as a function 
of the chemical potential. The calculations have been already shown in 
Fig.~\ref{paper:fig:IsotopesGognyRed}-\ref{paper:fig:IsotopesGognyRedB}. Here we make a zoom in the 
case of SLy4+SFRI for the four isotopic chains. On the same plot we show the calculation of Pure Neutron 
Matter (solid line). }
\label{paper:fig:CompareGapsPNM}
\end{figure}

We first show in Fig.~\ref{paper:fig:CompareGapsPNM} a zoom of  Figs.~\ref{paper:fig:IsotopesGognyRed} and 
\ref{paper:fig:IsotopesGognyRedB} focussed on the overflowing nuclear systems.
In calcium and nickel isotopes pairing correlations persists at overflow since the gap $\Delta_{UV}$ 
do not vanish,
while pairing correlations are almost suppressed at overflow 
in tin and lead isotopes.
The solid line corresponds to
the value of the pairing gap in uniform neutron matter for the densities of 
the gas.
By construction,
the gap $\Delta_{LCS}$ follows quite well the trend of the uniform neutron matter gap. However, in a realistic system, composed of a nucleus plus a gas, one 
has $\Delta_{UV} \ne \Delta _{LCS}$ and, thus,   $\Delta_{UV}$  shows clear 
differences with the gas due to the influence of the resonance states in calcium and nickel isotopes.
The understanding of overflowing systems, therefore, requires a better study 
of the pairing properties of 
nuclei immersed inside a gas at the limit of very low density. 
In the following we aim at decreasing the density of the gas to the lowest 
possible value.

Starting from an overflowing system with an external gas, the low density 
limit can be reached in two different 
manners: the first one is by increasing the size of the box for a fixed 
number of neutrons and the second one 
is by decreasing the total number of neutrons at fixed box size.
However, the  numerical calculations cannot be performed in boxes with sizes
larger than $80$~fm with SFRI and 150~fm with DDCI.
These limitations are due to the increasing number of partial waves as well as of 
the level density as the size of the box is increased.
The larger size of the box reached with DDCI is related to the lower CPU time and memory request to perform calculations compared to SFRI.

\begin{figure}
\begin{center}
\includegraphics[clip=,width=0.38\textwidth,angle=-90]{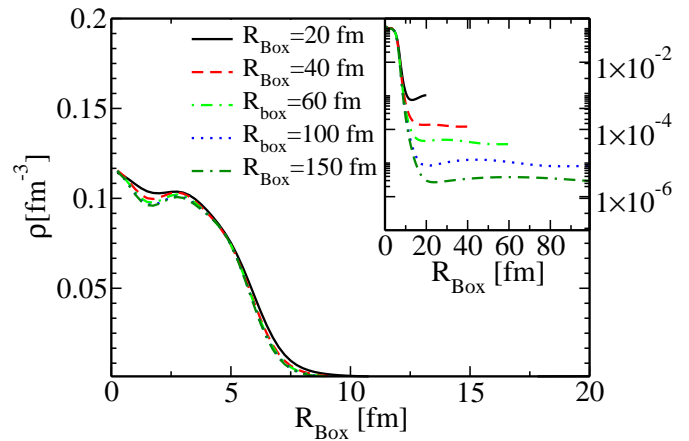}
\includegraphics[clip=,width=0.38\textwidth,angle=-90]{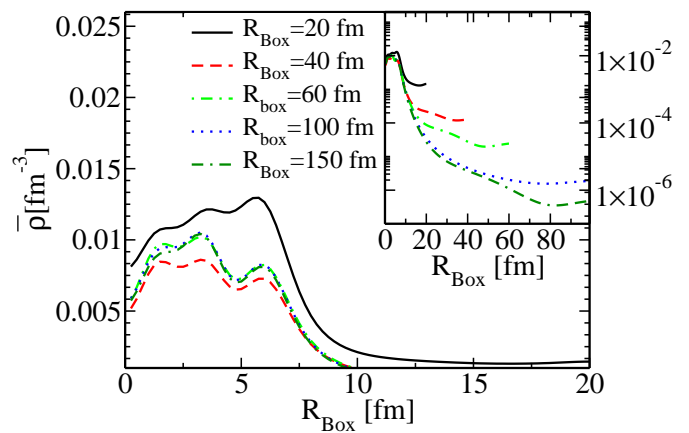}
\end{center}
\caption{(Colors online) Evolution of the density $\rho(r)$ (top panel) and abnormal density $\tilde{\rho}(r)$ (bottom panel) for 
$^{166}$Zr as a function of the box size for SLy4+SFRI. In the inset we show the  
semi-logarithmic scale.}
\label{paper:fig:abnZr166}
\end{figure}

The effect of increasing the size of the box is illustrated in Fig.~\ref{paper:fig:abnZr166} for the case of
$^{166}$Zr. The total density and anomalous density profiles are represented with a linear and
logarithmic scale.
There is an important reduction of the gas density for boxes going from 20 to 100~fm, while the 
reduction of the density going from 100 to 150~fm is quite marginal.
This is a limitation of this method which imposes to work with very large boxes to reach the
low density limit.

From the behavior of the density in $^{166}$Zr as a function of the box radius 
represented in
Fig.~\ref{paper:fig:abnZr166}, two regions can roughly be distinguished: one is the "bulk'' and the 
other the gas. 
Fixing an arbitrary limit $R_{lim}=10$~fm to separate the "bulk" from the "gas", the number of 
neutrons in the bulk can be estimated as,
\begin{equation}
N^{\text{bulk}}=\int_{0}^{R_{lim}} \rho_{n}(r) \, d^3r .
\end{equation}
We obtain that for $R_{box}=20$~fm, $N^{\text{bulk}}\approx 99$ neutrons and for
$R_{box}=100$~fm, $N^{\text{bulk}}\approx 87$.
The number of neutrons in the bulk decreases as a function of the box size  
having as a limit the isolated nucleus at the drip line. 
Since we perform constrained HFB calculation conserving the total number 
of neutrons, 
the particles evaporated from the bulk appear in the scattering states.

\begin{figure}
\begin{center}
\includegraphics[clip=,width=0.4\textwidth,angle=-90]{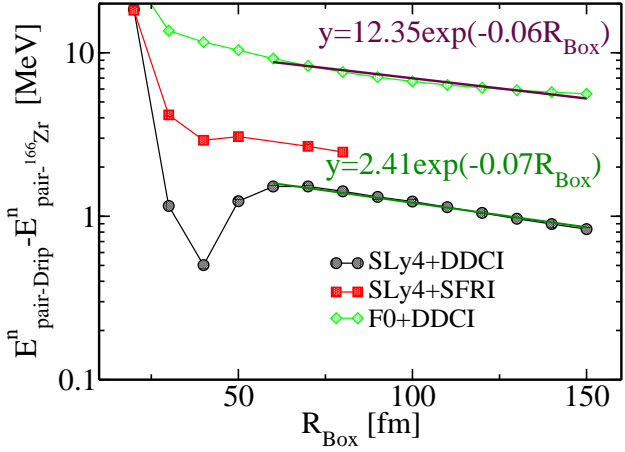}
\end{center}
\caption{(Colors online) Evolution of the neutron pairing energy for $^{166}$Zr as a function of the box size for different functionals. The dashed lines represent the pairing energies of the last bound nucleus in the chain as shown in 
Tab.~\ref{paper:tab:tab_zr_chain}. See text for details.}
\label{paper:fig:Zr166}
\end{figure}

\begin{table}%\footnotesize
\renewcommand{\arraystretch}{1.6}
\begin{center}
\begin{tabular}{c|cc|cc|cc}
\hline
\hline
&\multicolumn{2}{c|}{SLy4+DDCI  } &\multicolumn{2}{c|}{SLy4+SFRI } &\multicolumn{2}{c|}{F0+DDCI }\\
\hline
Nucleus & $\mu_{F}^{n}$ & $E_{pair}^{n}$ & $\mu_{F}^{n}$ & $E_{pair}^{n}$ & $\mu_{F}^{n}$& $E_{pair}^{n}$\\
\hline
 $^{130}$Zr & 0.05&-7.763 &0.06  &-4.643 &0.065&-6.309\\
 $^{128}$Zr &0.04 &-7.662 &0.05  & -4.521&0.054&-6.213\\
 $^{126}$Zr & -0.04&  -6.693& 0.03 & -4.271 &0.046& -6.073\\
 $^{124}$Zr & -0.16& -4.201& -0.03&-2.801  &0.002&-5.109\\
 $^{122}$Zr &-1.47& 0.000&-1.33 &0.000  &-1.13& 0.000\\
\hline
\hline
\end{tabular}
\caption{In this table we show the exact value of the neutron chemical potential and pairing energy, expressed in MeV, for Zr isotopes shown in Fig.\ref{paper:fig:ZrGognyRed}. }
\label{paper:tab:tab_zr_chain}
\end{center}
\end{table}

In Fig.~\ref{paper:fig:Zr166}, we  display by dots the difference 
between the neutron pairing energy of 
the drip-line nucleus and that of the overflowing nuclear system 
$^{166}$Zr,
$E^{n}_{pair}(^{X}\text{Zr})-E^{n}_{pair}(^{166}\text{Zr})$, as a function 
of the box size for the models SLy4+DDCI, SLy4+SFRI and F0+DDCI. 
The considered drip-line nuclei in our models 
are $^{126}$Zr for SLy4+DDCI, $^{124}$Zr for 
SLy4+SFRI, $^{122}$Zr for F0+DDCI.
The values of the pairing energy in these nuclei are extracted from Tab.~\ref{paper:tab:tab_zr_chain}.
The convergence to the asymptotic value is different for the models SLy4+DDCI and SLy4+SFRI.
We observe a fast convergence when going from a box of 20 fm to 40 fm for these two models. In such case we have seen that the particle density is sufficiently high to completely fill the resonances and thus such states do not contribute to pairing superfluidity.
The excess of pairing energy in those small boxes comes mainly from scattering states.
In fact when we go from $R_{box}=40$fm  to $R_{box}=50$~fm, the resonant state $f_{7/2}$ starts to depopulate and thus we can form Cooper pairs using such state. This explains why we have an increase of superfluidity.
Going from $R_{box}=50$~fm to $R_{box}=150$~fm, the convergence becomes very slow.
In this case the box is sufficiently large to decouple bound and scattering states, the residual pairing energy comes from the superfluidity of neutrons trapped into resonant states.
The behavior is different in case of F0+DDCI model since a different mean field produces a different single particle structure.

The asymptotic value is not reached in the larger boxes used in our calculations. Nevertheless, judging from the last numerical values, one may assume that the limit eventually goes to zero.

\begin{figure}
\begin{center}
\includegraphics[clip=,width=0.45\textwidth]{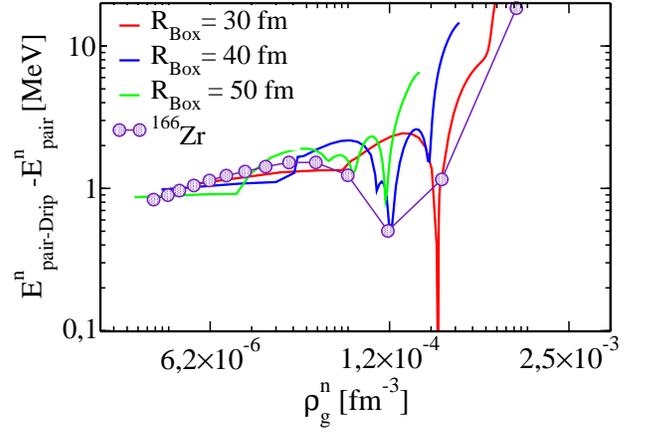}
\end{center}
\caption{(Colors online) The pairing energy as a function of the external density 
of the neutron gas, $\rho^n_g$. The solid lines are obtained by fixing a box and 
increasing the number of particles from $^{128}$Zr to $^{200}$Zr. 
The points are obtained by fixing the number of particles in $^{166}$Zr
and increasing the size of the box.The calculations are done using SLy4 and the DDDI pairing interaction }
\label{paper:fig:den}
\end{figure}

We also explore the alternative scenario where the size of the box is kept fixed and the number of neutrons 
is varied from 88 to 160. The results computed for boxes of size ranging from 30~fm to 50~fm are displayed by solid lines in 
Fig.~\ref{paper:fig:den}.
These results should be compared with the ones shown in Fig.~\ref{paper:fig:Zr166}.
To this end we show by points the pairing energy of the nucleus $^{166}$Zr computed within boxes of different sizes 
as in Fig.~\ref{paper:fig:Zr166}.
Again we observe in Fig.8 that for the case $R_{Box}=30$~fm we have a quick drop of superfluidity at 
$\rho^{n}_g\approx1.4\times10^{4}$ fm$^{-3}$ and then an increase.
This is the same phenomenon observed in Fig.~\ref{paper:fig:Zr166}, and it is due to a depopulation of the resonant state $f_{7/2}$, that when occupied does not contribute to superfluidity and when is half filled gives an important contribution to superfluidity.
In conclusion, the two  methods employed in this work to reach the low density 
limit in the
gas show that the overflowing systems tends to the limit of the drip-line nucleus.
The drip-line nucleus is therefore an important reference to understand and analyze the properties
of the overflowing systems.

\begin{figure}
\begin{center}
\includegraphics[clip=,width=0.4\textwidth,angle=-90]{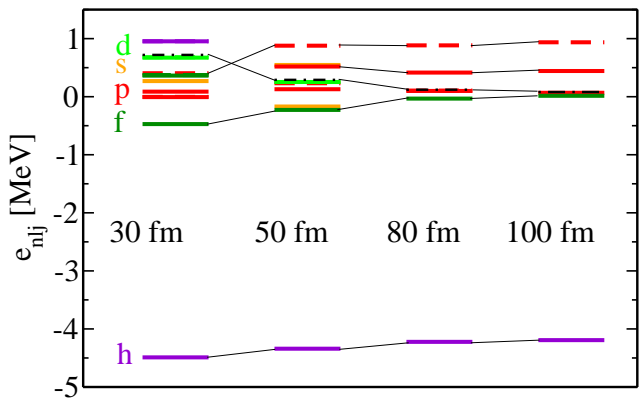}
\end{center}
\caption{(Colors online) We show the evolution of the canonical single particle levels, $e_{nlj}$, of neutrons for the 
nucleus $^{166}$Zr calculated using the SLy4 functional and a contact pairing force as  a function 
of the box size. The dashed black line indicates the position of the Fermi energy.}
\label{paper:fig:levZr166}
\end{figure}

In Fig.~\ref{paper:fig:levZr166}, we show the evolution of the canonical single particle states for neutrons 
as a function of the box for $^{166}$Zr.
In this case, to simplify the picture, we decided to represent only the most important levels.
When the size of the box increases, the level scheme is modified. 
However, the resonance levels remain practically at the same position independently of the size 
of the box, as in the case of weakly bound nuclei~\cite{Pei2011}.
With the help of this criterium we can identify the resonant state $f_{7/2}$ 
located very close to the Fermi energy.
Using the SLy4 plus DDCI interaction and performing the HFB calculation in a box of $R_{box}=100$~fm, we find
that the occupancy of the $f_{7/2}$ resonant state is $\approx 4.16$ neutrons and the one of the $p_{3/2}$ state
is $\approx 0.88$ neutrons.
From Tab.~\ref{paper:tab:tab_zr_chain}, we see that the corresponding last bound nucleus of the Zr-chain
is $^{126}$Zr. If we look at the canonical levels of this nucleus, we find a complete occupancy 
up to the state $h_{11/2}$ at $e_{h_{11/2}}=-4.15$ MeV, where we have a shell closure 
(N=82). There are two resonant states at  $e_{p_{3/2}}=0.653 $ MeV with 
occupation $v^{2}_{p_{3/2}}=0.11$ (corresponding to 0.44 neutrons) and  $e_{f_{7/2}}=0.15 $ MeV with occupation 
$v^{2}_{f_{7/2}}=0.39$ (corresponding to 2.73 neutrons).
This means that we can find a stable nucleus although some of the particles stay in a resonant state 
close to zero.

\begin{figure}
\begin{center}
\includegraphics[clip=,width=0.4\textwidth,angle=-90]{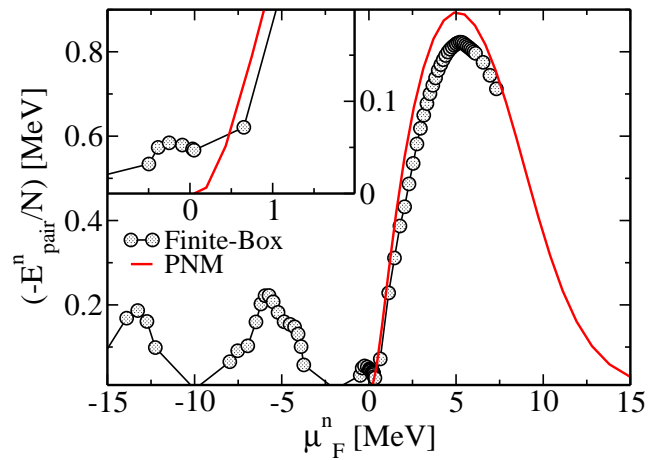}
\end{center}
\caption{(Colors online) We compare the pairing energy per neutron calculated using the WS 
approximation for Mo isotopes (circles) calculated as a function of the neutron Fermi energy 
$e^{n}_{F}$. We observe that we can have both positive and negative values. For positive 
values we compare the results with a calculation of Pure Neutron Matter (PNM), solid line. 
The inset shows the behavior of the transition among bound and unbound nuclei. 
SLy4+SFRI.}
\label{paper:fig:eee}
\end{figure}

In Fig.~\ref{paper:fig:eee}, we show the pairing energy per particle for Mo isotopes as a function of the neutron chemical potential $\mu_F^{n}$, using SLy4+SFRI model.
The calculations have been performed for a fixed number of protons (Z=42) and fixed box radius ($R_{box}=40$ fm), 
similarly to what has been done by Grasso \emph{et al.}~\cite{Grasso2008}.
We can observe in such a way the transition from bound nuclei to the gas+nucleus system.
The results are compared with the analogous calculation done in neutron matter. We observe that for small positive values of $\mu^{n}_{F}$ the cluster+gas system has a bigger pairing energy than the homogeneous system.
Such difference is due to the resonance structure as explained in the previous sections.
When we reach high density regions ($\mu^{n}_{F}\gtrsim 2$MeV), we see that the presence of the cluster reduces the pairing correlation.
Such phenomenon has been already discussed by many authors concerning Wigner-Seitz calculations~\cite{Grill2011,Pastore2011,Pizzochero2002,Baldo2006,Chamel2010b,Pastore2012}

\begin{figure*}
\begin{center}
\includegraphics[clip=,width=0.35\textwidth,angle=-90]{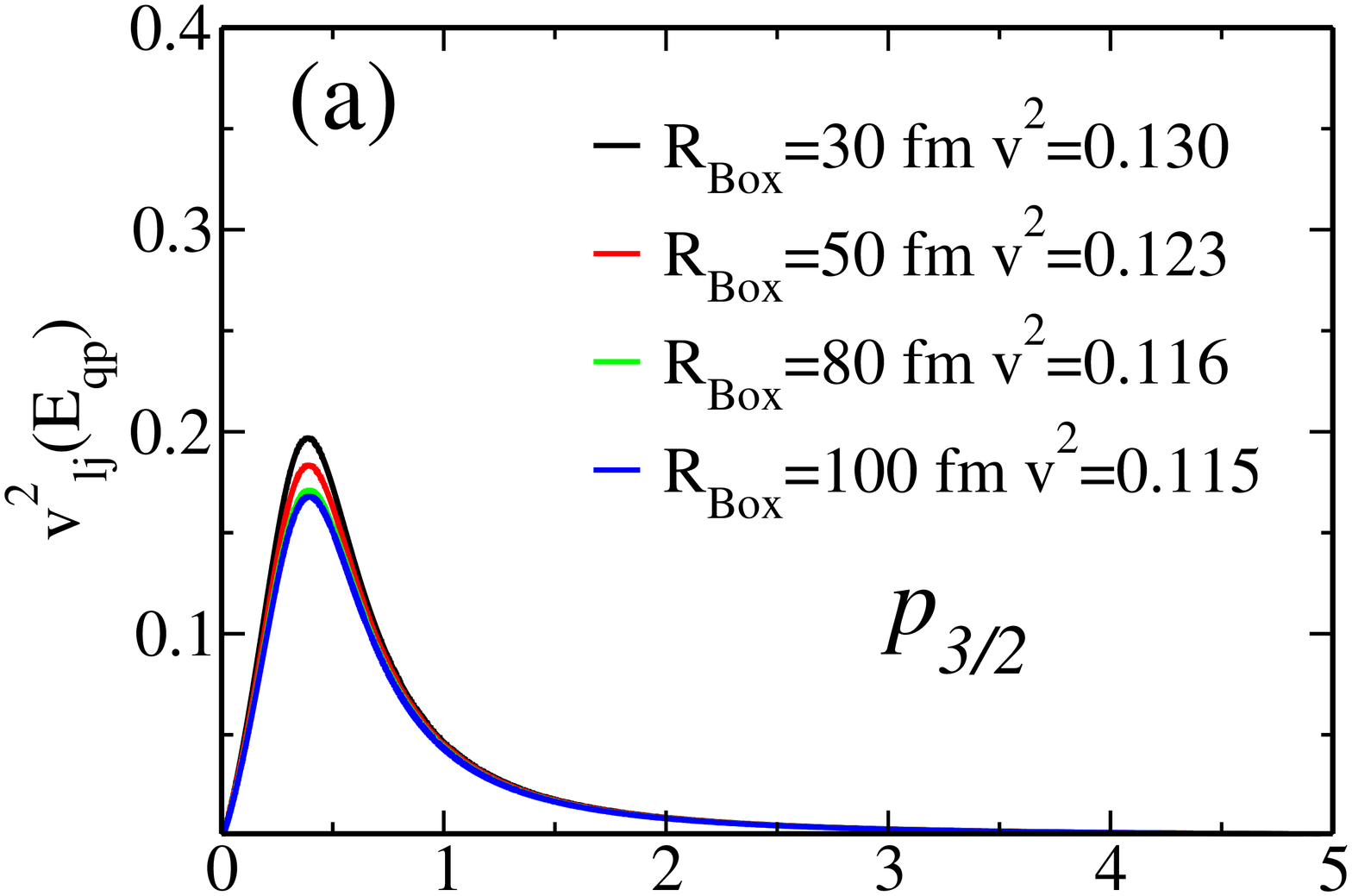}
\includegraphics[clip=,width=0.35\textwidth,angle=-90]{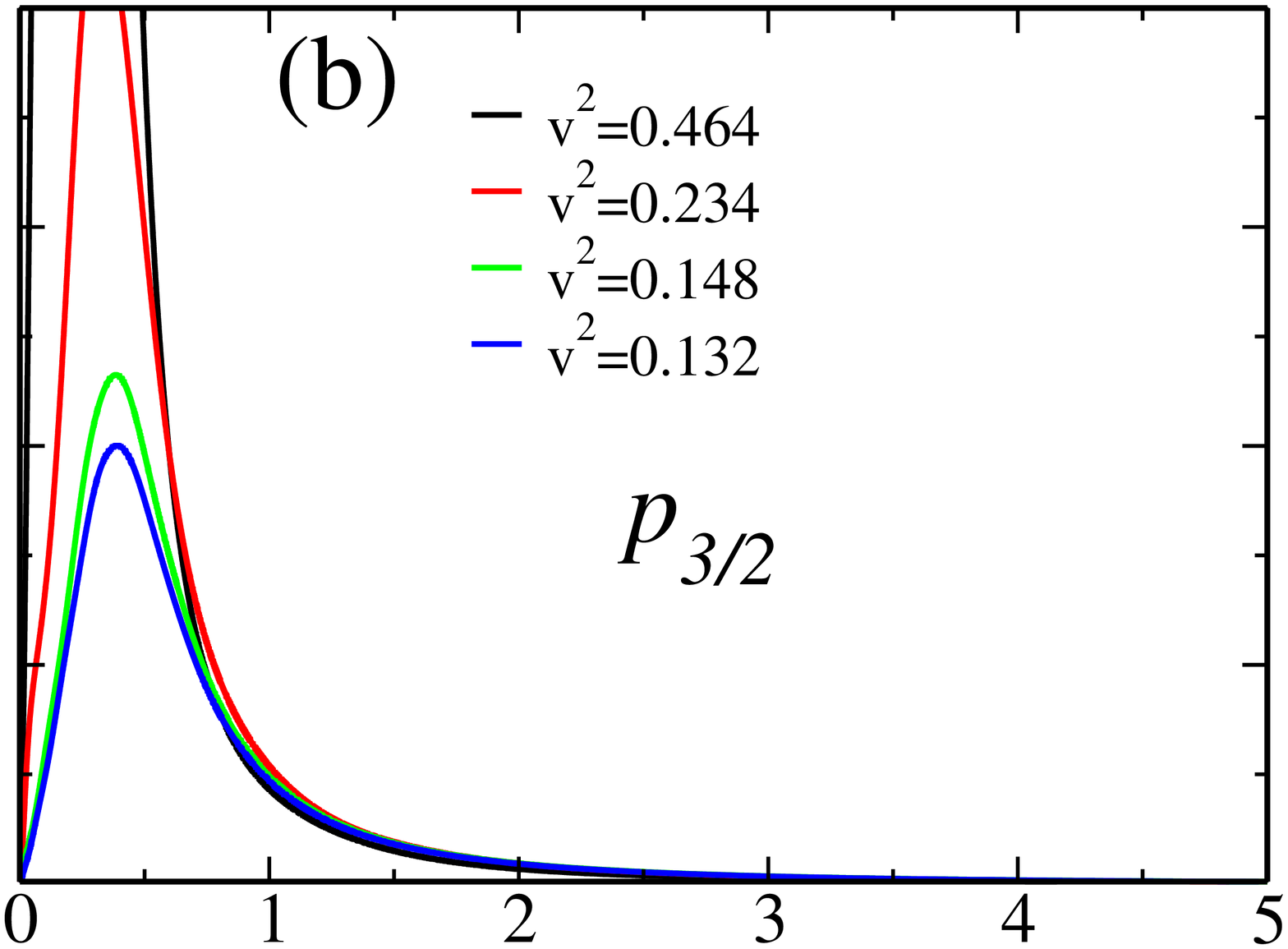}\\
\includegraphics[clip=,width=0.35\textwidth,angle=-90]{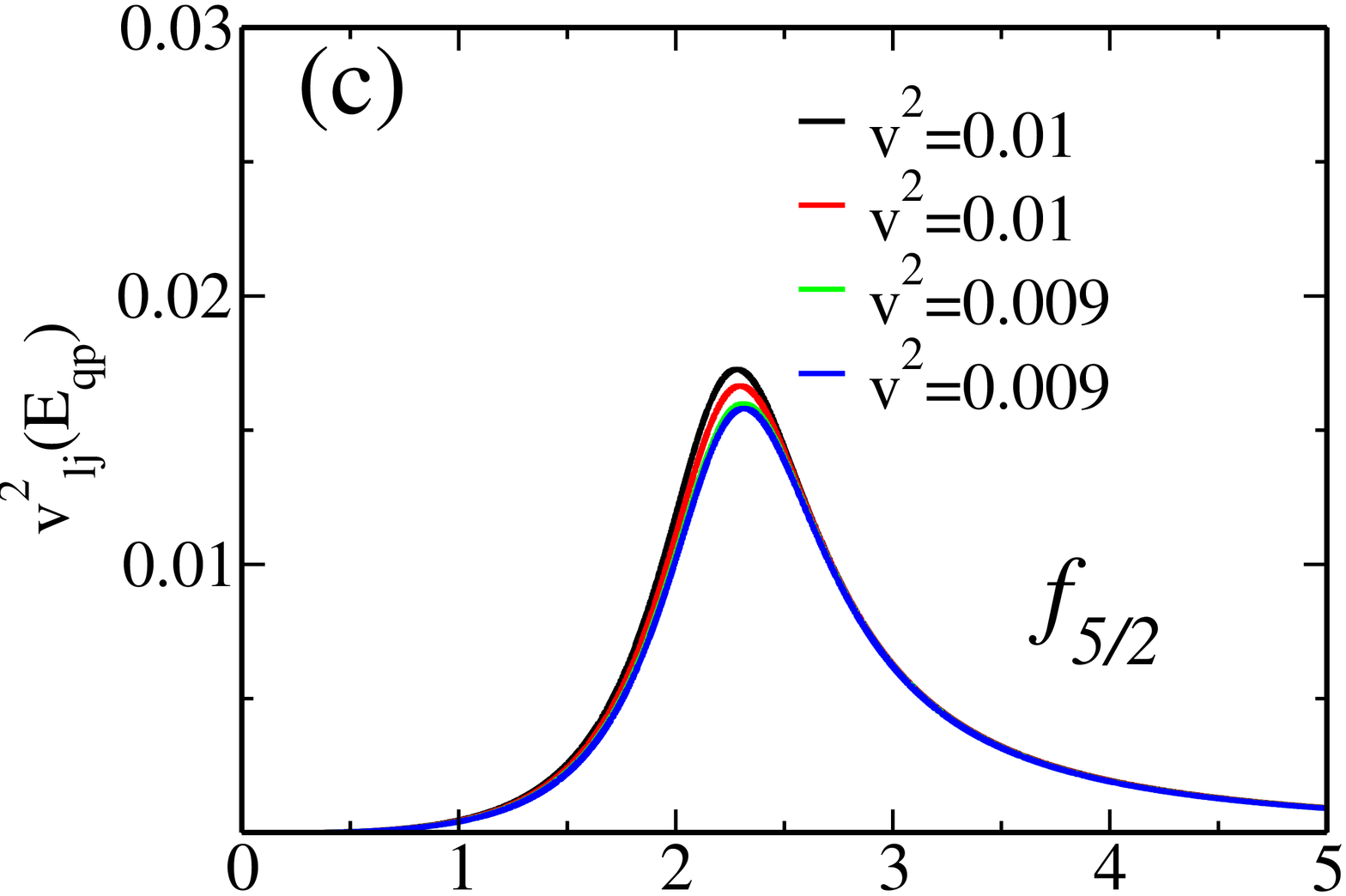}
\includegraphics[clip=,width=0.35\textwidth,angle=-90]{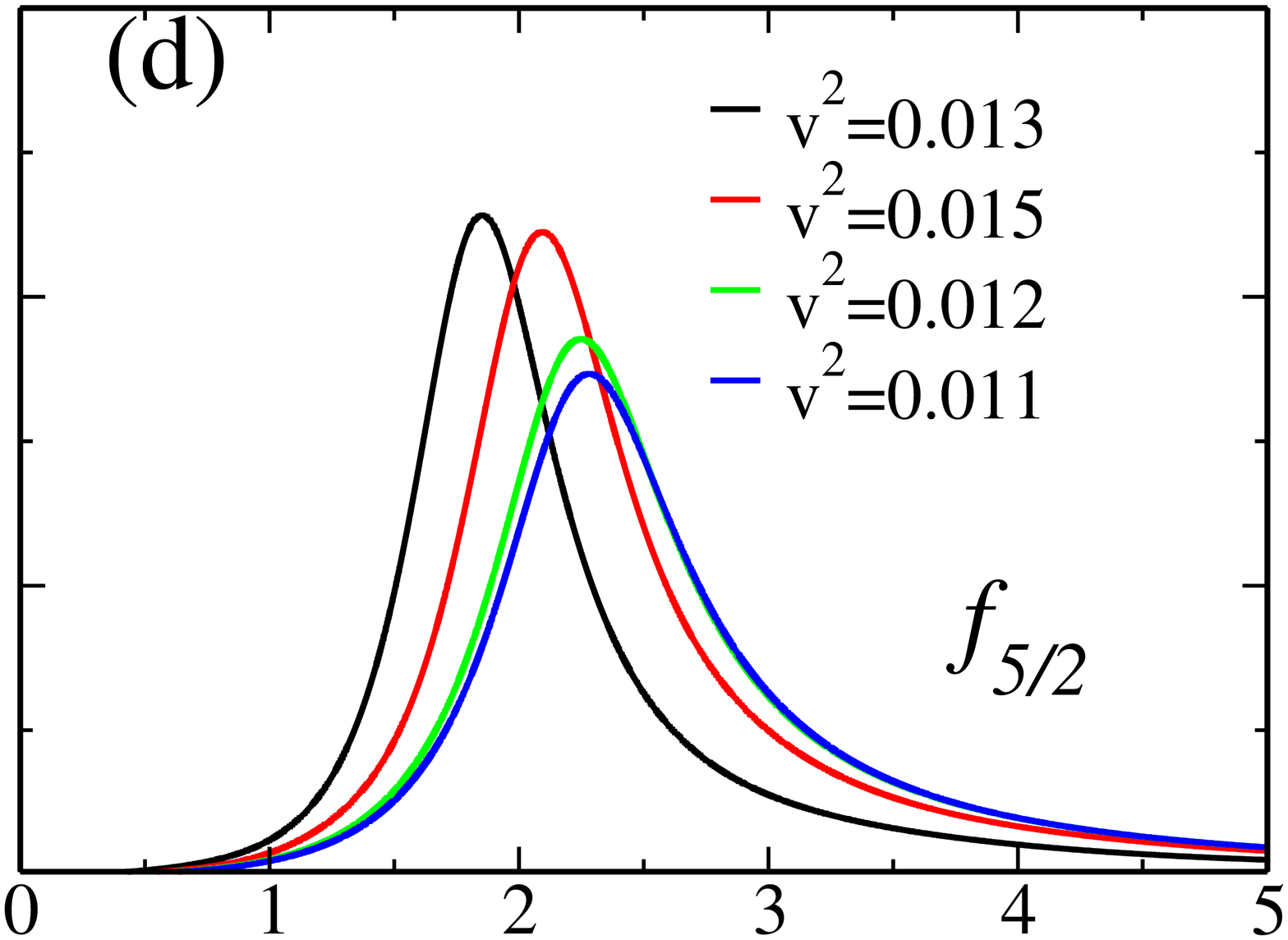}\\
\includegraphics[clip=,width=0.35\textwidth,angle=-90]{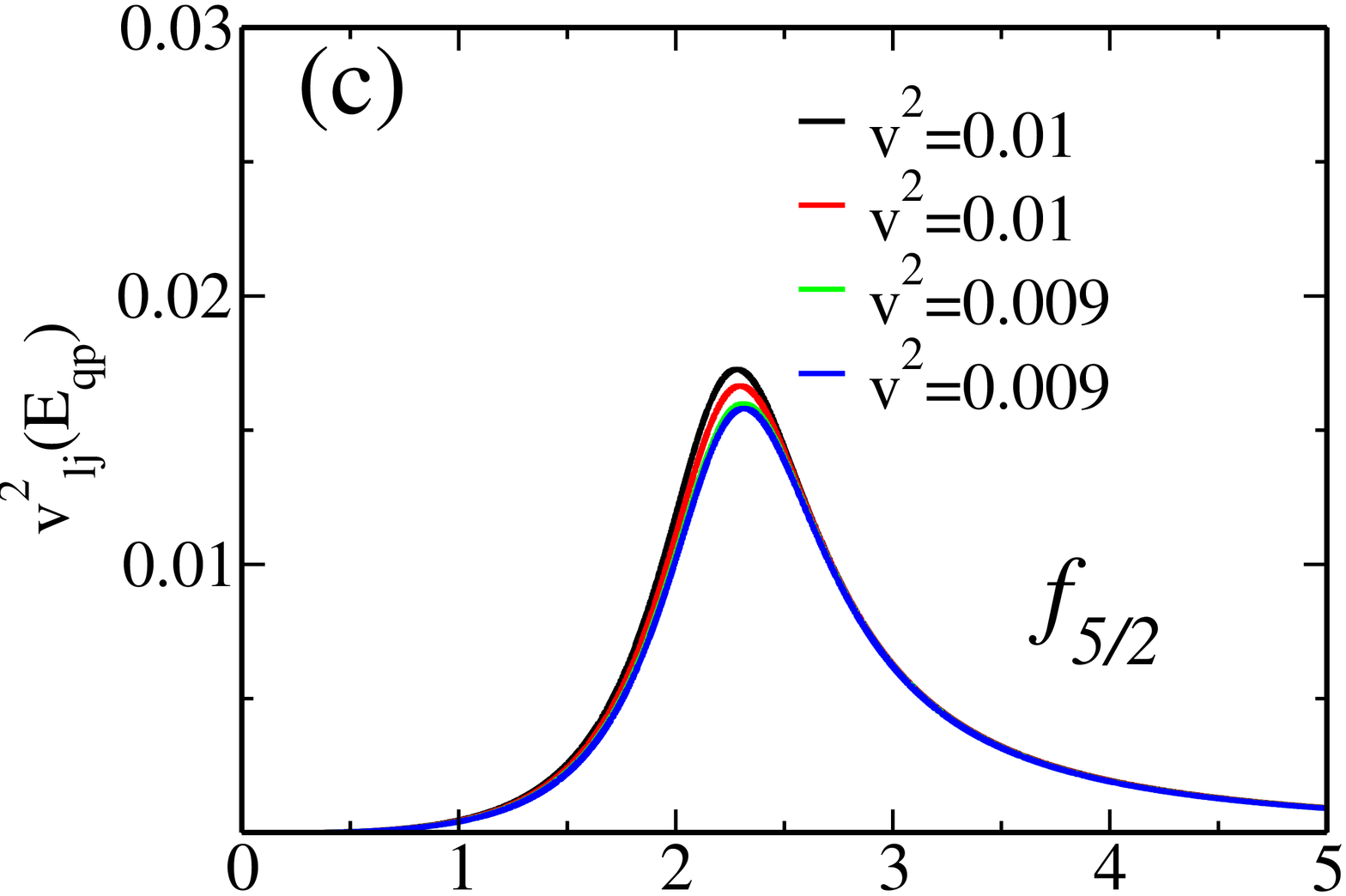}
\includegraphics[clip=,width=0.35\textwidth,angle=-90]{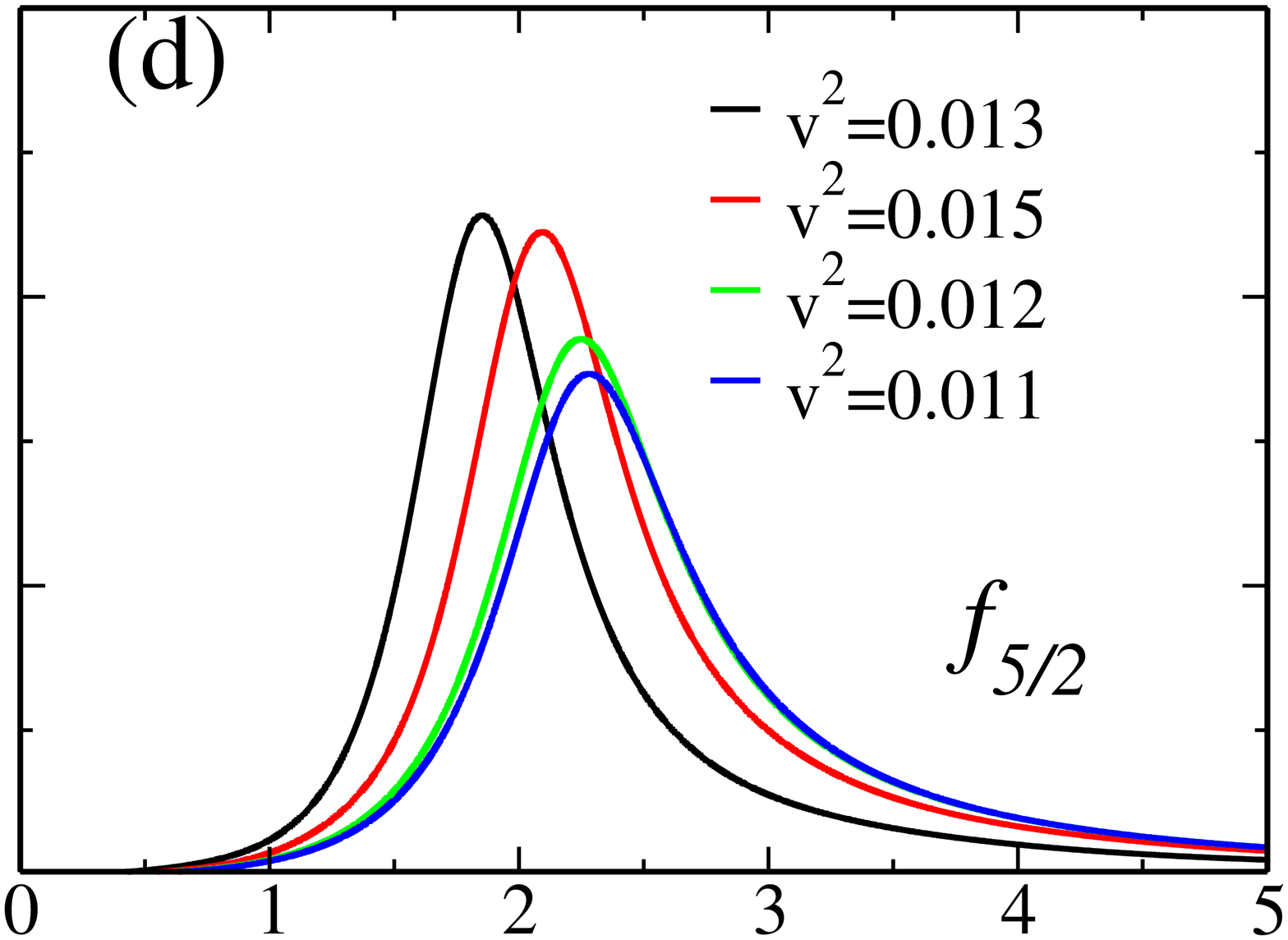}
\end{center}
\caption{(Colors online) We show the occupation probability $V^{2}_{lj}(E)$ for neutrons as a function of the 
quasi-particle energy for some given quantum numbers for $^{128}$Zr (left) and $^{166}$Zr 
(right) calculated using SLy4+DDCI functional starting from Eq.~(\ref{paper:eq:HFB-r}) and for 4 given values of the size of the box in which we perform the calculations. 
See text for more details.}
\label{paper:fig:resonancesZr}
\end{figure*}

\subsection{Detailed analysis of the resonant states}

In the previous sections, the role of  resonant states have been stressed  in order to understand the
transition between nuclei and overflowing systems.
To better describe within a theoretical framework the resonant states we decided to solve the HFB equations in r-space 
treating in a proper way the continuum (without discretization).

Defining, from the fully converged solution of Eq.~(\ref{paper:eq:HFBeq}), the hamiltonian $h(R)$ and the pairing field
$\Delta^q(R)$ in the following way,
\begin{eqnarray}
h(R)&=&\frac{\hbar^2}{2m^*_q}\left(\frac{d^2}{dR^2}-\frac{l(l+1)}{R^2}\right)-W^q(R)\nonumber\\
&&+\left(\frac{\hbar^2}{2m^*_q}\right)^\prime\frac{d}{dR}, \\
\Delta^{q}(R)&=&-\frac{V_{0}}{2}\left[1-\eta\left(\frac{\rho_{b}(R)}{\rho_{0}} \right)^\alpha \right]\nonumber\\
&&\times \sum_{nlj}(2j+1)U^{nlj,q}(R)V^{nlj,q}(R) ,
\label{eq:pairfield}
\end{eqnarray}
where $W^{q}(R)$ is the central potential and  $m^*_q(R)$ is the effective mass.
Eq.~(\ref{eq:pairfield}) is valid only in the case of a DDCI~(\ref{pairing_int_contact}) while
in the case of a finite range interaction, the pairing field is, in principle, 
a function of two variables, see for instance Eqs.~(\ref{eq:deltarrp}) and (\ref{eq21}).
The pairing field $\Delta^{q}(R)$ entering into Eq.~(\ref{paper:eq:HFB-r}) is, in such case,
defined as $\Delta^{q}(R)\equiv\Delta^{q}_{\text{LOC}}(R)$, 
where the local pairing field, $\Delta^{q}_{\text{LOC}}(R)$, is given by
\begin{equation}
\Delta^{q}_{\text{LOC}} (R) \equiv \Delta^{q}(R,k^{q}_F(R)),
\label{eq:locpairing}
\end{equation}
and the local Fermi momentum $k_{F}(R)$ is given by 
\begin{equation}\label{Fermi:mom}
 \hbar^2k_{F}^2(R)=2m^{*}_q(R)(\mu_{F}^{q}-W^{q}(R))).
\end{equation}
where $\mu_{F}^{q}$ is the chemical potential also fixed by previous calculations.

Eq.~(\ref{paper:eq:HFBeq}) can now be expressed in coordinate space, following 
Hamamoto~\emph{et al.}~\cite{Hamamoto2003}, 
\begin{eqnarray}
\left(h(R)+\mu_{F}^{q}+E^{q}_{qp}\right)U^{lj,q}(R,E) -\Delta^{q}(R)V^{lj,q}(R,E)&=&0,\nonumber \\
\left(h(R)+\mu_{F}^{q}-E^{q}_{qp}\right)V^{lj,q}(R,E) +\Delta^{q}(R)U^{lj,q}(R,E)&=&0, \nonumber \\
\label{paper:eq:HFB-r}
\end{eqnarray}
where the hamiltonian $h(R)$ and the pairing field $\Delta^q(R)$ are fixed from the
converged solution of Eq.~(\ref{paper:eq:HFBeq}).

In Eq.~(\ref{paper:eq:HFB-r}), the $U^{lj,q}(R,E)$ and $V^{lj,q}(R,E)$ amplitudes are the radial components of the 
quasi-particle wavefunctions~(\ref{eq:uvbasis}), see Ref.~\cite{Hamamoto2003} 
and references therein for  more details.
Since only a single iteration  is used to solve Eqs.~(\ref{paper:eq:HFB-r}), there is a small
lack of consistency in  this last calculation.
It is, however, the price to pay for a proper treatment of the continuum states without discretization.

We recall that in this case each quasi-particle energy $E$ is an acceptable solution and, therefore,
we can define an occupation probability as a function of the quasi-particle energy $E$ as
\begin{equation}
V^{2}_{lj}(E)=(2j+1)\int dR R^{2} V^{2}_{lj}(E,R).
\end{equation}
In Fig.~\ref{paper:fig:resonancesZr}, we represent some of the solutions of Eq.~(\ref{paper:eq:HFB-r}) 
for neutrons in $^{128}$Zr (left panels) and $^{166}$Zr (right panels) nuclei using the SLy4 Skyrme functional 
plus pairing with DDCI~(\ref{pairing_int_contact}).
The label R$_{box}$ in the upper left panel of Fig.~\ref{paper:fig:resonancesZr} 
 stands for the size of the box in which the central potential $W^{q}(R)$ and the pairing field 
 $\Delta^{q}(R)$, which enter in Eq.~(\ref{paper:eq:HFB-r}), have been obtained.
As expected, we observe that the energy of resonant states, such as $p_{3/2}$, $f_{5/2}$, $f_{7/2}$, in 
the drip line nucleus $^{128}$Zr, panels (a), (c) and (e) of Fig.~\ref{paper:fig:resonancesZr},  remain 
mostly constant as the size of the box R$_\mathrm{Box}$ increases. 
The situation can, however, be different in systems composed of a nucleus and a neutron gas, such as 
 $^{166}$Zr, since the density of the gas and the chemical potential are functions of R$_\mathrm{Box}$.
We notice, indeed, that the occupation of the resonant state $p_{3/2}$
for $^{166}$Zr changes considerably when we change the size of the box, and this is due to the different position 
of the chemical potential compared to the position of the resonant state, see Fig.~\ref{paper:fig:levZr166}.
The position of the centroid energy of the resonant state $f_{5/2}$ slightly increases as the size of the 
box increases.
It reveals the sensitivity of the $f_{5/2}$ resonant state, being closed to the threshold energy, on the mean-field 
potential which is  recalculated for each R$_{box}$.
It is however clear from the comparison of the left and right panels in Fig.~\ref{paper:fig:resonancesZr} that the 
centroid energies of the resonant states $p_{3/2}$, $f_{5/2}$, $f_{7/2}$ in $^{166}$Zr (right panels) converge to 
the associated ones in $^{128}$Zr (left panels) as the size of the box increases.

We can as well define a total occupation probability in an energy region as
\begin{equation}
V^{2}_{lj}=\int_{0}^{E_{max}} d E V^{2}_{lj}(E).
\end{equation}
We display the value of this integral for different resonant states in $^{128}$Zr and $^{166}$Zr in the labels of
Fig.~\ref{paper:fig:resonancesZr}. These integrals have been calculated in an interval between 0 and $E_{max}$=5~MeV 
using different box sizes for $W^{q}(R)$ and $\Delta^{q}(R)$. 
Comparing the results on the left and right sides of Fig.~\ref{paper:fig:resonancesZr},  we observe that there is a small fraction of particle that stays 
trapped in the resonances, that is the reason for the slow convergence of the evaporating gas to the value of the pairing 
gap of the last bound nucleus.

From Fig.~\ref{paper:fig:resonancesZr} we clearly see the resonant character of the state $f_{7/2}$. 
Its centroid is located around $\approx$0.85 MeV and its width $\approx$ 300 KeV for both 
$^{128}$Zr and $^{166}$Zr.
The result is in good agreement with the canonical basis result shown in Fig.~\ref{paper:fig:levZr166}.
Since this level is located very close to the Fermi energy we can use the approximate 
formula~\cite{Hamamoto2003}
\begin{equation}
E\approx\sqrt{(e-\lambda)^{2}+\Delta^{2}}\approx \Delta
\end{equation}
The position of this peak mostly depends on the strength of the pairing force.
It explains the stability of the resonant state $f_{7/2}$ in $^{166}$Zr as the size of the box increases.

%%%%%%%%%%%%%%%%%%%%%%%%%%%%%%%%%%%%%%%%%%%%%
%%%%%%%%%%%%%%%%%%%%%%%%%%%%%%%%%%%%%%%%%%%%%

\section{Pairing in the crust of neutron stars}\label{WSsystem}

%%%%%%%%%%%%%%%%%%%%%%%%%%%%%%%%%%%%%%%%%%%%%
%%%%%%%%%%%%%%%%%%%%%%%%%%%%%%%%%%%%%%%%%%%%%sect

\begin{table*}%\footnotesize
\setlength{\tabcolsep}{.15in}
\renewcommand{\arraystretch}{1.6}
\begin{center}
\begin{tabular}{ccccccccccc}
\hline
\hline
$\rho_B$ & $\mu_n$ & A & Z & $\Delta_{LCS}$ & $\Delta_{UV}$ & $-E_{pair}/N$ & $e_{d\frac{3}{2}}^\mathrm{can.}$ & $n_{d\frac{3}{2}}$ & $e_{g\frac{7}{2}}^\mathrm{can.}$ & $n_{g\frac{7}{2}}$ \\
$\times$10$^{10}$ g.cm$^{-3}$ & MeV & & & MeV & MeV & MeV & MeV & & MeV & \\
\hline
32.85 & -0.202 & 88 & 28 & 0.75 & 0.57 & 0.070 &-0.16 & 0.47& 1.58 & 0.05 \\
26.35 & -0.605 & 86 & 28 & 0.51 & 0.44 & 0.040 &0.08 & 0.12& 1.88 & 0.02 \\
18.55 & -1.201 & 84 & 28 & 0.49 & 0.43 & 0.034 &0.33 &0.03 & 2.16 & 0.00 \\
13.90 & -1.637 & 82 & 28 & 0.79 & 0.60 & 0.030 & 0.59& 0.03& 2.46 & 0.00 \\
10.76 & -1.873 & 80 & 28 & 0.72 & 0.55 & 0.034 & 0.83 & 0.02&2.75 & 0.00  \\
6.617 & -3.625 & 78 & 28 & 0.00 & 0.00 & 0.000 & 0.89 & 0.00 &2.66 & 0.00\\
2.118 & -5.949 & 78 & 30 & 1.06 & 0.91 & 0.095 & 1.27&0.00& 2.39 & 0.00\\
\hline
\hline
\end{tabular}
\caption{Equation of state of the outer crust based on Douchin-Haensel EoS~\cite{Douchin2001a}.}
\label{paper:tab:eosoutercrust}
\end{center}
\end{table*}

\begin{table*}%\footnotesize
\setlength{\tabcolsep}{.15in}
\renewcommand{\arraystretch}{1.6}
\begin{center}
\begin{tabular}{cccccccc}
\hline
\hline
$\rho_B$ & $\mu_n$ & A & Z & $\Delta_{LCS}$ & $\Delta_{UV}$ & $-E_{pair}/N$  \\
g.cm$^{-3}$ & MeV & & & MeV & MeV & MeV \\
\hline
%1.334$\times$10$^{14}$ & &   982 &  32 & \\
8.01$\times$10$^{13}$ & 10.776 & 1500 &  40  & 1.33 & 1.14 & 0.132\\
3.43$\times$10$^{13}$ & 7.268  & 1800 &  50 & 2.23 & 1.84 & 0.636\\
1.49$\times$10$^{13}$ & 4.759 & 1350 &  50 & 1.58 & 1.42 & 0.750\\
9.68$\times$10$^{12}$ & 3.726& 1100 &  50 & 1.38& 1.12 & 0.711\\
6.26$\times$10$^{12}$ & 2.894&   950 &  50 & 1.04 & 0.84 & 0.602\\
2.66$\times$10$^{12}$ &1.671 &   500 &  40 & 0.55 & 0.44 & 0.372\\
1.47$\times$10$^{12}$ & 1.033&   320 &  40 & 0.29& 0.23 & 0.189 \\
1.00$\times$10$^{12}$ & 0.697&  250 &   40 &0.16 & 0.14 & 0.093 \\
6.69$\times$10$^{11}$ &0.472 &  200 &   40 & 0.07& 0.10 & 0.051\\
4.67$\times$10$^{11}$ & 0.305&  180 &   40 & 0.05 & 0.10 & 0.051\\
\hline
\hline
\end{tabular}
\caption{Equation of state of the inner crust based on Negele-Vautherin~\cite{Negele1973}.}
\label{paper:tab:eosincrust}
\end{center}
\end{table*}

The inner crust of neutron stars provides an excellent frame to apply the self-consistent mean-field
theory~\cite{Grill2011,Pastore2011,Chamel2010b,Book:Margueron2012}.
The inner crust extends from the drip density $\rho_{drip} \simeq 4\times 10^{11}$ g cm$^{-3}$, where the 
neutrons start to leave from the nuclei into free space, till $\rho \simeq 1.4\times 10^{14}$ g cm$^{-3}$, where 
the transition to uniform matter takes place. 
The inner crust of neutron stars is believed to be formed by a crystal lattice of nuclear clusters embedded 
in a low-density neutron gas and ultra-relativistic electrons. 
To describe crust matter, the Wigner-Seitz (WS) approximation is widely used. 
In this approximation the crust is divided into spherical cells, each one representing an inner crust region of a 
given average density. 
The WS cells are electrically neutral and the interaction among them is neglected in many cases. 
Since the seminal calculation of Negele and Vautherin in the inner crust of neutron stars~\cite{Negele1973}, 
more refined quantal calculations at HF or HFB 
level~\cite{Sandulescu2004c,Sandulescu2004b,Sandulescu2008a,Margueron2011,Pastore2011,Barranco1998,Baldo2007} 
of different degrees of complexity within the WS approximation have been performed. 
Also semiclassical models as the Constrained Liquid Drop Model~\cite{Douchin2000} and Thomas-Fermi 
(TF) calculations including pairing correlations~\cite{Vinas2011,Schuck2011,Onsi2008,Pearson2011}  
have been used to study the crust of neutrons stars.

In the following, we will analyze the pairing properties of the lattice in the crust of neutron stars, based, 
for the outer crust, on the Douchin-Haensel equation of state~\cite{Douchin2001a} and, for the inner crust, on the 
Negele-Vautherin one~\cite{Negele1973}.
The properties of the WS cells are listed in Table~\ref{paper:tab:eosoutercrust} for the
outer crust and Table~\ref{paper:tab:eosincrust} for the inner crust.
We have not recalculated the WS configurations which minimize the energy for each of the densities
that we considered and we have preferred, as a first step, to build the pairing correlations on
WS configurations obtained from previous minimizations.
Our choice is motivated by two reasons: first, we want to compare our results with other
published previously, as in Refs.~\cite{Sandulescu2004c,Margueron2011,Pastore2011}, and
second, we did not want to introduce self-consistency by varying at the same time the pairing
interaction and the energy which would have introduced an important non-linear effect in the search scheme.

\begin{figure}
\includegraphics[clip=,width=0.35\textwidth,angle=-90]{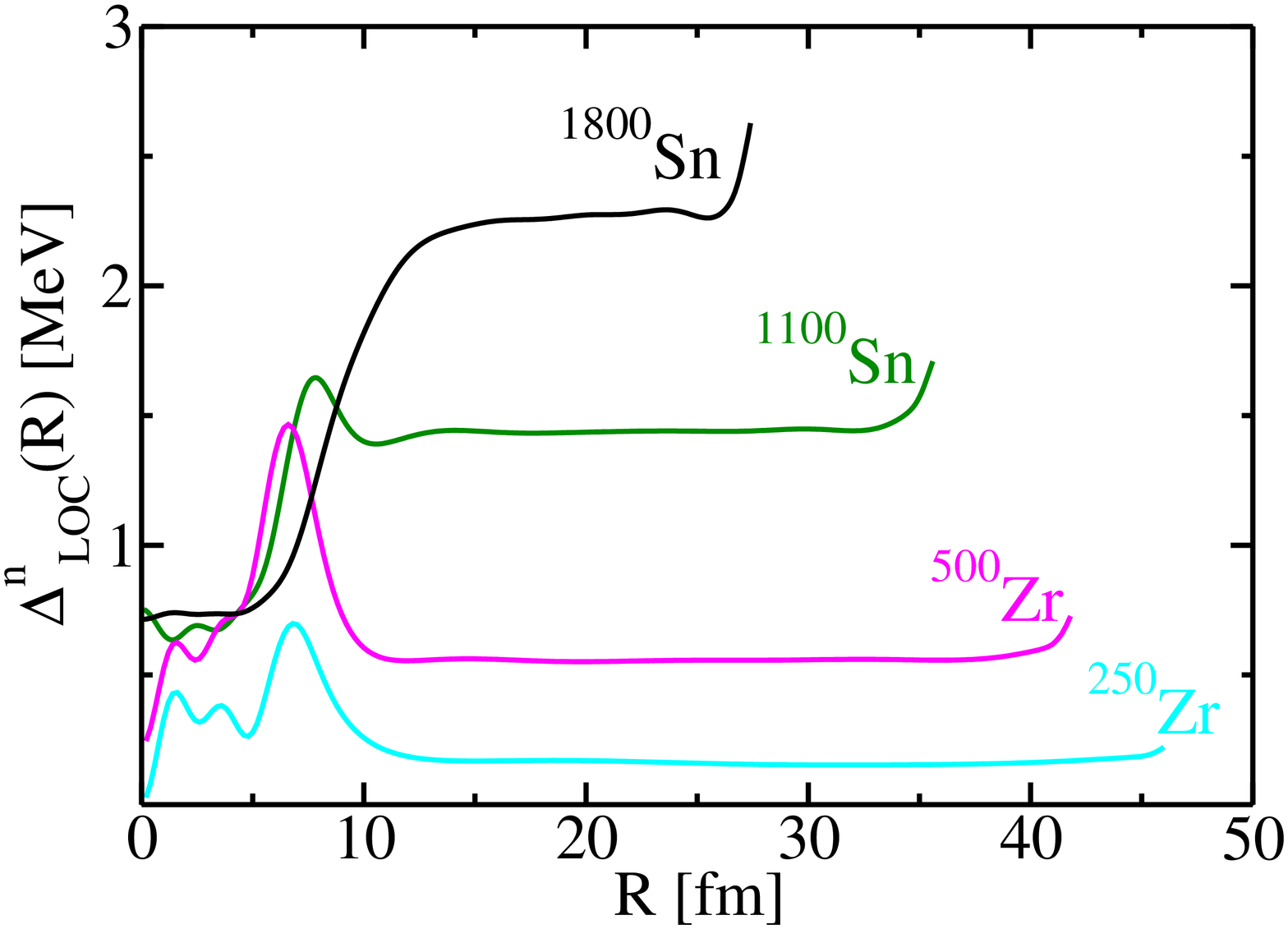}
\includegraphics[clip=,width=0.35\textwidth,angle=-90]{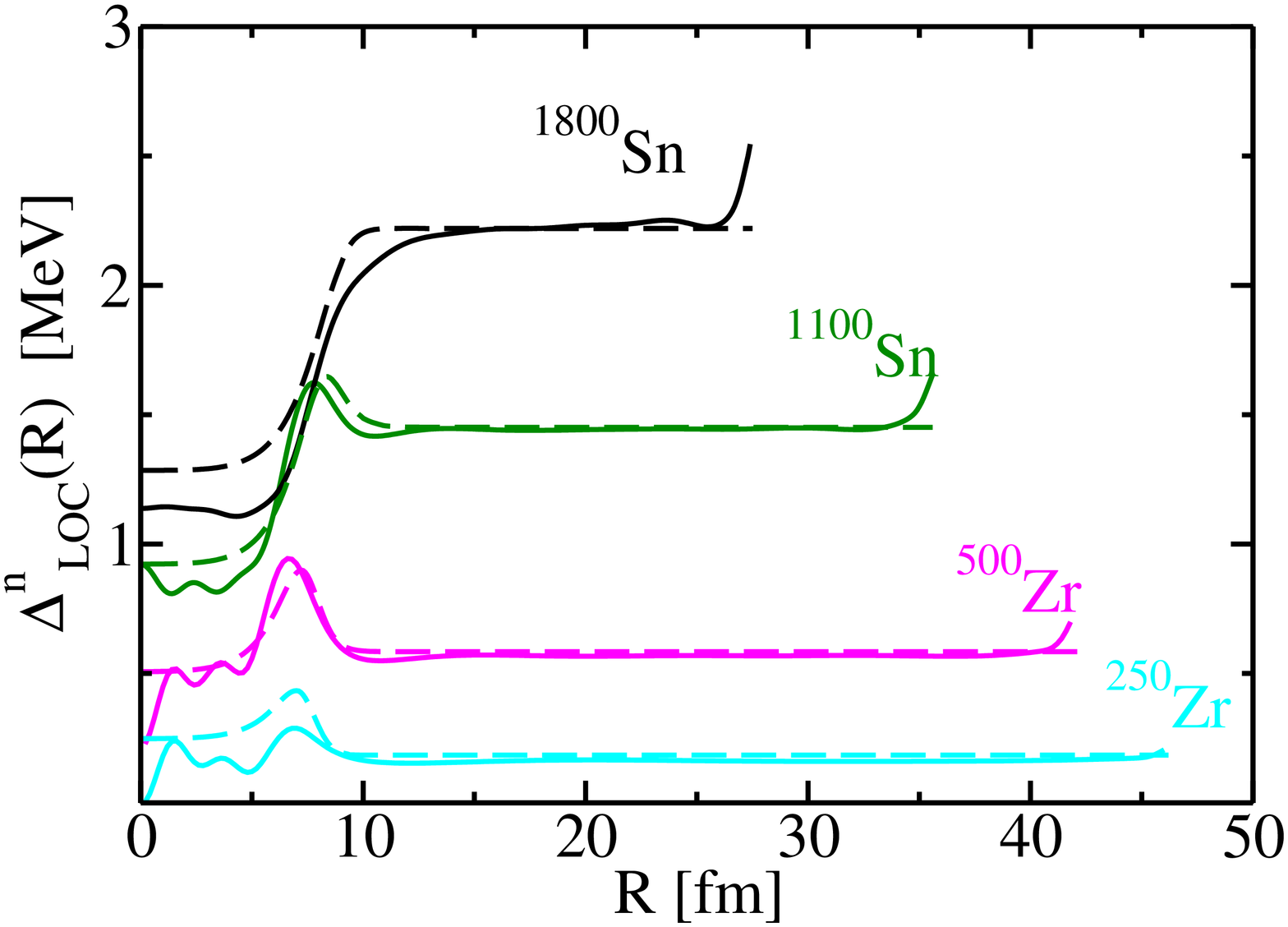}
\caption{(Colors online)
  Local neutron pairing gap $\Delta_{LOC}^{n}(R)$ for different WS cell calculated using the SLy4+SFRI. 
Lower panel: pairing gap as a function of neutron Fermi momentum, $k^{n}_{F}$, for the PNM and the WS cells.
See text for details.}
 \label{gapcrust1} 
\end{figure}

In Fig.~\ref{gapcrust1}, we represent the local pairing gap $\Delta^{n}_{LOC}(R)$, defined in 
Eq.~(\ref{eq:locpairing}), for some WS cells representative of the inner crust, which are:
$^{250}$Zr, $^{500}$Zr, $^{1100}$Sn, and $^{1800}$Sn.
On the top panel of this figure we show the results obtained by solving the full HFB equations given in 
Eq.~(\ref{paper:eq:HFBeq}), using the SLy4+SFRI model, 
while on the bottom panel we display the results obtained using the HF+BCS approximation, where only the
diagonal coupling among pairing matrix elements in Eq.(\ref{paper:eq:HFBeq})  are considered, by solid 
lines and the results computed with the TF+BCS approach~\cite{Vinas2011} by dashed lines.
For a presentation of the  Thomas-Fermi BCS approximation, see Appendix~\ref{app:TF} and references
therein.
We first discuss the HFB results shown on the top panel of Fig.~\ref{gapcrust1}.
The pairing field in the external region of the WS increases as the mass number of the WS cells increases.
This is a well known phenomenon which can qualitatively be understood from a local density approximation
in the very low density regime:
The density of the external neutron gas increases as the mass number of the cell goes up, and from
uniform matter calculations, it is known that the average pairing gap increases as a function of the
neutron density, for the cells considered in Fig.~\ref{gapcrust1}.
We observe that in the gas region both, HFB, HF+BCS, and TF+BCS approaches give the same 
value for the pairing field $\Delta^{n}_{LOC}(R)$.
It is indeed expected that HFB and BCS theories coincide in uniform matter~\cite{Book:Ring1980,Kucharek1989,Kucharek1989a}.

The peak of the pairing field at the surface of the clusters can also be roughly justified from a LDA and
it corresponds to the maximum of the pairing gap in neutron matter~\cite{Sandulescu2004b}. 
However, the peak in LDA is quantitatively much higher than in the HF+BCS and TF+BCS 
calculations~\cite{Pastore2008,Schuck2012}.

The behavior of the pairing field inside the cluster is more complex to understand.
From the HFB predictions, it is almost independent of the WS cells that we have considered, 
except for $^{250}$Zr.
In the case of $^{250}$Zr, the reduction of the pairing field compared to the other calculations is induced 
by a shell effect: a small increase of the pairing strength makes the pairing field inside this cluster 
identical to that of the other WS cells.
The independence  of the pairing field inside the cluster with respect to the outer gas density 
is typical of HFB theory as it can be seen in the upper panel of Fig.~\ref{gapcrust1}.
For the BCS approximations, shown on the bottom panel of Fig.~\ref{gapcrust1}, the pairing field inside
the cluster increases as the mass number of the cell goes up.
In the BCS approximations, the pairing properties of the gas and of the cluster are strongly coupled, 
as observed in the bottom part of Fig.~\ref{gapcrust1}, since pairing amplitudes $U$ and $V$ 
are diagonal in the HF basis.
The non-diagonal pairing matrix elements in HFB theory strongly reduce the coupling between the gas and 
the cluster, as far as the pairing correlations are concerned.

In Wigner-seitz cells, off-diagonal couplings in the pairing field play an important role in HFB theory, while they are
neglected in the BCS approximation.
Such a feature has been already remarked in finite nuclei~\cite{Dobaczewski2008a} where it was 
found that the use of the BCS approximation leads to a reduction of the gap compared to a full HFB solution. 
We leave a better analysis concerning the difference between HFB and BCS to a future work.

\begin{figure}
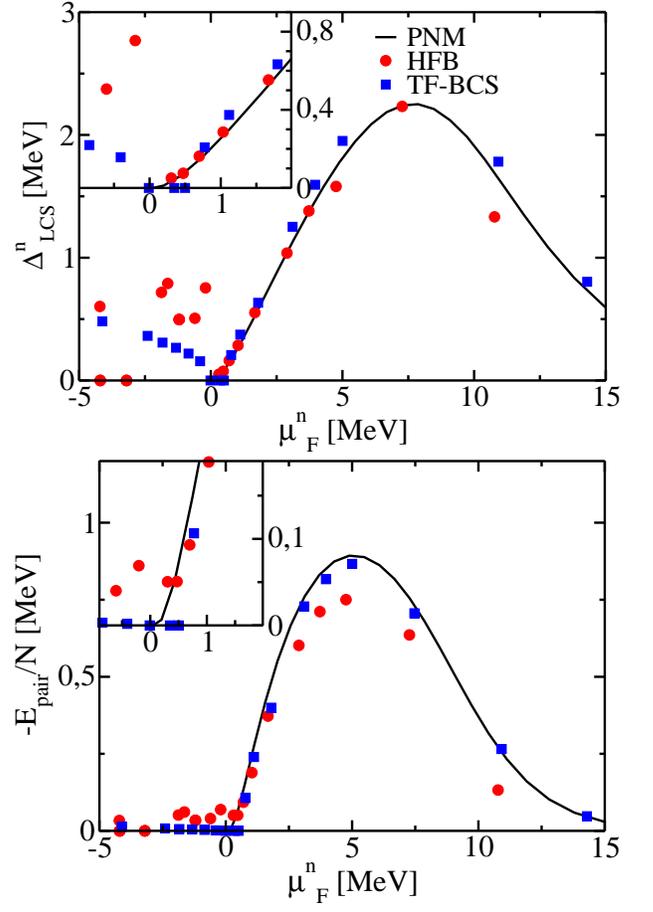

\includegraphics[clip=,width=0.45\textwidth]{13a.eps}
\includegraphics[clip=,width=0.45\textwidth]{13b.eps}
\caption{(Colors online) \label{gapcrust2} 
 Upper panel: pairing gap as a function of neutron Fermi momentum, $k^{n}_{F}$, for the PNM and the 
 WS cells for the model SLy4+SFRI. Lower panel: the pairing energy per neutron.
See text for details.}
\end{figure}

In the upper panel of Fig.~\ref{gapcrust2} we display the pairing gap at the Fermi energy
for a realistic sequence of WS cells from the outer to 
the inner crust as shown in Tab.~\ref{paper:tab:eosoutercrust} and \ref{paper:tab:eosincrust}, using 
HFB (red circles) and TF+BCS approximation (blue squares), 
while in the lower panel we show the pairing energy per neutron in the same scenario.
The pure neutron matter (PNM) solution of the BCS gap equation are also given in the two panels (solid line).
HFB and TF+BCS approaches give similar pairing gaps and pairing energies in the region of positive
chemical potential $\mu_F^n$.
In the region of negative chemical potential $\mu_F^n$, HFB gives larger pairing correlations than TF+BCS. 
This is basically due to the fact that the TF gap drops to zero at the neutron drip 
line~\cite{Vinas2011a} since in such semiclassical approaches shell effects 
are supressed.

%Notice that although in BCS approximation the off-diagonal pairing matrix elements are neglected, HF+BCS 
%calculations at the drip line do not give necessarily zero average gap if 
%quasi-bound levels owing the centrifugal barrier that simulate resonants states are included.
%See in this respect section 4.4 of Ref.[31] and M.Del Estal et al, PRC63,044321,(2001).

In the region of negative chemical potential $\mu_F^n$, which coincides with the outer-crust, the contribution of the
resonance states $d\frac{3}{2}$ and $g\frac{7}{2}$ are given in Tab.~\ref{paper:tab:eosoutercrust}.
It is shown that the resonance state $d\frac{3}{2}$ plays an important role as the chemical potential approaches
zero.
A qualitatively different behavior is observed around $\mu_F^n\approx 0$ between HFB and TF+BCS approaches,
as illustrated in Fig.~\ref{gapcrust2}.
According to the TF+BCS approach, the pairing correlations tend to vanish around $\mu_F^n\approx 0$, 
as just mentioned and as previously observed in Ref.~\cite{Schuck2011}.
In the HFB theory, where resonance states as well as off-diagonal pairing matrix coupling are included,
the pairing correlations around $\mu_F^n\approx 0$ are smaller than at stability but definitely not zero, 
as previously claimed in Ref.~\cite{Margueron2012}. 
 The fact that the TF approach gives a very much reduced gap at $\mu^n=0$ may indicate a feature of the macroscopic limit.

In the region of large positive chemical potentials, average quantities such as 
$\Delta^n_{LCS}$ and pairing energy are expected to be dominated by the neutron gas. 
However, differences in the local pairing gap inside the cluster (see Fig.~\ref{gapcrust1}) also induce 
some differences in the aforementioned average quantities (see Fig.~\ref{gapcrust2}).

The effect emphasized in the previous sections, 
e.g. the coupling to resonant states close to the drip-line, is not clearly seen in Fig.~\ref{gapcrust2}.
The reason is that  the change in neutron number is too sharp in the existing tables~\ref{paper:tab:eosoutercrust}
and \ref{paper:tab:eosincrust}, passing from $^{88}$Ni ($\mu_{F}^n<0$) to $^{180}$Zr ($\mu_{F}^n>0$).
The pairing persistence phenomenon concerns only the first neutrons that start to drip out, as we have seen 
previously.
It is now clear that a better investigation of  the transition from the outer to the inner crust will be necessary in the future,
using a smaller discretization on the values of the density of the equation of state.

%%%%%%%%%%%%%%%%%%%%%%%%%%%%%%%%%%%%%%%%%%%%%%%%%
%%%%%%%%%%%%%%%%%%%%%%%%%%%%%%%%%%%%%%%%%%%%%%%%%

\section{Conclusions}\label{conclusion}

%%%%%%%%%%%%%%%%%%%%%%%%%%%%%%%%%%%%%%%%%%%%%%%%%
%%%%%%%%%%%%%%%%%%%%%%%%%%%%%%%%%%%%%%%%%%%%%%%%%

In this work, we made a quite exhaustive study of pairing properties of nuclei around the neutron drip. 
In a LDA picture, one could have expected that neutron pairing is enhanced in such situations, 
since the neutron skin could be considered as a low density piece of neutron matter where pairing is 
at its maximum. However, nuclei are too small and the pairing force too weak for LDA being a good description. 
The reality is more complex as our investigations show. Globally pairing is certainly not enhanced going from stability to  the drip, 
rather it is reduced. However, the general feature is strongly hidden by nuclear shell effects.
The isovector dependence of the pairing gap extrapolated from measured nuclei masses is too strong, and
results as the consequence of an accumulation of closed shell nuclei at the border of the present
experimental knowledge.
From our theoretical calculations, we predict that exotic neutron rich nuclei beyond this border shall
exhibit a new raise of the pairing correlations compared to the present extrapolations yielding a much weaker average decrease to the drip than previously assumed~\cite{Yamagami2012}.

We should, moreover, mention that in this work our studies were restricted to spherical nuclei and that 
in this case and in most examples, the drip occurred at a magic, or close to magic neutron number with, 
naturally, reduced pairing correlations.
The reason for this is not entirely clear. 
Either the nuclei search to gain binding energy in approaching magicity towards the drip because gain 
in energy by pairing is weakened, or it is the other way round, i.e., pairing is reduced because anyway 
(spherical) nuclei drive to magicity at the drip.  
Furthermore, large scale nuclear mass calculations, such as for instance Gogny D1S web mass table 
(http://www-phynu.cea.fr/HFB-Gogny.htm) show that only about half of the nuclei at the neutron drip are
deformed, a ratio  much more in favor of sphericity than in the case of stable nuclei.
In the deformed cases pairing acquires more usual values. 
Nevertheless, looking at the values given in Gogny D1S web mass table,
pairing is certainly not enhanced  with respect to the stable region in such cases either.
These conclusions are based on theoretical predictions but it can be surmised that reality is not entirely different. 
In any case, the situation of pairing properties of nuclei around the neutron drip is overcast by very large shell 
fluctuations as can be seen from the various figures given in the main text. 
An additional feature which makes the situation complicated is the fact that there are resonances in the continuum 
which can be populated and which can have, in some cases, a sensible influence on the pairing properties of drip nuclei. 
This depends, for instance, on the precise position of the resonances. In particular, for cases where a strongly 
degenerate resonance level becomes located very closely to the chemical potential, pairing can become quite important 
like this is, e.g., the case for the Ca isotopes in Fig.~\ref{paper:fig:IsotopesGognyRed}.

We have also analyzed the pairing correlations in the crust of neutron stars described in the Wigner-Seitz (WS)
approach, which allowed us to study a scenario formed by nuclear clusters embedded in a low-density neutron gas. 
We see that in this situation the gap in the clusters is only weakly affected by the pairing in the neutron gas and that
the cluster and the gas behave as almost independent systems.
This result obtained from the state-of-the-art HFB theory is not reproduced by its BCS approximation, since 
BCS is less adapted for such systems as WS cells as compared with stable nuclei, at least in the case of low 
neutron density. 
When the density of the gas decreases in approaching the transition to the outer crust where all the neutrons are bound, 
the average pairing gap and the pairing energy in the WS cell decrease following the trend of the neutron matter. 
The pairing correlations, though reduced at the transition between the inner-crust and the outer-crust, 
remain, however, non-zero.
Our studies have revealed that in this situation there is a large qualitative difference between HFB theory and 
semiclassical TF+BCS approximations. We postpone a detailed investigation of this feature to a future work.

%%%%%%%%%%%%%%%%%%%%%%%%%%%%%%%%%%%%%%%%%%%%%%%%%
%%%%%%%%%%%%%%%%%%%%%%%%%%%%%%%%%%%%%%%%%%%%%%%%%

\acknowledgments

%%%%%%%%%%%%%%%%%%%%%%%%%%%%%%%%%%%%%%%%%%%%%%%%%
%%%%%%%%%%%%%%%%%%%%%%%%%%%%%%%%%%%%%%%%%%%%%%%%%

One of the authors (J. M.) thanks E. Khan for interesting discussions that have motivated this work.
We are grateful to L. Robledo for informations on the differences between HFB and BCS results
in finite nuclei.
A. P. acknowledges the hospitality of the Theory Group of the Institut de Physique Nucl\'eaire 
de Lyon during the two post-doctoral years at the University of Lyon and the interesting discussions with K. Bennaceur, T. Duguet, T. Lesinski and J. Meyer that helped in the development of the numerical code used in the present article.
This project was partially supported as well by the ANR SN2NS contract.
One of us (X.V.) acknowledges the support of the Consolider Ingenio 2010 Programme
CPAN CSD2007-00042, Grant No. FIS2011-24154 from MICINN and FEDER, and Grant
No. 2009SGR-1289 from Generalitat de Catalunya. 

\begin{appendix}
\section{Pairing field}\label{app:pairfield}

In this appendix we briefly describe the numerical procedure to obtain the pairing field in coordinate space.
Following ref.\cite{Pastore2008} we can write it as

\begin{eqnarray}\label{a:density:r} 
\Phi^{q}(\vec{r}_1,\vec{r}_2)=\sum_{\alpha\alpha'nlj} \frac{2j+1}{2} 
\left( U^{nlj,q}_{\alpha}V^{nlj,q}_{\alpha'}\right)\psi^{00}_{\alpha\alpha'lj}(\vec{r}_1,\vec{r}_2),\nonumber\\
\end{eqnarray}

\noindent where $\psi^{00}_{\alpha\alpha'lj}(\vec{r}_1,\vec{r}_2)=[\phi_{\alpha lj}(\vec{r}_1)\phi_{\alpha'lj}(\vec{r}_2)]_{00}$ is the wavefunction of two neutrons
coupled to $L=S=0$.
\begin{equation}
\psi^{00}_{\alpha \alpha 'lj}(\vec{r}_1,\vec{r}_2) = \frac{1}{4\pi}\phi_{nlj}(r_1)\phi_{n'lj}(r_2)P_l(\cos\theta_{12})\chi_{00},
\end{equation}

\noindent where $\chi_{00}$ is the total spin function of two particles coupled to S=0.
\noindent The wave function of the basis $\phi_{\alpha lj}(\vec{r}_1)$ is defined as

\begin{equation}
\phi_{\alpha,ljm}(\vec{r}) =\sum_{m_{l}m_{s}}C^{jm}_{lm_{l}\frac{1}{2}m_{s}}\frac{u_{\alpha,l}(r)}{r}Y_{lm_{l}}(\hat{r})\chi_{\frac{1}{2}m_{s}}
\end{equation}
where $u_{\alpha,l}(r)$ are defined in Eq.~(\ref{eq:uvbasis}).

As recently discussed in ref.\cite{Baroni2010} the $^{1}$S$_{0}$ component is by far the dominant one concerning calculations of pairing gaps at subnuclear densities.
We can thus immediately obtain the pairing field as

\begin{equation}
\Delta^{q}(\vec{r}_1,\vec{r}_2)=-v(|\vec{r}_1-\vec{r}_2|)\Phi^{q}(\vec{r}_1,\vec{r}_2)
\label{eq:deltarrp}
\end{equation}

\noindent where $v(|\vec{r}_1-\vec{r}_2|)$ is the pairing interaction.
Whose matrix elements  can be written as

\begin{equation}
 \Delta^{lj,q}_{\alpha\alpha'}= -\int d^{3}r_{1}\int d^{3}r_{2} \psi^{00}_{\alpha \alpha 'lj}(\vec{r}_1,\vec{r}_2)  v(|\vec{r}_1-\vec{r}_2|)\Phi^{q}(\vec{r}_1,\vec{r}_2).
\label{deltann}
\end{equation}
To define a local pairing field, we need to apply the Wigner transformation \cite{Book:Ring1980} and write the pairing field as $\Delta^{q}(R,k)$, where $R$ is the center of mass coordinate and $k$ is the relative momentum among two particles.

\section{Thomas-Fermi BCS approximation}\label{app:TF}

We present briefly the Thomas-Fermi BCS approximation~\cite{Vinas2011a,Vinas2011}. 
In such case  we take the gap in phase space \cite{Kucharek1989} at a momentum equal 
to the Fermi momentum as~\cite{Kucharek1989a}
\begin{equation}
\Delta({\bf R},{\bf k_F}) = -\int \frac{d {\bf k'}}{(2\pi)^3}
v({\bf k_F} - {\bf k'})
\kappa({\bf R},{\bf k'}).
\label{eq21}
\end{equation}

Within the TF approach to the pairing problem, the above formula  
can still be written in a different way. Quantally the 
anomalous density matrix in {\bf $r$}-space in the BCS approach is given by 
$\kappa({\bf r},{\bf r'}) =
\sum_n \kappa_n \langle{\bf r}|n\rangle \langle n|{\bf r'}\rangle$.
Therefore, after Wigner transformation, in the TF ($\hbar \to 0$) limit, one obtains

\begin{equation}
\kappa({\bf R},{\bf p}) =
 \int dE g^{TF}(E) \kappa(E) f_E({\bf R},{\bf p}),
\label{eq19}
\end{equation}
where $f_E({\bf R},{\bf p})$ is the normalized distribution function~\cite{Vinas2003}. Inserting
Eq.(\ref{eq19}) in Eq.(\ref{eq21}), using the fact that in this case
we can write the distribution function as $f_E({\bf R},{\bf p})= \delta(E - H_{cl})/g(E)$
with $H_{cl}$ the classical Hamiltonian, and performing the angular average of the pairing 
force $\tilde{v}(p,p') = \frac{1}{4 \pi} \int v({\bf p} - {\bf p'}) d\Omega$ (assuming that $\Delta({\bf 
R},{\bf p})$ and $\kappa({\bf R},{\bf p})$ are spherically symmetric in momentum space), the 
gap $\Delta({\bf R}, {\bf k=k}_F)$ can be finally recast as

\begin{eqnarray}
\Delta({\bf R},{\bf k_F}) = - \frac{1}{4 \pi^2}
\big(\frac{2m^*(\bf R)}{\hbar^2}\big)^2 \nonumber \\
\times \int dE \kappa(E) k_E ({\bf R})\tilde{v}(k_F, k_{E}({\bf R})),
\label{Delav}
\end{eqnarray}

\noindent
where $k_E({\bf R})=(\frac{2m^*({\bf R})}{\hbar^2}
(E - V({\bf R}))^{1/2}$ is the local Fermi momentum at
energy $E$.

\end{appendix}

\bibliography{biblio}

\begin{thebibliography}{70}
\expandafter\ifx\csname natexlab\endcsname\relax\def\natexlab#1{#1}\fi
\expandafter\ifx\csname bibnamefont\endcsname\relax
  \def\bibnamefont#1{#1}\fi
\expandafter\ifx\csname bibfnamefont\endcsname\relax
  \def\bibfnamefont#1{#1}\fi
\expandafter\ifx\csname citenamefont\endcsname\relax
  \def\citenamefont#1{#1}\fi
\expandafter\ifx\csname url\endcsname\relax
  \def\url#1{\texttt{#1}}\fi
\expandafter\ifx\csname urlprefix\endcsname\relax\def\urlprefix{URL }\fi
\providecommand{\bibinfo}[2]{#2}
\providecommand{\eprint}[2][]{\url{#2}}

\bibitem[{\citenamefont{Pillet et~al.}(2010)\citenamefont{Pillet, Sandulescu,
  Schuck, and Berger}}]{Pillet2010}
\bibinfo{author}{\bibfnamefont{N.}~\bibnamefont{Pillet}},
  \bibinfo{author}{\bibfnamefont{N.}~\bibnamefont{Sandulescu}},
  \bibinfo{author}{\bibfnamefont{P.}~\bibnamefont{Schuck}}, \bibnamefont{and}
  \bibinfo{author}{\bibfnamefont{J.-F.} \bibnamefont{Berger}},
  \bibinfo{journal}{Physical Review C} \textbf{\bibinfo{volume}{81}},
  \bibinfo{pages}{034307} (\bibinfo{year}{2010}).

\bibitem[{\citenamefont{Vi\~{n}as et~al.}(2010)\citenamefont{Vi\~{n}as, Schuck,
  and Pillet}}]{Vinas2010}
\bibinfo{author}{\bibfnamefont{X.}~\bibnamefont{Vi\~{n}as}},
  \bibinfo{author}{\bibfnamefont{P.}~\bibnamefont{Schuck}}, \bibnamefont{and}
  \bibinfo{author}{\bibfnamefont{N.}~\bibnamefont{Pillet}},
  \bibinfo{journal}{Physical Review C} \textbf{\bibinfo{volume}{82}},
  \bibinfo{pages}{034314} (\bibinfo{year}{2010}), ISSN
  \bibinfo{issn}{0556-2813},
  \urlprefix\url{http://link.aps.org/doi/10.1103/PhysRevC.82.034314}.

\bibitem[{\citenamefont{Pastore et~al.}(2011)\citenamefont{Pastore, Baroni, and
  Losa}}]{Pastore2011}
\bibinfo{author}{\bibfnamefont{A.}~\bibnamefont{Pastore}},
  \bibinfo{author}{\bibfnamefont{S.}~\bibnamefont{Baroni}}, \bibnamefont{and}
  \bibinfo{author}{\bibfnamefont{C.}~\bibnamefont{Losa}},
  \bibinfo{journal}{Physical Review C} \textbf{\bibinfo{volume}{84}},
  \bibinfo{pages}{065807} (\bibinfo{year}{2011}), ISSN
  \bibinfo{issn}{0556-2813},
  \urlprefix\url{http://link.aps.org/doi/10.1103/PhysRevC.84.065807}.

\bibitem[{\citenamefont{Pastore et~al.}(2008)\citenamefont{Pastore, Barranco,
  Broglia, and Vigezzi}}]{Pastore2008}
\bibinfo{author}{\bibfnamefont{A.}~\bibnamefont{Pastore}},
  \bibinfo{author}{\bibfnamefont{F.}~\bibnamefont{Barranco}},
  \bibinfo{author}{\bibfnamefont{R.~A.} \bibnamefont{Broglia}},
  \bibnamefont{and} \bibinfo{author}{\bibfnamefont{E.}~\bibnamefont{Vigezzi}},
  \bibinfo{journal}{Physical Review C} \textbf{\bibinfo{volume}{78}},
  \bibinfo{pages}{024315} (\bibinfo{year}{2008}), ISSN
  \bibinfo{issn}{0556-2813},
  \urlprefix\url{http://link.aps.org/doi/10.1103/PhysRevC.78.024315}.

\bibitem[{\citenamefont{Margueron et~al.}(2008)\citenamefont{Margueron, Sagawa,
  and Hagino}}]{Margueron2008b}
\bibinfo{author}{\bibfnamefont{J.}~\bibnamefont{Margueron}},
  \bibinfo{author}{\bibfnamefont{H.}~\bibnamefont{Sagawa}}, \bibnamefont{and}
  \bibinfo{author}{\bibfnamefont{K.}~\bibnamefont{Hagino}},
  \bibinfo{journal}{Physical Review C} \textbf{\bibinfo{volume}{77}},
  \bibinfo{pages}{054309} (\bibinfo{year}{2008}).

\bibitem[{\citenamefont{Khan et~al.}(2009)\citenamefont{Khan, Grasso, and
  Margueron}}]{Khan2009}
\bibinfo{author}{\bibfnamefont{E.}~\bibnamefont{Khan}},
  \bibinfo{author}{\bibfnamefont{M.}~\bibnamefont{Grasso}}, \bibnamefont{and}
  \bibinfo{author}{\bibfnamefont{J.}~\bibnamefont{Margueron}},
  \bibinfo{journal}{Physical Review C} \textbf{\bibinfo{volume}{80}},
  \bibinfo{pages}{044328} (\bibinfo{year}{2009}), ISSN
  \bibinfo{issn}{0556-2813},
  \urlprefix\url{http://link.aps.org/doi/10.1103/PhysRevC.80.044328}.

\bibitem[{\citenamefont{Khan et~al.}(2010)\citenamefont{Khan, Margueron,
  Col\`{o}, Hagino, and Sagawa}}]{Khan2010}
\bibinfo{author}{\bibfnamefont{E.}~\bibnamefont{Khan}},
  \bibinfo{author}{\bibfnamefont{J.}~\bibnamefont{Margueron}},
  \bibinfo{author}{\bibfnamefont{G.}~\bibnamefont{Col\`{o}}},
  \bibinfo{author}{\bibfnamefont{K.}~\bibnamefont{Hagino}}, \bibnamefont{and}
  \bibinfo{author}{\bibfnamefont{H.}~\bibnamefont{Sagawa}},
  \bibinfo{journal}{Physical Review C} \textbf{\bibinfo{volume}{82}},
  \bibinfo{pages}{024322} (\bibinfo{year}{2010}).

\bibitem[{\citenamefont{Pllumbi et~al.}(2011)\citenamefont{Pllumbi, Grasso,
  Beaumel, Khan, Margueron, and van~de Wiele}}]{Pllumbi2011}
\bibinfo{author}{\bibfnamefont{E.}~\bibnamefont{Pllumbi}},
  \bibinfo{author}{\bibfnamefont{M.}~\bibnamefont{Grasso}},
  \bibinfo{author}{\bibfnamefont{D.}~\bibnamefont{Beaumel}},
  \bibinfo{author}{\bibfnamefont{E.}~\bibnamefont{Khan}},
  \bibinfo{author}{\bibfnamefont{J.}~\bibnamefont{Margueron}},
  \bibnamefont{and} \bibinfo{author}{\bibfnamefont{J.}~\bibnamefont{van~de
  Wiele}}, \bibinfo{journal}{Physical Review C} \textbf{\bibinfo{volume}{83}},
  \bibinfo{pages}{034613} (\bibinfo{year}{2011}), ISSN
  \bibinfo{issn}{0556-2813},
  \urlprefix\url{http://link.aps.org/doi/10.1103/PhysRevC.83.034613}.

\bibitem[{\citenamefont{Potel et~al.}(2011)\citenamefont{Potel, Barranco,
  Marini, Idini, Vigezzi, and Broglia}}]{Potel2011}
\bibinfo{author}{\bibfnamefont{G.}~\bibnamefont{Potel}},
  \bibinfo{author}{\bibfnamefont{F.}~\bibnamefont{Barranco}},
  \bibinfo{author}{\bibfnamefont{F.}~\bibnamefont{Marini}},
  \bibinfo{author}{\bibfnamefont{A.}~\bibnamefont{Idini}},
  \bibinfo{author}{\bibfnamefont{E.}~\bibnamefont{Vigezzi}}, \bibnamefont{and}
  \bibinfo{author}{\bibfnamefont{R.~A.} \bibnamefont{Broglia}},
  \bibinfo{journal}{Physical Review Letters} \textbf{\bibinfo{volume}{107}},
  \bibinfo{pages}{092501} (\bibinfo{year}{2011}), ISSN
  \bibinfo{issn}{0031-9007},
  \urlprefix\url{http://link.aps.org/doi/10.1103/PhysRevLett.107.092501}.

\bibitem[{\citenamefont{Sandulescu et~al.}(2005)\citenamefont{Sandulescu,
  Schuck, and Vi\~{n}as}}]{Sandulescu2005}
\bibinfo{author}{\bibfnamefont{N.}~\bibnamefont{Sandulescu}},
  \bibinfo{author}{\bibfnamefont{P.}~\bibnamefont{Schuck}}, \bibnamefont{and}
  \bibinfo{author}{\bibfnamefont{X.}~\bibnamefont{Vi\~{n}as}},
  \bibinfo{journal}{Physical Review C} \textbf{\bibinfo{volume}{71}},
  \bibinfo{pages}{054303} (\bibinfo{year}{2005}).

\bibitem[{\citenamefont{Schuck and Vi\~{n}as}(2011)}]{Schuck2011}
\bibinfo{author}{\bibfnamefont{P.}~\bibnamefont{Schuck}} \bibnamefont{and}
  \bibinfo{author}{\bibfnamefont{X.}~\bibnamefont{Vi\~{n}as}},
  \bibinfo{journal}{Physical Review Letters} \textbf{\bibinfo{volume}{107}},
  \bibinfo{pages}{205301} (\bibinfo{year}{2011}).

\bibitem[{\citenamefont{Viverit et~al.}(2001)\citenamefont{Viverit, Giorgini,
  Pitaevskii, and Stringari}}]{Viverit2001}
\bibinfo{author}{\bibfnamefont{L.}~\bibnamefont{Viverit}},
  \bibinfo{author}{\bibfnamefont{S.}~\bibnamefont{Giorgini}},
  \bibinfo{author}{\bibfnamefont{L.}~\bibnamefont{Pitaevskii}},
  \bibnamefont{and}
  \bibinfo{author}{\bibfnamefont{S.}~\bibnamefont{Stringari}},
  \bibinfo{journal}{Physical Review A} \textbf{\bibinfo{volume}{63}},
  \bibinfo{pages}{033603} (\bibinfo{year}{2001}), ISSN
  \bibinfo{issn}{1050-2947},
  \urlprefix\url{http://link.aps.org/doi/10.1103/PhysRevA.63.033603}.

\bibitem[{\citenamefont{Margueron et~al.}(2011)\citenamefont{Margueron, Fortin,
  Grill, Page, and Sandulescu}}]{Margueron2011}
\bibinfo{author}{\bibfnamefont{J.}~\bibnamefont{Margueron}},
  \bibinfo{author}{\bibfnamefont{M.}~\bibnamefont{Fortin}},
  \bibinfo{author}{\bibfnamefont{F.}~\bibnamefont{Grill}},
  \bibinfo{author}{\bibfnamefont{D.}~\bibnamefont{Page}}, \bibnamefont{and}
  \bibinfo{author}{\bibfnamefont{N.}~\bibnamefont{Sandulescu}},
  \bibinfo{journal}{Journal of Physics: Conference Series}
  \textbf{\bibinfo{volume}{321}}, \bibinfo{pages}{012031}
  (\bibinfo{year}{2011}), ISSN \bibinfo{issn}{1742-6596},
  \urlprefix\url{http://stacks.iop.org/1742-6596/321/i=1/a=012031?key=crossref.37059220200cb37b3f0daf9012f39934}.

\bibitem[{\citenamefont{Pastore}(2012)}]{Pastore2012}
\bibinfo{author}{\bibfnamefont{A.}~\bibnamefont{Pastore}},
  \bibinfo{journal}{Physical Review C} \textbf{\bibinfo{volume}{86}},
  \bibinfo{pages}{065802} (\bibinfo{year}{2012}), ISSN
  \bibinfo{issn}{0556-2813},
  \urlprefix\url{http://link.aps.org/doi/10.1103/PhysRevC.86.065802}.

\bibitem[{\citenamefont{Lesinski et~al.}(2009)\citenamefont{Lesinski, Duguet,
  Bennaceur, and Meyer}}]{Lesinski2009}
\bibinfo{author}{\bibfnamefont{T.}~\bibnamefont{Lesinski}},
  \bibinfo{author}{\bibfnamefont{T.}~\bibnamefont{Duguet}},
  \bibinfo{author}{\bibfnamefont{K.}~\bibnamefont{Bennaceur}},
  \bibnamefont{and} \bibinfo{author}{\bibfnamefont{J.}~\bibnamefont{Meyer}},
  \bibinfo{journal}{The European Physical Journal A}
  \textbf{\bibinfo{volume}{40}}, \bibinfo{pages}{121} (\bibinfo{year}{2009}),
  ISSN \bibinfo{issn}{1434-6001},
  \urlprefix\url{http://www.springerlink.com/index/10.1140/epja/i2009-10780-y}.

\bibitem[{\citenamefont{Bender et~al.}(2000)\citenamefont{Bender, Rutz,
  Reinhard, and Maruhn}}]{Bender2000}
\bibinfo{author}{\bibfnamefont{M.}~\bibnamefont{Bender}},
  \bibinfo{author}{\bibfnamefont{K.}~\bibnamefont{Rutz}},
  \bibinfo{author}{\bibfnamefont{P.-G.} \bibnamefont{Reinhard}},
  \bibnamefont{and} \bibinfo{author}{\bibfnamefont{J.~A.}
  \bibnamefont{Maruhn}}, \bibinfo{journal}{European Physical Journal A}
  \textbf{\bibinfo{volume}{8}}, \bibinfo{pages}{59} (\bibinfo{year}{2000}),
  \urlprefix\url{http://epja.edpsciences.org/articles/epja/abs/2000/05/epja3102/epja3102.html}.

\bibitem[{\citenamefont{Ring and Schuck}(1980)}]{Book:Ring1980}
\bibinfo{author}{\bibfnamefont{P.}~\bibnamefont{Ring}} \bibnamefont{and}
  \bibinfo{author}{\bibfnamefont{P.}~\bibnamefont{Schuck}},
  \emph{\bibinfo{title}{{The Nuclear Many-Body Problem}}}
  (\bibinfo{publisher}{Springer-Verlag}, \bibinfo{year}{1980}).

\bibitem[{\citenamefont{Lesinski}(2008)}]{Thesis:Lesinski2008}
\bibinfo{author}{\bibfnamefont{T.}~\bibnamefont{Lesinski}}, Ph.D. thesis,
  \bibinfo{school}{Universit\'{e} Claude Bernard Lyon 1}
  (\bibinfo{year}{2008}).

\bibitem[{\citenamefont{Bertsch and Esbensen}(1991)}]{Bertsch1991}
\bibinfo{author}{\bibfnamefont{G.~F.} \bibnamefont{Bertsch}} \bibnamefont{and}
  \bibinfo{author}{\bibfnamefont{H.}~\bibnamefont{Esbensen}},
  \bibinfo{journal}{Annals of Physics} \textbf{\bibinfo{volume}{209}},
  \bibinfo{pages}{327} (\bibinfo{year}{1991}),
  \urlprefix\url{http://www.osti.gov/energycitations/product.biblio.jsp?osti\_id=5244244}.

\bibitem[{\citenamefont{Garrido et~al.}(1999)\citenamefont{Garrido, Sarriguren,
  {Moya De Guerra}, and Schuck}}]{Garrido1999}
\bibinfo{author}{\bibfnamefont{E.}~\bibnamefont{Garrido}},
  \bibinfo{author}{\bibfnamefont{P.}~\bibnamefont{Sarriguren}},
  \bibinfo{author}{\bibfnamefont{E.}~\bibnamefont{{Moya De Guerra}}},
  \bibnamefont{and} \bibinfo{author}{\bibfnamefont{P.}~\bibnamefont{Schuck}},
  \bibinfo{journal}{Physical Review C} \textbf{\bibinfo{volume}{60}},
  \bibinfo{pages}{064312} (\bibinfo{year}{1999}),
  \urlprefix\url{http://link.aps.org/doi/10.1103/PhysRevC.60.064312}.

\bibitem[{\citenamefont{Grill et~al.}(2011)\citenamefont{Grill, Margueron, and
  Sandulescu}}]{Grill2011}
\bibinfo{author}{\bibfnamefont{F.}~\bibnamefont{Grill}},
  \bibinfo{author}{\bibfnamefont{J.}~\bibnamefont{Margueron}},
  \bibnamefont{and}
  \bibinfo{author}{\bibfnamefont{N.}~\bibnamefont{Sandulescu}},
  \bibinfo{journal}{Physical Review C} \textbf{\bibinfo{volume}{84}},
  \bibinfo{pages}{065801} (\bibinfo{year}{2011}), \eprint{1107.4275},
  \urlprefix\url{http://arxiv.org/abs/1107.4275}.

\bibitem[{\citenamefont{Duguet}(2004)}]{Duguet2004}
\bibinfo{author}{\bibfnamefont{T.}~\bibnamefont{Duguet}},
  \bibinfo{journal}{Physical Review C} \textbf{\bibinfo{volume}{69}},
  \bibinfo{pages}{054317} (\bibinfo{year}{2004}).

\bibitem[{\citenamefont{Tian et~al.}(2009)\citenamefont{Tian, Ma, and
  Ring}}]{Tian2009}
\bibinfo{author}{\bibfnamefont{Y.}~\bibnamefont{Tian}},
  \bibinfo{author}{\bibfnamefont{Z.~Y.} \bibnamefont{Ma}}, \bibnamefont{and}
  \bibinfo{author}{\bibfnamefont{P.}~\bibnamefont{Ring}},
  \bibinfo{journal}{Physics Letters B} \textbf{\bibinfo{volume}{676}},
  \bibinfo{pages}{44} (\bibinfo{year}{2009}), ISSN \bibinfo{issn}{0370-2693},
  \urlprefix\url{http://dx.doi.org/10.1016/j.physletb.2009.04.067}.

\bibitem[{\citenamefont{Chabanat
  et~al.}(1998{\natexlab{a}})\citenamefont{Chabanat, Bonche, Haensel, Meyer,
  and Schaeffer}}]{Chabanat1998a}
\bibinfo{author}{\bibfnamefont{E.}~\bibnamefont{Chabanat}},
  \bibinfo{author}{\bibfnamefont{P.}~\bibnamefont{Bonche}},
  \bibinfo{author}{\bibfnamefont{P.}~\bibnamefont{Haensel}},
  \bibinfo{author}{\bibfnamefont{J.}~\bibnamefont{Meyer}}, \bibnamefont{and}
  \bibinfo{author}{\bibfnamefont{R.}~\bibnamefont{Schaeffer}},
  \bibinfo{journal}{Nuclear Physics A} \textbf{\bibinfo{volume}{635}},
  \bibinfo{pages}{231} (\bibinfo{year}{1998}{\natexlab{a}}).

\bibitem[{\citenamefont{Duguet et~al.}(2001)\citenamefont{Duguet, Bonche,
  Heenen, and Meyer}}]{Duguet2001b}
\bibinfo{author}{\bibfnamefont{T.}~\bibnamefont{Duguet}},
  \bibinfo{author}{\bibfnamefont{P.}~\bibnamefont{Bonche}},
  \bibinfo{author}{\bibfnamefont{P.-H.} \bibnamefont{Heenen}},
  \bibnamefont{and} \bibinfo{author}{\bibfnamefont{J.}~\bibnamefont{Meyer}},
  \bibinfo{journal}{Physical Review C} \textbf{\bibinfo{volume}{65}},
  \bibinfo{pages}{014311} (\bibinfo{year}{2001}).

\bibitem[{\citenamefont{Audi et~al.}(2003)\citenamefont{Audi, Wapstra, and
  Thibault}}]{Audi2003a}
\bibinfo{author}{\bibfnamefont{G.}~\bibnamefont{Audi}},
  \bibinfo{author}{\bibfnamefont{A.~H.} \bibnamefont{Wapstra}},
  \bibnamefont{and} \bibinfo{author}{\bibfnamefont{C.}~\bibnamefont{Thibault}},
  \bibinfo{journal}{Nuclear Physics A} \textbf{\bibinfo{volume}{729}},
  \bibinfo{pages}{337} (\bibinfo{year}{2003}).

\bibitem[{\citenamefont{Dobaczewski and Nazarewicz}(2002)}]{Dobaczewski2002b}
\bibinfo{author}{\bibfnamefont{J.}~\bibnamefont{Dobaczewski}} \bibnamefont{and}
  \bibinfo{author}{\bibfnamefont{W.}~\bibnamefont{Nazarewicz}},
  \bibinfo{journal}{Progress of Theoretical Physics Supplement}
  \textbf{\bibinfo{volume}{146}}, \bibinfo{pages}{70} (\bibinfo{year}{2002}).

\bibitem[{\citenamefont{Lesinski et~al.}(2012)\citenamefont{Lesinski, Hebeler,
  Duguet, and Schwenk}}]{Lesinski2012}
\bibinfo{author}{\bibfnamefont{T.}~\bibnamefont{Lesinski}},
  \bibinfo{author}{\bibfnamefont{K.}~\bibnamefont{Hebeler}},
  \bibinfo{author}{\bibfnamefont{T.}~\bibnamefont{Duguet}}, \bibnamefont{and}
  \bibinfo{author}{\bibfnamefont{A.}~\bibnamefont{Schwenk}},
  \bibinfo{journal}{Journal of Physics G} \textbf{\bibinfo{volume}{39}},
  \bibinfo{pages}{015108} (\bibinfo{year}{2012}), ISSN
  \bibinfo{issn}{0954-3899},
  \urlprefix\url{http://stacks.iop.org/0954-3899/39/i=1/a=015108?key=crossref.cf5f714761eebdfa044fb89076247a79}.

\bibitem[{\citenamefont{Bertsch et~al.}(2009)\citenamefont{Bertsch, Bertulani,
  Nazarewicz, Schunck, and Stoitsov}}]{Bertsch2009}
\bibinfo{author}{\bibfnamefont{G.~F.} \bibnamefont{Bertsch}},
  \bibinfo{author}{\bibfnamefont{C.~A.} \bibnamefont{Bertulani}},
  \bibinfo{author}{\bibfnamefont{W.}~\bibnamefont{Nazarewicz}},
  \bibinfo{author}{\bibfnamefont{N.}~\bibnamefont{Schunck}}, \bibnamefont{and}
  \bibinfo{author}{\bibfnamefont{M.~V.} \bibnamefont{Stoitsov}},
  \bibinfo{journal}{Physical Review C} \textbf{\bibinfo{volume}{79}},
  \bibinfo{pages}{034306} (\bibinfo{year}{2009}).

\bibitem[{\citenamefont{Schunck et~al.}(2010)\citenamefont{Schunck,
  Dobaczewski, McDonnell, Mor\'{e}, Nazarewicz, Sarich, and
  Stoitsov}}]{Schunck2010}
\bibinfo{author}{\bibfnamefont{N.}~\bibnamefont{Schunck}},
  \bibinfo{author}{\bibfnamefont{J.}~\bibnamefont{Dobaczewski}},
  \bibinfo{author}{\bibfnamefont{J.}~\bibnamefont{McDonnell}},
  \bibinfo{author}{\bibfnamefont{J.}~\bibnamefont{Mor\'{e}}},
  \bibinfo{author}{\bibfnamefont{W.}~\bibnamefont{Nazarewicz}},
  \bibinfo{author}{\bibfnamefont{J.}~\bibnamefont{Sarich}}, \bibnamefont{and}
  \bibinfo{author}{\bibfnamefont{M.~V.} \bibnamefont{Stoitsov}},
  \bibinfo{journal}{Physical Review C} \textbf{\bibinfo{volume}{81}},
  \bibinfo{pages}{024316} (\bibinfo{year}{2010}), ISSN
  \bibinfo{issn}{0556-2813},
  \urlprefix\url{http://link.aps.org/doi/10.1103/PhysRevC.81.024316}.

\bibitem[{\citenamefont{Margueron et~al.}(2009)\citenamefont{Margueron,
  Goriely, Grasso, Col\`{o}, and Sagawa}}]{Margueron2009a}
\bibinfo{author}{\bibfnamefont{J.}~\bibnamefont{Margueron}},
  \bibinfo{author}{\bibfnamefont{S.}~\bibnamefont{Goriely}},
  \bibinfo{author}{\bibfnamefont{M.}~\bibnamefont{Grasso}},
  \bibinfo{author}{\bibfnamefont{G.}~\bibnamefont{Col\`{o}}}, \bibnamefont{and}
  \bibinfo{author}{\bibfnamefont{H.}~\bibnamefont{Sagawa}},
  \bibinfo{journal}{Journal of Physics G: Nuclear and Particle Physics}
  \textbf{\bibinfo{volume}{36}}, \bibinfo{pages}{125103}
  (\bibinfo{year}{2009}), ISSN \bibinfo{issn}{0954-3899},
  \urlprefix\url{http://stacks.iop.org/0954-3899/36/i=12/a=125103?key=crossref.b69e6380174e1f3f47631968df14b4cf}.

\bibitem[{\citenamefont{Dobaczewski et~al.}(2001)\citenamefont{Dobaczewski,
  Magierski, Nazarewicz, Satuła, and Szymanski}}]{Dobaczewski2001b}
\bibinfo{author}{\bibfnamefont{J.}~\bibnamefont{Dobaczewski}},
  \bibinfo{author}{\bibfnamefont{P.}~\bibnamefont{Magierski}},
  \bibinfo{author}{\bibfnamefont{W.}~\bibnamefont{Nazarewicz}},
  \bibinfo{author}{\bibfnamefont{W.}~\bibnamefont{Satuła}}, \bibnamefont{and}
  \bibinfo{author}{\bibfnamefont{Z.}~\bibnamefont{Szymanski}},
  \bibinfo{journal}{Physical Review} \textbf{\bibinfo{volume}{63}},
  \bibinfo{pages}{024308} (\bibinfo{year}{2001}).

\bibitem[{\citenamefont{Hebeler et~al.}(2009)\citenamefont{Hebeler, Duguet,
  Lesinski, and Schwenk}}]{Hebeler2009}
\bibinfo{author}{\bibfnamefont{K.}~\bibnamefont{Hebeler}},
  \bibinfo{author}{\bibfnamefont{T.}~\bibnamefont{Duguet}},
  \bibinfo{author}{\bibfnamefont{T.}~\bibnamefont{Lesinski}}, \bibnamefont{and}
  \bibinfo{author}{\bibfnamefont{A.}~\bibnamefont{Schwenk}},
  \bibinfo{journal}{Physical Review C} \textbf{\bibinfo{volume}{80}},
  \bibinfo{pages}{044321} (\bibinfo{year}{2009}), ISSN
  \bibinfo{issn}{0556-2813},
  \urlprefix\url{http://link.aps.org/doi/10.1103/PhysRevC.80.044321}.

\bibitem[{\citenamefont{Barranco et~al.}(1999)\citenamefont{Barranco, Broglia,
  Gori, Vigezzi, Bortignon, and Terasaki}}]{Barranco1999}
\bibinfo{author}{\bibfnamefont{F.}~\bibnamefont{Barranco}},
  \bibinfo{author}{\bibfnamefont{R.~A.} \bibnamefont{Broglia}},
  \bibinfo{author}{\bibfnamefont{G.}~\bibnamefont{Gori}},
  \bibinfo{author}{\bibfnamefont{E.}~\bibnamefont{Vigezzi}},
  \bibinfo{author}{\bibfnamefont{P.~F.} \bibnamefont{Bortignon}},
  \bibnamefont{and} \bibinfo{author}{\bibfnamefont{J.}~\bibnamefont{Terasaki}},
  \bibinfo{journal}{Physical Review Letters} \textbf{\bibinfo{volume}{83}},
  \bibinfo{pages}{2147} (\bibinfo{year}{1999}).

\bibitem[{\citenamefont{Pastore et~al.}(2009)\citenamefont{Pastore, Barranco,
  Broglia, and Vigezzi}}]{Pastore2009}
\bibinfo{author}{\bibfnamefont{A.}~\bibnamefont{Pastore}},
  \bibinfo{author}{\bibfnamefont{F.}~\bibnamefont{Barranco}},
  \bibinfo{author}{\bibfnamefont{R.~A.} \bibnamefont{Broglia}},
  \bibnamefont{and} \bibinfo{author}{\bibfnamefont{E.}~\bibnamefont{Vigezzi}},
  \bibinfo{journal}{Journal of Physics: Conference Series}
  \textbf{\bibinfo{volume}{168}}, \bibinfo{pages}{012015}
  (\bibinfo{year}{2009}), ISSN \bibinfo{issn}{1742-6596},
  \urlprefix\url{http://stacks.iop.org/1742-6596/168/i=1/a=012015?key=crossref.17c345e2421980e61211c47610c2a01a}.

\bibitem[{\citenamefont{Chabanat et~al.}(1997)\citenamefont{Chabanat, Bonche,
  Haensel, Meyer, and Schaeffer}}]{Chabanat1997}
\bibinfo{author}{\bibfnamefont{E.}~\bibnamefont{Chabanat}},
  \bibinfo{author}{\bibfnamefont{P.}~\bibnamefont{Bonche}},
  \bibinfo{author}{\bibfnamefont{P.}~\bibnamefont{Haensel}},
  \bibinfo{author}{\bibfnamefont{J.}~\bibnamefont{Meyer}}, \bibnamefont{and}
  \bibinfo{author}{\bibfnamefont{R.}~\bibnamefont{Schaeffer}},
  \bibinfo{journal}{Nuclear Physics A} \textbf{\bibinfo{volume}{627}},
  \bibinfo{pages}{710} (\bibinfo{year}{1997}).

\bibitem[{\citenamefont{Chabanat
  et~al.}(1998{\natexlab{b}})\citenamefont{Chabanat, Bonche, Haensel, Meyer,
  and Schaeffer}}]{Chabanat1998b}
\bibinfo{author}{\bibfnamefont{E.}~\bibnamefont{Chabanat}},
  \bibinfo{author}{\bibfnamefont{P.}~\bibnamefont{Bonche}},
  \bibinfo{author}{\bibfnamefont{P.}~\bibnamefont{Haensel}},
  \bibinfo{author}{\bibfnamefont{J.}~\bibnamefont{Meyer}}, \bibnamefont{and}
  \bibinfo{author}{\bibfnamefont{R.}~\bibnamefont{Schaeffer}},
  \bibinfo{journal}{Nuclear Physics A} \textbf{\bibinfo{volume}{643}},
  \bibinfo{pages}{441} (\bibinfo{year}{1998}{\natexlab{b}}).

\bibitem[{\citenamefont{Lesinski et~al.}(2006)\citenamefont{Lesinski,
  Bennaceur, Duguet, and Meyer}}]{Lesinski2006}
\bibinfo{author}{\bibfnamefont{T.}~\bibnamefont{Lesinski}},
  \bibinfo{author}{\bibfnamefont{K.}~\bibnamefont{Bennaceur}},
  \bibinfo{author}{\bibfnamefont{T.}~\bibnamefont{Duguet}}, \bibnamefont{and}
  \bibinfo{author}{\bibfnamefont{J.}~\bibnamefont{Meyer}},
  \bibinfo{journal}{Physical Review C} \textbf{\bibinfo{volume}{74}},
  \bibinfo{pages}{044315} (\bibinfo{year}{2006}).

\bibitem[{\citenamefont{Yamagami and Shimizu}(2008)}]{yamagami2008}
\bibinfo{author}{\bibfnamefont{M.}~\bibnamefont{Yamagami}} \bibnamefont{and}
  \bibinfo{author}{\bibfnamefont{Y.~R.} \bibnamefont{Shimizu}},
  \bibinfo{journal}{Physical Review C} \textbf{\bibinfo{volume}{77}},
  \bibinfo{pages}{064319} (\bibinfo{year}{2008}), ISSN
  \bibinfo{issn}{0556-2813},
  \urlprefix\url{http://link.aps.org/doi/10.1103/PhysRevC.77.064319}.

\bibitem[{\citenamefont{Yamagami et~al.}(2012)\citenamefont{Yamagami,
  Margueron, Sagawa, and Hagino}}]{Yamagami2012}
\bibinfo{author}{\bibfnamefont{M.}~\bibnamefont{Yamagami}},
  \bibinfo{author}{\bibfnamefont{J.}~\bibnamefont{Margueron}},
  \bibinfo{author}{\bibfnamefont{H.}~\bibnamefont{Sagawa}}, \bibnamefont{and}
  \bibinfo{author}{\bibfnamefont{K.}~\bibnamefont{Hagino}},
  \bibinfo{journal}{Physical Review C} \textbf{\bibinfo{volume}{86}},
  \bibinfo{pages}{034333} (\bibinfo{year}{2012}), ISSN
  \bibinfo{issn}{0556-2813},
  \urlprefix\url{http://link.aps.org/doi/10.1103/PhysRevC.86.034333}.

\bibitem[{\citenamefont{Vogel et~al.}(1984)\citenamefont{Vogel, Jonson, and
  Hansen}}]{Vogel1984}
\bibinfo{author}{\bibfnamefont{P.}~\bibnamefont{Vogel}},
  \bibinfo{author}{\bibfnamefont{B.}~\bibnamefont{Jonson}}, \bibnamefont{and}
  \bibinfo{author}{\bibfnamefont{P.~G.} \bibnamefont{Hansen}},
  \bibinfo{journal}{Physics Letters B} \textbf{\bibinfo{volume}{139}},
  \bibinfo{pages}{227} (\bibinfo{year}{1984}).

\bibitem[{\citenamefont{Gambacurta et~al.}(2011)\citenamefont{Gambacurta, Li,
  Col\`{o}, Lombardo, {Van Giai}, and Zuo}}]{Gambacurta2011}
\bibinfo{author}{\bibfnamefont{D.}~\bibnamefont{Gambacurta}},
  \bibinfo{author}{\bibfnamefont{L.}~\bibnamefont{Li}},
  \bibinfo{author}{\bibfnamefont{G.}~\bibnamefont{Col\`{o}}},
  \bibinfo{author}{\bibfnamefont{U.}~\bibnamefont{Lombardo}},
  \bibinfo{author}{\bibfnamefont{N.}~\bibnamefont{{Van Giai}}},
  \bibnamefont{and} \bibinfo{author}{\bibfnamefont{W.}~\bibnamefont{Zuo}},
  \bibinfo{journal}{Physical Review C} \textbf{\bibinfo{volume}{84}},
  \bibinfo{pages}{024301} (\bibinfo{year}{2011}), ISSN
  \bibinfo{issn}{0556-2813},
  \urlprefix\url{http://link.aps.org/doi/10.1103/PhysRevC.84.024301}.

\bibitem[{\citenamefont{Margueron and Khan}(2012)}]{Margueron2012}
\bibinfo{author}{\bibfnamefont{J.}~\bibnamefont{Margueron}} \bibnamefont{and}
  \bibinfo{author}{\bibfnamefont{E.}~\bibnamefont{Khan}},
  \bibinfo{journal}{Physical Review C} \textbf{\bibinfo{volume}{86}},
  \bibinfo{pages}{065801} (\bibinfo{year}{2012}), ISSN
  \bibinfo{issn}{0556-2813},
  \urlprefix\url{http://link.aps.org/doi/10.1103/PhysRevC.86.065801}.

\bibitem[{\citenamefont{Lombardo and Schulze}(2001)}]{Book:Lombardo2001}
\bibinfo{author}{\bibfnamefont{U.}~\bibnamefont{Lombardo}} \bibnamefont{and}
  \bibinfo{author}{\bibfnamefont{H.-J.} \bibnamefont{Schulze}}, in
  \emph{\bibinfo{booktitle}{Physics of Neutron stars interiors}}, edited by
  \bibinfo{editor}{\bibfnamefont{D.}~\bibnamefont{Blaschke}},
  \bibinfo{editor}{\bibfnamefont{A.}~\bibnamefont{Sedrakian}},
  \bibnamefont{and} \bibinfo{editor}{\bibfnamefont{N.~K.}
  \bibnamefont{Glendenning}} (\bibinfo{publisher}{Springer Berlin Heidelberg},
  \bibinfo{address}{Berlin, Heidelberg}, \bibinfo{year}{2001}),
  chap.~\bibinfo{chapter}{2}, pp. \bibinfo{pages}{30--53}, ISBN
  \bibinfo{isbn}{978-3-540-42340-9}, \eprint{0012209},
  \urlprefix\url{http://arxiv.org/abs/astro-ph/0012209
  http://www.springerlink.com/index/10.1007/3-540-44578-1}.

\bibitem[{\citenamefont{Bohr and Mottelson}(1998)}]{Book:Bohr1998}
\bibinfo{author}{\bibfnamefont{A.}~\bibnamefont{Bohr}} \bibnamefont{and}
  \bibinfo{author}{\bibfnamefont{B.~R.} \bibnamefont{Mottelson}},
  \emph{\bibinfo{title}{{Nuclear Structure vol. II}}}
  (\bibinfo{publisher}{World Scientific}, \bibinfo{year}{1998}).

\bibitem[{\citenamefont{Pei et~al.}(2011)\citenamefont{Pei, Kruppa, and
  Nazarewicz}}]{Pei2011}
\bibinfo{author}{\bibfnamefont{J.~C.} \bibnamefont{Pei}},
  \bibinfo{author}{\bibfnamefont{A.~T.} \bibnamefont{Kruppa}},
  \bibnamefont{and}
  \bibinfo{author}{\bibfnamefont{W.}~\bibnamefont{Nazarewicz}},
  \bibinfo{journal}{Physical Review C} \textbf{\bibinfo{volume}{84}},
  \bibinfo{pages}{024311} (\bibinfo{year}{2011}), ISSN
  \bibinfo{issn}{0556-2813},
  \urlprefix\url{http://link.aps.org/doi/10.1103/PhysRevC.84.024311}.

\bibitem[{\citenamefont{Grasso et~al.}(2008)\citenamefont{Grasso, Khan,
  Margueron, and {Van Giai}}}]{Grasso2008}
\bibinfo{author}{\bibfnamefont{M.}~\bibnamefont{Grasso}},
  \bibinfo{author}{\bibfnamefont{E.}~\bibnamefont{Khan}},
  \bibinfo{author}{\bibfnamefont{J.}~\bibnamefont{Margueron}},
  \bibnamefont{and} \bibinfo{author}{\bibfnamefont{N.}~\bibnamefont{{Van
  Giai}}}, \bibinfo{journal}{Nuclear Physics A} \textbf{\bibinfo{volume}{807}},
  \bibinfo{pages}{1} (\bibinfo{year}{2008}), ISSN \bibinfo{issn}{03759474},
  \urlprefix\url{http://linkinghub.elsevier.com/retrieve/pii/S0375947408004491}.

\bibitem[{\citenamefont{Pizzochero et~al.}(2002)\citenamefont{Pizzochero,
  Barranco, Vigezzi, and Broglia}}]{Pizzochero2002}
\bibinfo{author}{\bibfnamefont{P.~A.~M.} \bibnamefont{Pizzochero}},
  \bibinfo{author}{\bibfnamefont{F.}~\bibnamefont{Barranco}},
  \bibinfo{author}{\bibfnamefont{E.}~\bibnamefont{Vigezzi}}, \bibnamefont{and}
  \bibinfo{author}{\bibfnamefont{R.~A.} \bibnamefont{Broglia}},
  \bibinfo{journal}{The Astrophysical Journal} \textbf{\bibinfo{volume}{569}},
  \bibinfo{pages}{381} (\bibinfo{year}{2002}).

\bibitem[{\citenamefont{Baldo et~al.}(2006)\citenamefont{Baldo, Saperstein, and
  Tolokonnikov}}]{Baldo2006}
\bibinfo{author}{\bibfnamefont{M.}~\bibnamefont{Baldo}},
  \bibinfo{author}{\bibfnamefont{E.~E.} \bibnamefont{Saperstein}},
  \bibnamefont{and} \bibinfo{author}{\bibfnamefont{S.~V.}
  \bibnamefont{Tolokonnikov}}, \bibinfo{journal}{Nuclear Physics A}
  \textbf{\bibinfo{volume}{775}}, \bibinfo{pages}{235} (\bibinfo{year}{2006}).

\bibitem[{\citenamefont{Chamel et~al.}(2010)\citenamefont{Chamel, Goriely,
  Pearson, and Onsi}}]{Chamel2010b}
\bibinfo{author}{\bibfnamefont{N.}~\bibnamefont{Chamel}},
  \bibinfo{author}{\bibfnamefont{S.}~\bibnamefont{Goriely}},
  \bibinfo{author}{\bibfnamefont{J.~M.} \bibnamefont{Pearson}},
  \bibnamefont{and} \bibinfo{author}{\bibfnamefont{M.}~\bibnamefont{Onsi}},
  \bibinfo{journal}{Physical Review C} \textbf{\bibinfo{volume}{81}},
  \bibinfo{pages}{045804} (\bibinfo{year}{2010}), ISSN
  \bibinfo{issn}{0556-2813},
  \urlprefix\url{http://link.aps.org/doi/10.1103/PhysRevC.81.045804}.

\bibitem[{\citenamefont{Hamamoto and Mottelson}(2003)}]{Hamamoto2003}
\bibinfo{author}{\bibfnamefont{I.}~\bibnamefont{Hamamoto}} \bibnamefont{and}
  \bibinfo{author}{\bibfnamefont{B.~R.} \bibnamefont{Mottelson}},
  \bibinfo{journal}{Physical Review C} \textbf{\bibinfo{volume}{68}},
  \bibinfo{pages}{034312} (\bibinfo{year}{2003}), ISSN
  \bibinfo{issn}{0556-2813},
  \urlprefix\url{http://link.aps.org/doi/10.1103/PhysRevC.68.034312}.

\bibitem[{\citenamefont{Douchin and Haensel}(2001)}]{Douchin2001a}
\bibinfo{author}{\bibfnamefont{F.}~\bibnamefont{Douchin}} \bibnamefont{and}
  \bibinfo{author}{\bibfnamefont{P.}~\bibnamefont{Haensel}},
  \bibinfo{journal}{Astronomy and Astrophysics} \textbf{\bibinfo{volume}{380}},
  \bibinfo{pages}{151} (\bibinfo{year}{2001}).

\bibitem[{\citenamefont{Negele and Vautherin}(1973)}]{Negele1973}
\bibinfo{author}{\bibfnamefont{J.~W.} \bibnamefont{Negele}} \bibnamefont{and}
  \bibinfo{author}{\bibfnamefont{D.}~\bibnamefont{Vautherin}},
  \bibinfo{journal}{Nuclear Physics A} \textbf{\bibinfo{volume}{207}},
  \bibinfo{pages}{298} (\bibinfo{year}{1973}).

\bibitem[{\citenamefont{Margueron and Sandulescu}(2012)}]{Book:Margueron2012}
\bibinfo{author}{\bibfnamefont{J.}~\bibnamefont{Margueron}} \bibnamefont{and}
  \bibinfo{author}{\bibfnamefont{N.}~\bibnamefont{Sandulescu}}, in
  \emph{\bibinfo{booktitle}{Neutron Star crust}}, edited by
  \bibinfo{editor}{\bibfnamefont{C.~A.} \bibnamefont{Bertulani}}
  \bibnamefont{and}
  \bibinfo{editor}{\bibfnamefont{J.}~\bibnamefont{Piekarewicz}}
  (\bibinfo{publisher}{World Scientific}, \bibinfo{year}{2012}), pp.
  \bibinfo{pages}{68--86}.

\bibitem[{\citenamefont{Sandulescu et~al.}(2004)\citenamefont{Sandulescu, {Van
  Giai}, and Liotta}}]{Sandulescu2004c}
\bibinfo{author}{\bibfnamefont{N.}~\bibnamefont{Sandulescu}},
  \bibinfo{author}{\bibfnamefont{N.}~\bibnamefont{{Van Giai}}},
  \bibnamefont{and} \bibinfo{author}{\bibfnamefont{R.~J.}
  \bibnamefont{Liotta}}, \bibinfo{journal}{Physical Review C}
  \textbf{\bibinfo{volume}{69}}, \bibinfo{pages}{045802}
  (\bibinfo{year}{2004}).

\bibitem[{\citenamefont{Sandulescu}(2004)}]{Sandulescu2004b}
\bibinfo{author}{\bibfnamefont{N.}~\bibnamefont{Sandulescu}},
  \bibinfo{journal}{Physical Review C} \textbf{\bibinfo{volume}{70}},
  \bibinfo{pages}{025801} (\bibinfo{year}{2004}), ISSN
  \bibinfo{issn}{0556-2813},
  \urlprefix\url{http://link.aps.org/doi/10.1103/PhysRevC.70.025801}.

\bibitem[{\citenamefont{Sandulescu}(2008)}]{Sandulescu2008a}
\bibinfo{author}{\bibfnamefont{N.}~\bibnamefont{Sandulescu}},
  \bibinfo{journal}{The European Physical Journal Special Topics}
  \textbf{\bibinfo{volume}{156}}, \bibinfo{pages}{265} (\bibinfo{year}{2008}),
  ISSN \bibinfo{issn}{1951-6355},
  \urlprefix\url{http://www.springerlink.com/index/10.1140/epjst/e2008-00624-0}.

\bibitem[{\citenamefont{Barranco et~al.}(1998)\citenamefont{Barranco, Broglia,
  Esbensen, and Vigezzi}}]{Barranco1998}
\bibinfo{author}{\bibfnamefont{F.}~\bibnamefont{Barranco}},
  \bibinfo{author}{\bibfnamefont{R.~A.} \bibnamefont{Broglia}},
  \bibinfo{author}{\bibfnamefont{H.}~\bibnamefont{Esbensen}}, \bibnamefont{and}
  \bibinfo{author}{\bibfnamefont{E.}~\bibnamefont{Vigezzi}},
  \bibinfo{journal}{Physical Review C} \textbf{\bibinfo{volume}{58}},
  \bibinfo{pages}{1257} (\bibinfo{year}{1998}), ISSN \bibinfo{issn}{0556-2813},
  \urlprefix\url{http://link.aps.org/doi/10.1103/PhysRevC.58.1257}.

\bibitem[{\citenamefont{Baldo et~al.}(2007)\citenamefont{Baldo, Saperstein, and
  Tolokonnikov}}]{Baldo2007}
\bibinfo{author}{\bibfnamefont{M.}~\bibnamefont{Baldo}},
  \bibinfo{author}{\bibfnamefont{E.~E.} \bibnamefont{Saperstein}},
  \bibnamefont{and} \bibinfo{author}{\bibfnamefont{S.~V.}
  \bibnamefont{Tolokonnikov}}, \bibinfo{journal}{European Physical Journal A}
  \textbf{\bibinfo{volume}{32}}, \bibinfo{pages}{97} (\bibinfo{year}{2007}),
  ISSN \bibinfo{issn}{1434-6001},
  \urlprefix\url{http://www.springerlink.com/index/10.1140/epja/i2006-10356-5
  http://epja.edpsciences.org/articles/epja/abs/2007/05/10050\_2007\_Article\_100288/10050\_2007\_Article\_100288.html}.

\bibitem[{\citenamefont{Douchin and Haensel}(2000)}]{Douchin2000}
\bibinfo{author}{\bibfnamefont{F.}~\bibnamefont{Douchin}} \bibnamefont{and}
  \bibinfo{author}{\bibfnamefont{P.}~\bibnamefont{Haensel}},
  \bibinfo{journal}{Physics Letters B} \textbf{\bibinfo{volume}{485}},
  \bibinfo{pages}{107} (\bibinfo{year}{2000}),
  \urlprefix\url{http://linkinghub.elsevier.com/retrieve/pii/S0370269300006729}.

\bibitem[{\citenamefont{Vi\~{n}as
  et~al.}(2011{\natexlab{a}})\citenamefont{Vi\~{n}as, Schuck, and
  Farine}}]{Vinas2011}
\bibinfo{author}{\bibfnamefont{X.}~\bibnamefont{Vi\~{n}as}},
  \bibinfo{author}{\bibfnamefont{P.}~\bibnamefont{Schuck}}, \bibnamefont{and}
  \bibinfo{author}{\bibfnamefont{M.}~\bibnamefont{Farine}},
  \bibinfo{journal}{Journal of Physics: Conference Series}
  \textbf{\bibinfo{volume}{321}}, \bibinfo{pages}{012024}
  (\bibinfo{year}{2011}{\natexlab{a}}), ISSN \bibinfo{issn}{1742-6596},
  \urlprefix\url{http://stacks.iop.org/1742-6596/321/i=1/a=012024?key=crossref.f1f3e72558970d2a054ae611b3d9c24f}.

\bibitem[{\citenamefont{Onsi et~al.}(2008)\citenamefont{Onsi, Dutta, Chatri,
  Goriely, Chamel, and Pearson}}]{Onsi2008}
\bibinfo{author}{\bibfnamefont{M.}~\bibnamefont{Onsi}},
  \bibinfo{author}{\bibfnamefont{A.}~\bibnamefont{Dutta}},
  \bibinfo{author}{\bibfnamefont{H.}~\bibnamefont{Chatri}},
  \bibinfo{author}{\bibfnamefont{S.}~\bibnamefont{Goriely}},
  \bibinfo{author}{\bibfnamefont{N.}~\bibnamefont{Chamel}}, \bibnamefont{and}
  \bibinfo{author}{\bibfnamefont{J.~M.} \bibnamefont{Pearson}},
  \bibinfo{journal}{Physical Review C} \textbf{\bibinfo{volume}{77}},
  \bibinfo{pages}{065805} (\bibinfo{year}{2008}), ISSN
  \bibinfo{issn}{0556-2813},
  \urlprefix\url{http://link.aps.org/doi/10.1103/PhysRevC.77.065805}.

\bibitem[{\citenamefont{Pearson et~al.}(2011)\citenamefont{Pearson, Goriely,
  and Chamel}}]{Pearson2011}
\bibinfo{author}{\bibfnamefont{J.~M.} \bibnamefont{Pearson}},
  \bibinfo{author}{\bibfnamefont{S.}~\bibnamefont{Goriely}}, \bibnamefont{and}
  \bibinfo{author}{\bibfnamefont{N.}~\bibnamefont{Chamel}},
  \bibinfo{journal}{Physical Review C} \textbf{\bibinfo{volume}{83}},
  \bibinfo{pages}{065810} (\bibinfo{year}{2011}), ISSN
  \bibinfo{issn}{0556-2813},
  \urlprefix\url{http://link.aps.org/doi/10.1103/PhysRevC.83.065810}.

\bibitem[{\citenamefont{Kucharek
  et~al.}(1989{\natexlab{a}})\citenamefont{Kucharek, Ring, Schuck, Bengtsson,
  and Girod}}]{Kucharek1989}
\bibinfo{author}{\bibfnamefont{H.}~\bibnamefont{Kucharek}},
  \bibinfo{author}{\bibfnamefont{P.}~\bibnamefont{Ring}},
  \bibinfo{author}{\bibfnamefont{P.}~\bibnamefont{Schuck}},
  \bibinfo{author}{\bibfnamefont{R.}~\bibnamefont{Bengtsson}},
  \bibnamefont{and} \bibinfo{author}{\bibfnamefont{M.}~\bibnamefont{Girod}},
  \bibinfo{journal}{Physics Letters B} \textbf{\bibinfo{volume}{216}},
  \bibinfo{pages}{249} (\bibinfo{year}{1989}{\natexlab{a}}).

\bibitem[{\citenamefont{Kucharek
  et~al.}(1989{\natexlab{b}})\citenamefont{Kucharek, Ring, and
  Schuck}}]{Kucharek1989a}
\bibinfo{author}{\bibfnamefont{H.}~\bibnamefont{Kucharek}},
  \bibinfo{author}{\bibfnamefont{P.}~\bibnamefont{Ring}}, \bibnamefont{and}
  \bibinfo{author}{\bibfnamefont{P.}~\bibnamefont{Schuck}},
  \bibinfo{journal}{Zeitschrift f\"{u}r Physik A}
  \textbf{\bibinfo{volume}{334}}, \bibinfo{pages}{119}
  (\bibinfo{year}{1989}{\natexlab{b}}).

\bibitem[{\citenamefont{Schuck and Vi\~{n}as}(2012)}]{Schuck2012}
\bibinfo{author}{\bibfnamefont{P.}~\bibnamefont{Schuck}} \bibnamefont{and}
  \bibinfo{author}{\bibfnamefont{X.}~\bibnamefont{Vi\~{n}as}}, in
  \emph{\bibinfo{booktitle}{50 Years of Nuclear BCS}}, edited by
  \bibinfo{editor}{\bibfnamefont{W.~S. P. C.~s.} \bibnamefont{{R.A.Broglia and
  V.Zelevinsky Editors}}} (\bibinfo{year}{2012}), p. \bibinfo{pages}{212}.

\bibitem[{\citenamefont{Dobaczewski et~al.}(1996)\citenamefont{Dobaczewski,
  Nazarewicz, Werner, Berger, Chinn, and Decharg\'{e}}}]{Dobaczewski2008a}
\bibinfo{author}{\bibfnamefont{J.}~\bibnamefont{Dobaczewski}},
  \bibinfo{author}{\bibfnamefont{W.}~\bibnamefont{Nazarewicz}},
  \bibinfo{author}{\bibfnamefont{T.~R.} \bibnamefont{Werner}},
  \bibinfo{author}{\bibfnamefont{J.-F.} \bibnamefont{Berger}},
  \bibinfo{author}{\bibfnamefont{C.~R.} \bibnamefont{Chinn}}, \bibnamefont{and}
  \bibinfo{author}{\bibfnamefont{J.}~\bibnamefont{Decharg\'{e}}},
  \bibinfo{journal}{Physical Review C} \textbf{\bibinfo{volume}{53}},
  \bibinfo{pages}{2809} (\bibinfo{year}{1996}).

\bibitem[{\citenamefont{Vi\~{n}as
  et~al.}(2011{\natexlab{b}})\citenamefont{Vi\~{n}as, Schuck, and
  Farine}}]{Vinas2011a}
\bibinfo{author}{\bibfnamefont{X.}~\bibnamefont{Vi\~{n}as}},
  \bibinfo{author}{\bibfnamefont{P.}~\bibnamefont{Schuck}}, \bibnamefont{and}
  \bibinfo{author}{\bibfnamefont{M.}~\bibnamefont{Farine}},
  \bibinfo{journal}{International Journal of Modern Physics E}
  \textbf{\bibinfo{volume}{20}}, \bibinfo{pages}{399}
  (\bibinfo{year}{2011}{\natexlab{b}}), ISSN \bibinfo{issn}{0218-3013},
  \urlprefix\url{http://www.worldscientific.com/doi/abs/10.1142/S0218301311017788}.

\bibitem[{\citenamefont{Baroni et~al.}(2010)\citenamefont{Baroni, Macchiavelli,
  and Schwenk}}]{Baroni2010}
\bibinfo{author}{\bibfnamefont{S.}~\bibnamefont{Baroni}},
  \bibinfo{author}{\bibfnamefont{A.~O.} \bibnamefont{Macchiavelli}},
  \bibnamefont{and} \bibinfo{author}{\bibfnamefont{A.}~\bibnamefont{Schwenk}},
  \bibinfo{journal}{Physical Review C} \textbf{\bibinfo{volume}{81}},
  \bibinfo{pages}{064308} (\bibinfo{year}{2010}), ISSN
  \bibinfo{issn}{0556-2813},
  \urlprefix\url{http://link.aps.org/doi/10.1103/PhysRevC.81.064308}.

\bibitem[{\citenamefont{Vi\~{n}as et~al.}(2003)\citenamefont{Vi\~{n}as, Schuck,
  Farine, and Centelles}}]{Vinas2003}
\bibinfo{author}{\bibfnamefont{X.}~\bibnamefont{Vi\~{n}as}},
  \bibinfo{author}{\bibfnamefont{P.}~\bibnamefont{Schuck}},
  \bibinfo{author}{\bibfnamefont{M.}~\bibnamefont{Farine}}, \bibnamefont{and}
  \bibinfo{author}{\bibfnamefont{M.}~\bibnamefont{Centelles}},
  \bibinfo{journal}{Physical Review C} \textbf{\bibinfo{volume}{67}},
  \bibinfo{pages}{054307} (\bibinfo{year}{2003}), ISSN
  \bibinfo{issn}{0556-2813},
  \urlprefix\url{http://link.aps.org/doi/10.1103/PhysRevC.67.054307}.

\end{thebibliography}

\end{document}